%% file: higher-reps.tex
\documentclass[11pt,a4paper]{article}
\usepackage{jheppub}
\pdfoutput=1

\input{preamble}

\usepackage{braket}

\title{Higher representations for extended operators}
\author{Thomas Bartsch, Mathew Bullimore, Andrea Grigoletto}
\affiliation{Department of Mathematical Sciences, Durham University, \\
Upper Mountjoy, Stockton Road, Durham, DH1 3LE, United Kingdom}
\emailAdd{thomas.d.bartsch@durham.ac.uk}
\emailAdd{mathew.r.bullimore@durham.ac.uk}
\emailAdd{andrea.grigoletto@durham.ac.uk}
\date{}

\abstract{It is known that local operators in quantum field theory transform in representations of ordinary global symmetry groups. The purpose of this paper is to generalise this statement to extended operators such as line and surface defects. We explain that $(n-1)$-dimensional operators transform in $n$-representations of a finite invertible or group-like symmetry and thoroughly explore this statement for $n = 1,2,3$. We therefore propose higher representation theory as the natural framework to describe the action of symmetries on the extended operator content in quantum field theory.}

\begin{document}
\maketitle

%%%%%%%%%%%%%%%%%%%%%%%%%%%%%%%%%%%%%%%%%%%%%%%%%%%
%%%%%%%%%%%%%%%%%%%%%%%%%%%%%%%%%%%%%%%%%%%%%%%%%%%
%%%%%%%%%%%%%%%%%%%%%%%%%%%%%%%%%%%%%%%%%%%%%%%%%%%
%%%%%%%%%%%%%%%%%%%%%%%%%%%%%%%%%%%%%%%%%%%%%%%%%%%
%%%%%%%%%%%%%%%%%%%%%%%%%%%%%%%%%%%%%%%%%%%%%%%%%%%
%%%%%%%%%%%%%%%%%%%%%%%%%%%%%%%%%%%%%%%%%%%%%%%%%%%

\section{Introduction}
\label{sec:intro}

It is well-known that local operators in quantum field theory transform in representations of an ordinary finite group symmetry $G$. The purpose of this paper is to generalise this statement to general finite invertible symmetries - see for instance~\cite{Gaiotto:2014kfa,Sharpe:2015mja,Bhardwaj:2017xup,Tachikawa:2017gyf,Benini:2018reh,Hsin:2018vcg, GarciaEtxebarria:2019caf,Eckhard:2019jgg,Bergman:2020ifi, Cordova:2020tij,DelZotto:2020esg,Bhardwaj:2020phs,Hsin:2020nts,Morrison:2020ool,Gukov:2020btk,Yu:2020twi,Bhardwaj:2021wif,Lee:2021crt,Apruzzi:2021vcu,Bhardwaj:2021pfz} - and their action on local and extended operators. 

The general proposal is as follows: $(n-1)$-dimensional extended operators transform in $n$-representations of finite higher-group symmetry $\mathcal{G}$. For example, line defects transform in 2-representations and surface defects in 3-representations. Importantly, this statement is independent of the ambient spacetime dimension $D$ as well as 't Hooft anomalies for the higher group symmetry $\mathcal{G}$. 

\begin{figure}[h]
	\centering
	\includegraphics[height=2.1cm]{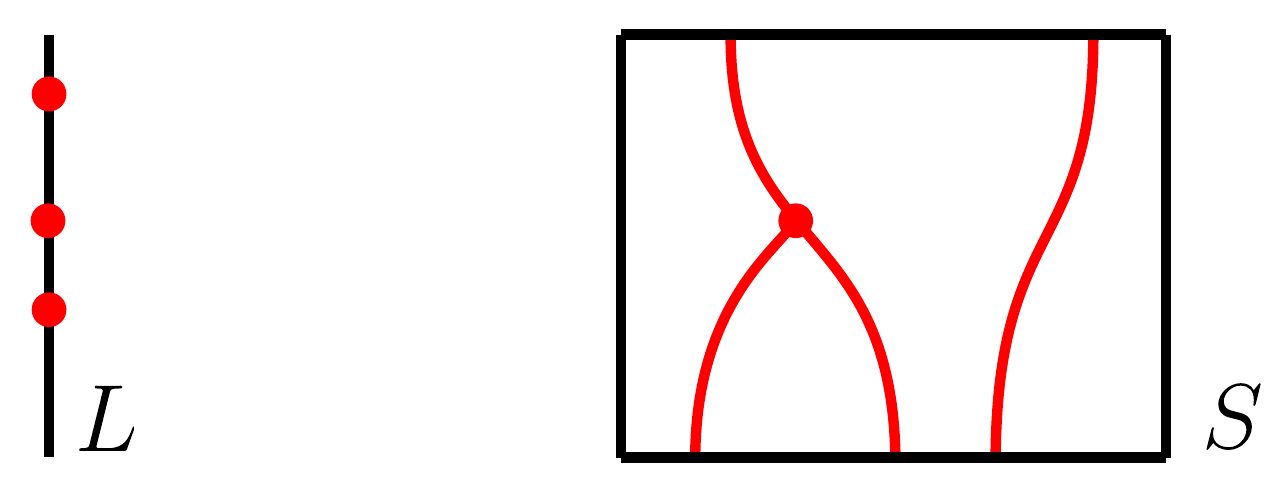}
	\vspace{-5pt}
	\caption{}
	\label{fig:introduction-2}
\end{figure}

The appearance of higher representations reflects the richer structure of topological defects supported on extended operators. While ordinary linear representations correspond to group actions on vector spaces, $n$-representations correspond to group actions on $n$-vector spaces, which encode the topological defects on an $(n-1)$-dimensional defect. Here are some examples:
\begin{itemize}
\item 2-vector spaces are finite semi-simple associative algebras and describe the topological local operators on a one-dimensional line defect $L$.
\item 3-vector spaces are multi-fusion categories and describe the topological line defects on a two-dimensional surface defect $S$. 
\end{itemize}
These examples are illustrated schematically in figure \ref{fig:introduction-2}. 

While the appearance of higher categorical structures is natural, the classification of higher representations often takes a concrete form that reflects clear physical phenomena. Let us first consider one-dimensional line defects in a theory with a finite group symmetry $G$. The irreducible 2-representations are classified by~\cite{GANTER20082268,ELGUETA200753,OSORNO2010369}:
\begin{enumerate}
\item A subgroup $H \subset G$.
\item A group cohomology class $c \in H^2(H,U(1))$.
\end{enumerate}
Line defects transforming in this 2-representation preserve the subgroup $H \subset G$ and support a 't Hooft anomaly captured by the SPT phase associated to $c$. As a consequence, they end on local operators transforming in projective representations of $H$ with Schur multiplier $c$. This has a straightforward extension to include 1-form and 2-group symmetries, which is spelled out in the main draft and provides an organising principle for many of the results presented in~\cite{Bhardwaj:2022dyt,Delmastro:2022pfo,Brennan:2022tyl,Bhardwaj:2023zix}.

Moving on to two-dimensional surface defects in a theory with finite group symmetry $G$, we find that irreducible 3-representations are classified by:
\begin{enumerate}
\item A subgroup $H \subset G$.
\item A fusion category $\mathsf{C}$ together with a homomorphism $\rho : H \to \text{Aut}(\mathsf{C})$ and a group cohomology class $f \in H^2(H,Z(\mathsf{C})^\times)$.
\item A group cohomology class $c \in H^3(H,U(1))$.
\end{enumerate} 
Surface defects transforming in this 3-representation preserve $H \subset G$ and support topological lines captured by $\mathsf{C}$. The remaining information in the second point determines the action of $H \subset G$ on the topological lines supported on the surface defect, while the third point describes a defect 't Hooft anomaly. We note that one-dimensional 3-representations with $H = G$ were classified in~\cite{etingof2010fusion}.

Throughout the paper, we consider a number of equivalent perspectives. The most elementary is to utilise the properties of topological symmetry defects generating the finite $n$-group symmetries and and their consistent interactions with extended operators to derive the data of $n$-representations. This leads fairly directly to the concrete data that classifies $n$-representations as presented above.

\begin{figure}[h]
	\centering
	\includegraphics[height=3cm]{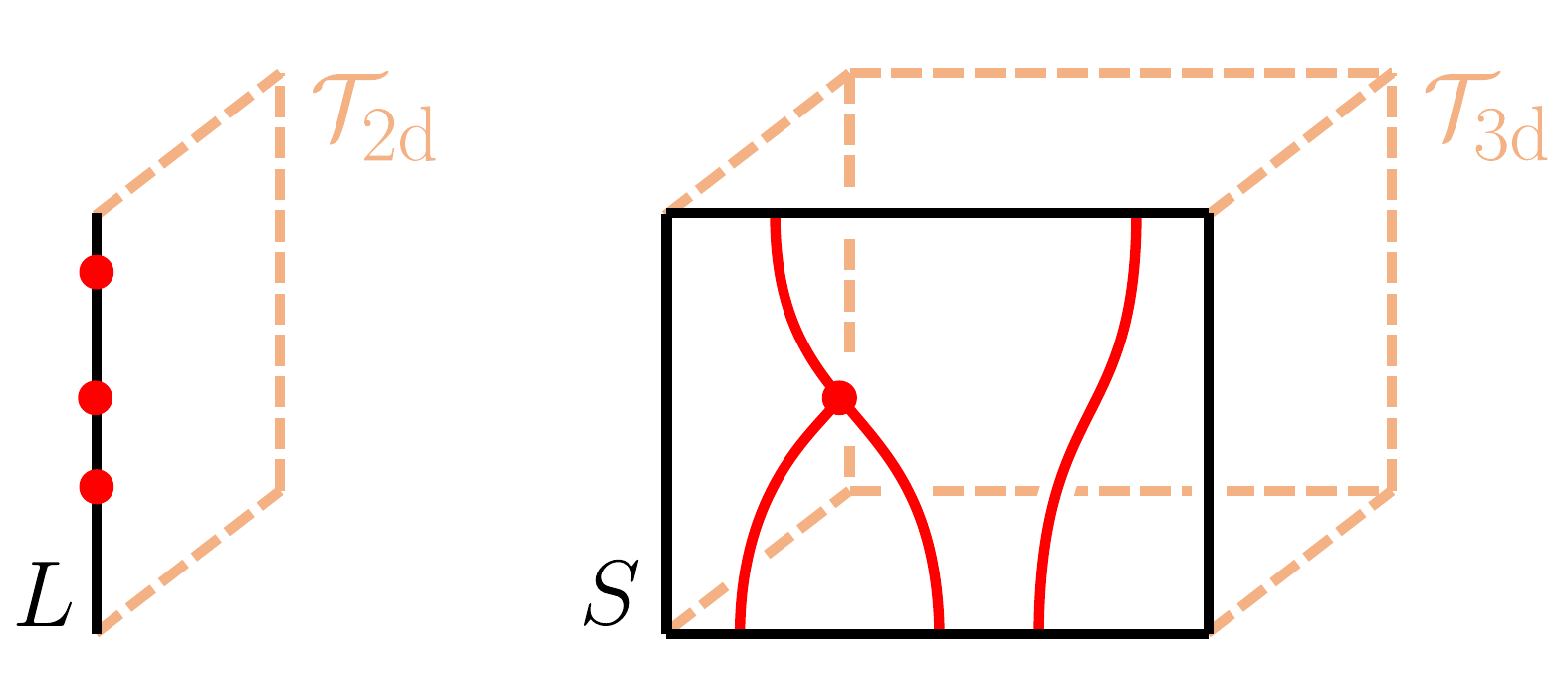}
	\vspace{-5pt}
	\caption{}
	\label{fig:introduction-3}
\end{figure}

However, in order to uncover the fundamental reason why $n$-representations appear, we also consider a more abstract approach. In this approach we image $(n-1)$-dimensional extended operators as gapped boundary conditions for an auxiliary attached $n$-dimensional TQFT, illustrated schematically in figure~\ref{fig:introduction-3}. The consistent interaction with topological symmetry defects then demands a $\cG$-equivariant structure on the $n$-dimensional TQFT, which is a functor
\be
\cG \to n \mathsf{Vec}
\ee
into the fusion $n$-category of finite-dimensional $n$-vector spaces. This reproduces the abstract mathematical definition of an $n$-representation. 

The attached $n$-dimensional TQFT may be regarded as an $n$-dimensional analogue of a background Wilson line. Indeed, upon gauging $\cG$, the attached $n$-dimensional TQFT becomes an $n$-dimensional analogue of a dynamical Wilson line: the topological defects generating the symmetry category $n \mathsf{Rep}(\cG)$, as described in~\cite{Bartsch:2022mpm,Bhardwaj:2022kot,Bartsch:2022ytj,Bhardwaj:2022lsg}. The $(n-1)$-dimensional charged operators then become twisted sector operators.

While this framework describes the transformation properties of any generic non-topological $(n-1)$-dimensional extended operators in terms of $n$-representations, it can be applied to extended topological objects which themselves generate a higher group $\mathcal{H}$. This can be used to capture mixed 't Hooft anomalies between $\mathcal{G}$ and $\mathcal{H}$. Therefore, the understanding of higher representation theory of higher groups provides a tool for the study of discrete 't Hooft anomalies.

Finally, in this paper we deal exclusively with charged extended operators supporting only bosonic topological operators. We expect a natural extension to $n$-representations on fermionic $n$-vector spaces: super-vector spaces, semi-simple super-algebras, super multi-fusion categories, etc., corresponding to extended operators supporting fermionic topological operators. Alternatively, a more abstract construction is in terms of $\cG$-equivariant fermionic $n$-dimensional TQFTs admitting gapped boundary conditions. We hope to return to this in future work.

\subsection{Outline}

The structure of the paper follows the increasing dimension of extended operators:
\begin{itemize}
\item In section~\ref{sec:locals}, we recap of the well-known result that local operators transform in representations of a finite symmetry group and frame this in a manner that allows for generalisations.
\item In section~\ref{sec:lines-groups}, we show that line operators transform in 2-representations of finite group symmetries. 
\item In section~\ref{sec:lines-2groups}, we generalise the previous section to show that line operators transform in 2-representations of finite 2-group symmetries. 
\item In section~\ref{sec:surfaces}, we consider how surface defects transform in 3-representations of finite group symmetries. The reader will be glad to hear that we stop here. 
\end{itemize}
In each section, we follow a consistent pattern. We first provide an elementary derivation of the data of $n$-representations using properties of topological symmetry defects and how they interact with $(n-1)$-dimensional extended operators. We then provide an equivalent description in terms of induction of $n$-representations. We then introduce a more categorical perspective by introducing auxiliary attached $n$-dimensional TQFTs, and then conclude with elementary examples from gauge theory and beyond. In particular, we hope the physically minded reader can appreciate all of the main points without touching the higher categorical framework if they wish.

\emph{Note added: after this work was completed but before submission, reference~\cite{Bhardwaj:2023wzd} appeared with overlapping results.}

\section{Local operators and 1-representations}
\label{sec:locals}

In this section, we revisit the well-known result that local operators transform in irreducible representations of a finite symmetry group. We begin by reviewing standard arguments using topological defects, as presented in~\cite{Gaiotto:2014kfa}, and proceed by reformulating this in a more categorical context that admits a smooth generalisation to extended operators. All of these considerations hold in any dimension $D$ and independently of possible 't Hooft anomalies for the involved symmetry groups.

\subsection{Elementary perspective}
\label{ssec:local-elementary}

\begin{figure}[h]
	\centering
	\includegraphics[height=3.4cm]{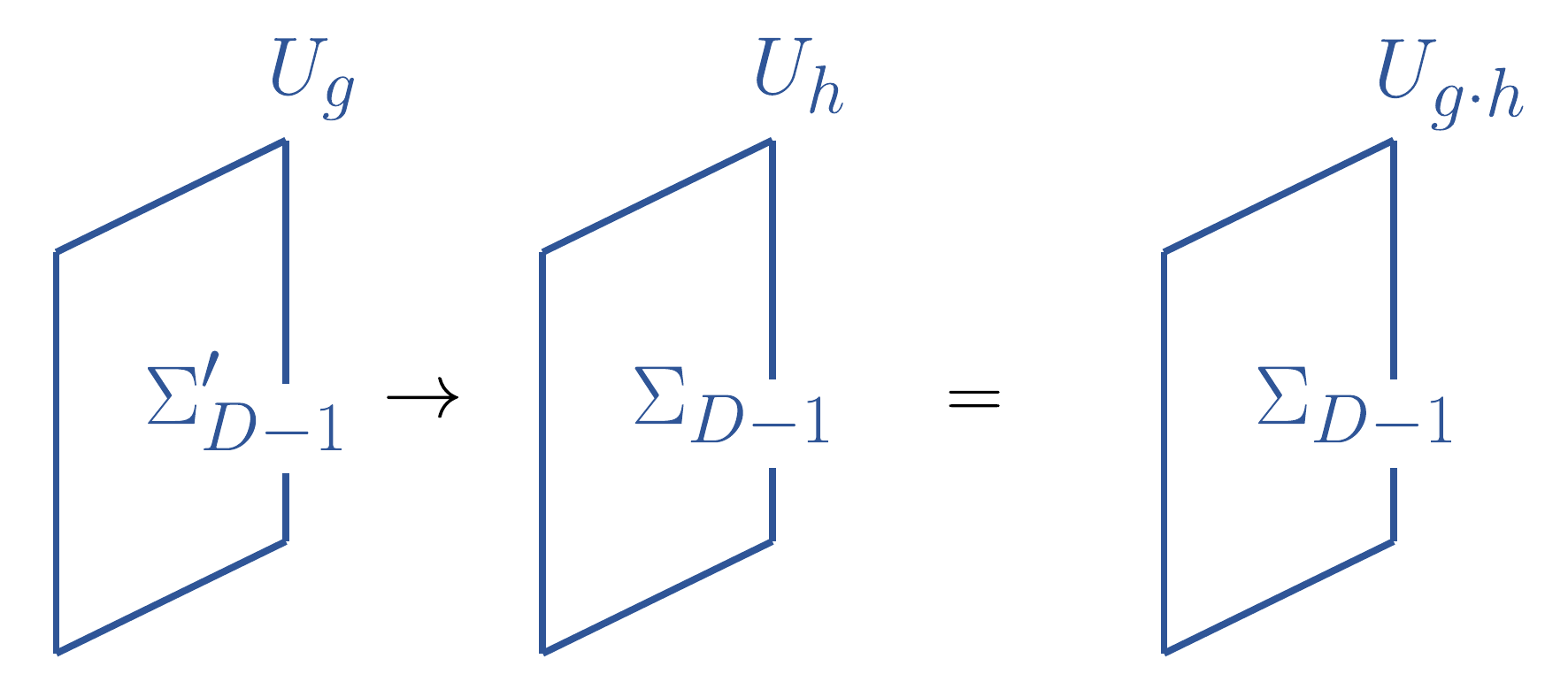}
	\vspace{-5pt}
	\caption{}
	\label{fig:0-form-fusion}
\end{figure}

In a $D$-dimensional quantum field theory with finite symmetry group $G$, there are topological defects $U_g(\Sigma_{D-1})$ labelled by group elements $g \in G$ and supported on codimension-one sub-manifolds $\Sigma_{D-1}$. These defects fuse according to the group law of $G$, as illustrated in figure \ref{fig:0-form-fusion}. 

\begin{figure}[h]
	\centering
	\includegraphics[height=3.8cm]{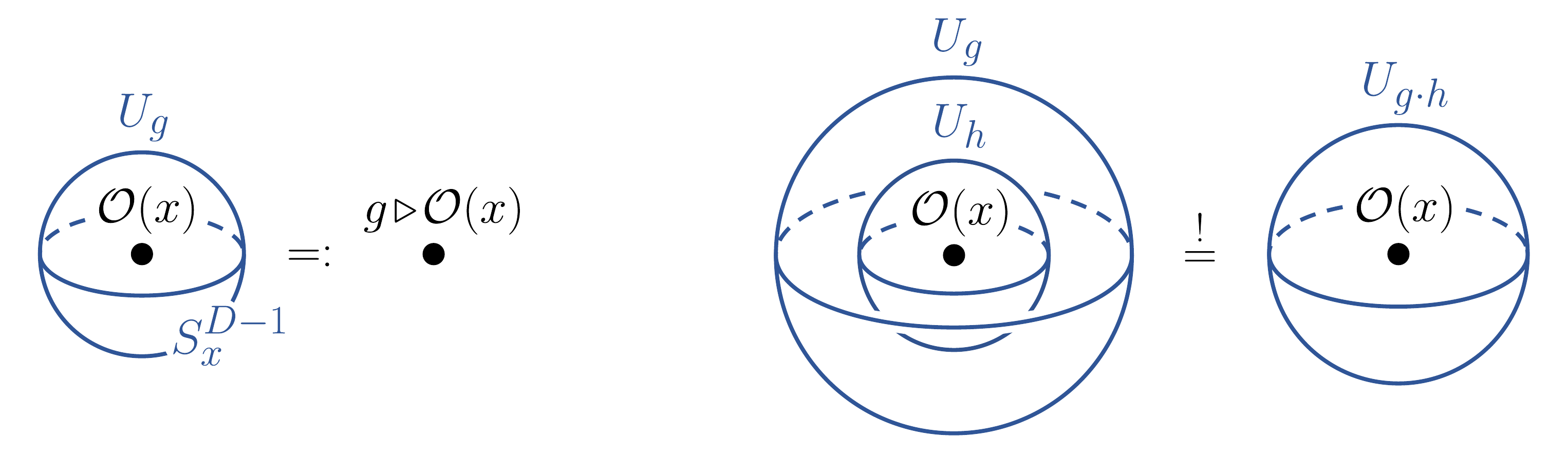}
	\vspace{-5pt}
	\caption{}
	\label{fig:local-action}
\end{figure}

We define the action of group elements $g \in G$ on local operators supported at a point $x$ by linking with the associated topological defect
\begin{equation}
g \,\triangleright\, \cO(x) \; := \; U_g(S^{D-1}_x) \, \cO(x) \, ,
\label{eq:action-operators}
\end{equation}
where $S^{D-1}_x$ denotes a small sphere linking the point $x$. The consecutive action of group elements $g,h \in G$ satisfies
\begin{equation} \label{eq:action-compatibility}
g \triangleright (h \triangleright \mathcal{O}(x)) \; = \; (g \cdot h) \,\triangleright\, \mathcal{O}(x) \, ,
\end{equation}
as a consequence of the fusion law of the topological defects. These properties are illustrated in figure \ref{fig:local-action}.

The collection of local operators at a given space-time point will typically generate an infinite-dimensional complex vector space. However, we may restrict ourselves to the study of finite-dimensional subspaces $V$ by fixing a local operator $\mathcal{O}(x)$ and defining
\begin{equation}
V \; := \; \text{span}_{\mathbb{C}}\lbrace \, g \triangleright \mathcal{O}(x) \; | \; g \in G \, \rbrace \, ,
\end{equation}
which is finite-dimensional due to the finiteness of $G$. The action (\ref{eq:action-operators}) of group elements on local operators then induces an irreducible linear representation $\rho : G \to \text{End}(V)$. In this way, local operators transform in irreducible representations of $G$.

We could of course choose a basis $\mathcal{O}_i(x)$ of local operators indexed by $i=1,...,n$. This then sets up an isomorphism $V \cong \mathbb{C}^n$ and presents the action of $G$ on the basis vectors by
\begin{equation}
g \,\triangleright\, \mathcal{O}_i(x) \; = \; \rho(g)_i{}^j \, \cO_j(x) \, ,
\end{equation}
where the $(n \times n)$-matrices $\rho(g) \in \text{GL}_n(\mathbb{C})$ satisfy
\begin{equation}
\rho(g)_i{}^j \, \rho(h)_j{}^k \; = \; \rho(g \cdot h)_i{}^k \, .
\end{equation} 
This description suffers from a redundancy that corresponds to choosing a different basis $\mathcal{O}'_i(x) = \tau_i^{\;j} \, \mathcal{O}_j(x)$. The resulting representation matrices $\rho'(g)$ are then related to the original ones by $\rho'(g) = \tau \cdot \rho(g) \cdot \tau^{-1}$ and thus define an equivalent irreducible representation.

\begin{figure}[h]
	\centering
	\includegraphics[height=5.5cm]{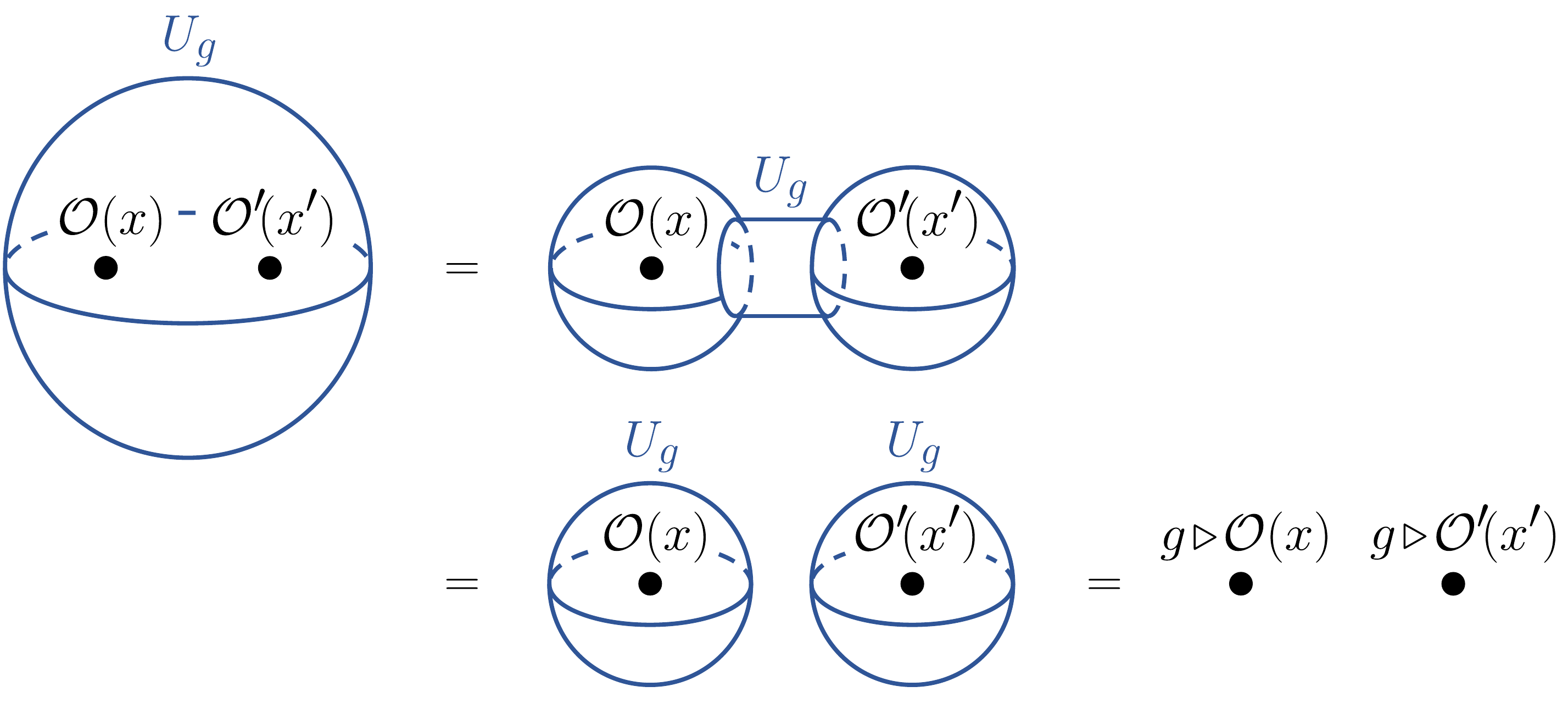}
	\vspace{-5pt}
	\caption{}
	\label{fig:tensor-product-local-action}
\end{figure}

So far we considered the action on local operators supported at a fixed spacetime point. More generally, we may consider collections of local operators supported at distinct points. For example, consider $\mathcal{O}(x)\, \mathcal{O}'(x')$ with $x \neq x'$. A group element $g \in G$ acts by linking with a sufficiently large sphere. The sequence of topological operations illustrated in figure \ref{fig:tensor-product-local-action} then reveals that this action is given by 
\begin{equation}\label{eq:tensor-action}
g \,\triangleright\, \big[ \mathcal{O}(x) \, \mathcal{O}'(x') \big] \; = \; \big[ g \triangleright \mathcal{O}(x) \big]  \big[ g \triangleright \mathcal{O}'(x')\big] \, .
\end{equation}
A straightforward argument shows that if $\cO(x)$ transforms in a representation $\rho$ and $\cO'(x')$ transforms in a representation $\rho'$ of $G$ then $\cO(x)\, \cO(x')$ transforms in the tensor product representation $\rho \otimes \rho'$. This extends in an obvious way to collections of operators at distinct points.

\subsection{Categorical perspective}

\begin{figure}[h]
	\centering
	\includegraphics[height=2.6cm]{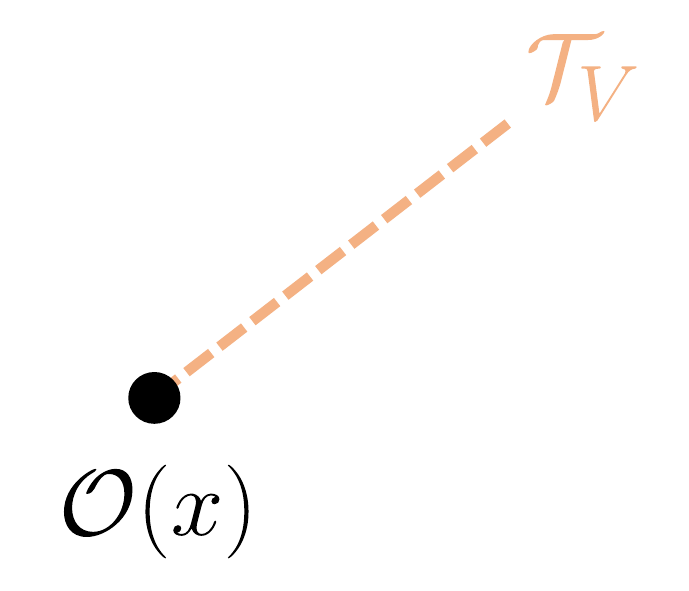}
	\vspace{-5pt}
	\caption{}
	\label{fig:1d-tqft-1}
\end{figure}

It is illuminating to reformulate the findings of the previous subsection in an alternative but equivalent way that provides a more categorical perspective. To do so, we again consider the vector space $V$ of genuine local operators supported at a spacetime point $x$ and regard it as the space of gapped boundary conditions for an auxiliary framed 1d topological quantum field theory\footnote{This can be viewed as a quantum mechanical system with zero Hamiltonian.} (TQFT) $\mathcal{T}_V$ as illustrated in figure \ref{fig:1d-tqft-1}.

The collection of framed 1d TQFTs forms a fusion category:
\begin{itemize}
\item Objects are 1d framed TQFTs, which are completely determined by their space of states or equivalently  gapped boundary conditions. The objects are thus identified with finite-dimensional complex vector spaces $V$.
\item Morphisms are topological interfaces between 1d framed TQFTs, which are linear maps $\varphi: V \to V'$ between vector spaces of gapped boundary conditions. In particular, topological local operators are endomorphisms $\varphi : V \to V$.
\end{itemize}
In other words, we can identify the category of framed 1d TQFTs with the category $\mathsf{Vec}$ of finite-dimensional complex vector spaces. The fusion of objects corresponds to the stacking of 1d TQFTs: $\cT_V \otimes \cT_W \, \cong \, \cT_{V \otimes W}$.

\begin{figure}[h]
	\centering
	\includegraphics[height=3.4cm]{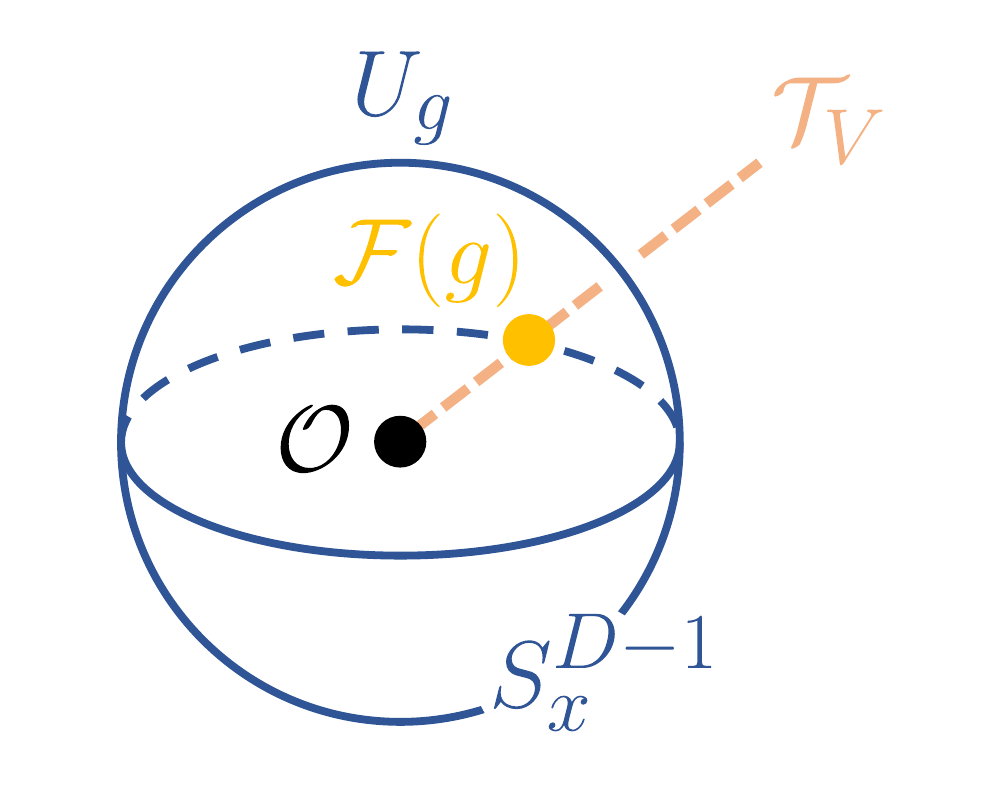}
	\vspace{-5pt}
	\caption{}
	\label{fig:1d-tqft-2}
\end{figure}

Using this picture, the action of the symmetry group $G$ on local operators $\mathcal{O} \in V$ is translated into a $G$-equivariant structure on the attached 1d framed TQFT $\mathcal{T}_V$. We will explain further below that this is nothing but a background topological Wilson line for the symmetry group $G$.

Concretely, consider linking a local operator $\mathcal{O}(x)$ with a topological defect $U_g(S^{D-1}_x)$. This now requires a choice of topological intersection between the topological defect $U_g$ and the 1d TQFT $\mathcal{T}_V$, as illustrated in figure \ref{fig:1d-tqft-2}. This may be regarded as a topological local operator or interface acting on topological boundary conditions in $\mathcal{T}_V$, which is a linear map $\cF(g): V \to V$.

\begin{figure}[h]
	\centering
	\includegraphics[height=3.8cm]{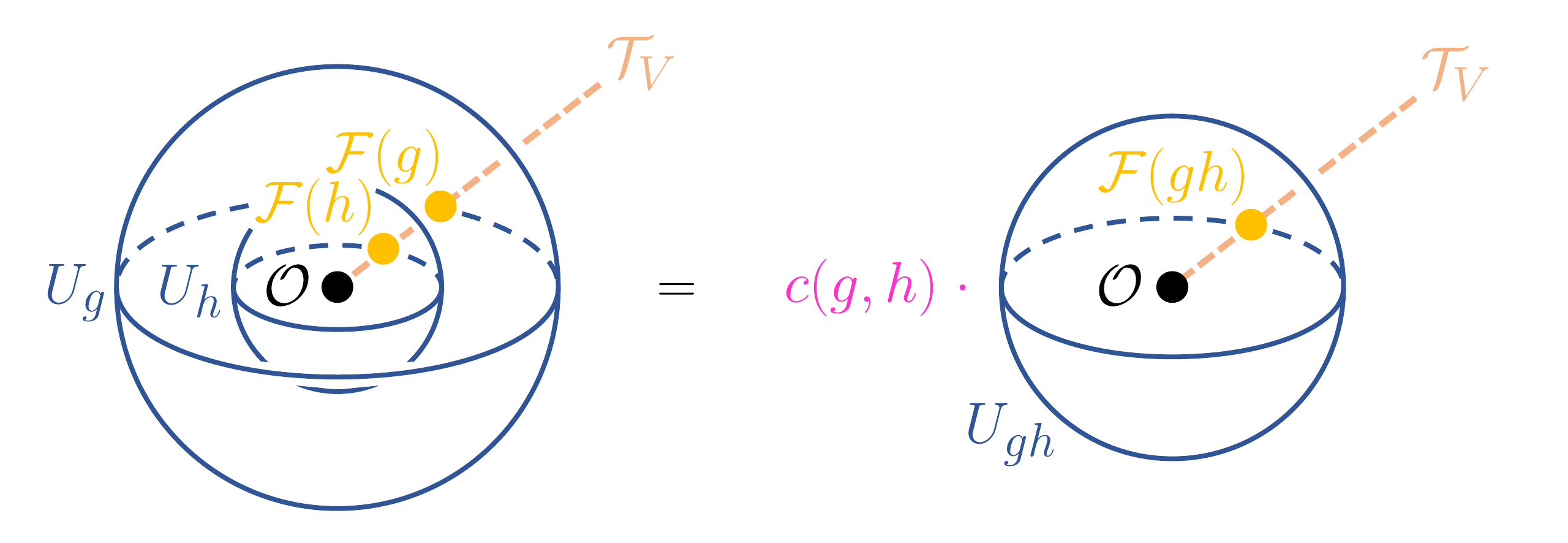}
	\vspace{-5pt}
	\caption{}
	\label{fig:1d-tqft-4}
\end{figure}

The topological intersections $\mathcal{F}(g)$ could respect the group law only projectively in the sense that intersecting with two symmetry defects $g,h \in G$ consecutively is equivalent to intersecting it with their fusion up to a multiplicative phase as illustrated in figure \ref{fig:1d-tqft-4}. The collection of such phases $c(g,h)$ may be regarded as a map
\begin{equation}
c: \; G \times G \; \to \; U(1) \, .
\end{equation}
For this to be compatible with the intersection of three symmetry defects $g,h,k \in G$, it needs to satisfy the 2-cocycle condition
\begin{equation}
(dc)(g,h,k) \; := \; \frac{c(h,k) \cdot c(g,hk)}{c(gh,k) \cdot c(g,h)} \; \stackrel{!}{=} \; 1 
\end{equation}
and defines a group cohomology class $[c] \in H^2(G,U(1))$.  If $[c] = 1$, the linear maps $\cF(g)$ can be renormalised to satisfy the compatibility condition
\begin{equation}\label{eq:hom-compatibility}
\mathcal{F}(g) \, \circ \, \mathcal{F}(h) \; \stackrel{!}{=} \; \mathcal{F}(g \cdot h) \, .
\end{equation}  
If $[c] \neq 1$, $\mathcal{T}_V$ would have to be regarded as the boundary of a 2d SPT phase. The class $[c] \in H^2(G,U(1))$ is therefore the obstruction for the maps $\mathcal{F}(g)$ to define consistent intersections compatible with the fusion of symmetry defects in the bulk. We require that the obstruction vanishes, $[c] = 1$.

Now shrinking the sphere implements the action of the group element $g$ on the corresponding local operator as before. The utility of this approach is that the data of the representation is abstracted away from the local operators and onto a $G$-equivariant structure on the 1d framed TQFT $\mathcal{T}_V$. 

In summary, the action of the symmetry group $G$ on genuine local operators in this framework can be described by the following data:
\begin{enumerate}
\item A 1d framed TQFT $\mathcal{T}_V$.
\item Topological interfaces $\mathcal{F}(g)$ for each group element $g \in G$ satisfying
 \begin{equation}
\mathcal{F}(g) \, \circ \, \mathcal{F}(h) \; = \; \mathcal{F}(g \cdot h) \, .
\end{equation}  
\end{enumerate}
This collection of data specifies a functor
\begin{equation}
\mathcal{F}: \; \widehat{G} \, \to \, \mathsf{Vec} \, ,
\end{equation}
where $\widehat{G}$ is the category with a single object $\ast$ and endomorphisms $\text{End}_{\widehat{G}}(\ast) = G$. The collection of such functors itself forms a category whose
\begin{itemize}
\item objects are functors $\mathcal{F}: \widehat{G} \to \mathsf{Vec}$,
\item morphisms are natural transformations $\alpha: \mathcal{F} \Rightarrow \mathcal{F}'$.
\end{itemize}
We denote this category by $[\widehat{G},\mathsf{Vec}]$ and recognise it as the category of finite-dimensional representations of $G$, 
\begin{equation}\label{eq:category-of-representations-of-G}
[\widehat{G}, \mathsf{Vec}] \, = \, \mathsf{Rep}(G) \, ,
\end{equation}
or, equivalently, the category of $G$-equivariant 1d framed TQFTs. This reproduces the result from the previous subsection that genuine local operators transform in finite-dimensio-nal representations of the symmetry group $G$. For a more detailed account of why the two categories in (\ref{eq:category-of-representations-of-G}) are equivalent see appendix \ref{app:1-representations}.

\begin{figure}[h]
	\centering
	\includegraphics[height=3.8cm]{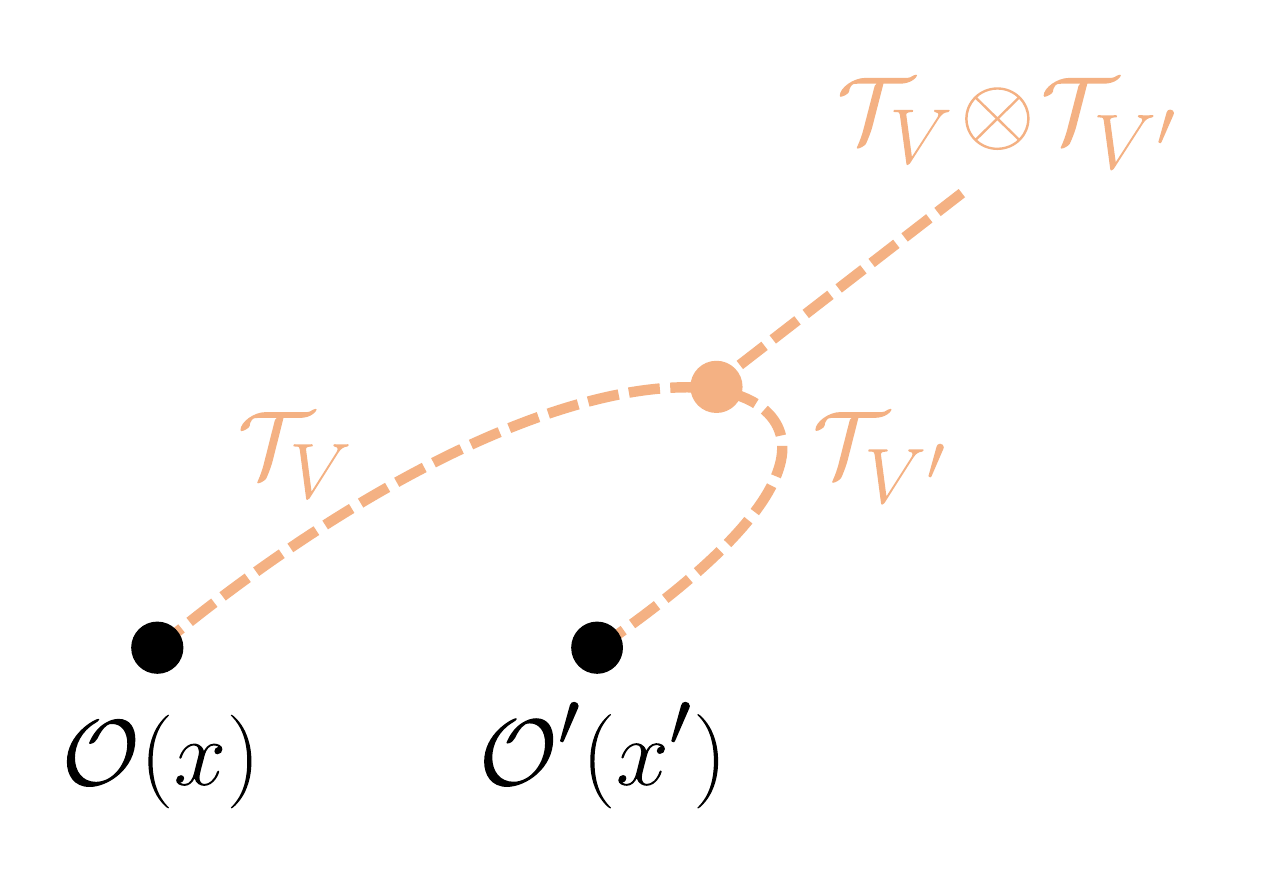}
	\vspace{-5pt}
	\caption{}
	\label{fig:1d-tqft-3}
\end{figure}

The transformation of two genuine local operators $\mathcal{O}(x)$ and $\mathcal{O}'(x')$ at separated spacetime points $x$ and $x'$ can be reframed as the stacking of the associated $G$-equivariant TQFTs $\mathcal{T}_V$ and $\mathcal{T}_{V'}$, which induces a fusion structure on the category $[\widehat{G},\mathsf{Vec}]$ as illustrated in figure \ref{fig:1d-tqft-3}. This corresponds to the ordinary tensor product of representations and intertwiners, so we reproduce the standard result that local operators at separated spacetime points transform in tensor product representations of the symmetry group $G$.

\subsubsection{Background Wilson lines and Gauging}

The introduction of an auxiliary G-equivariant 1d TQFT naturally uncovers a categorical perspective on representation theory but may seem somewhat formal. However, we will now explain that it admits a clean physical interpretation of a background Wilson line for the symmetry group $G$.

To understand this perspective, introduce a classical background field corresponding to flat connection $\mathsf{A}$ on a $G$-bundle. Correlation functions of a genuine local operator at a point $x$ transforming in a non-trivial irreducible representation $\rho$ of $G$ are not invariant under background gauge transformations. However, they are invariant after attaching the operator to a background Wilson line
\be
W_\rho(\gamma_x) \, = \, \text{Tr}_\rho (\text{Hol}_\gamma(\mathsf{A})) \, , 
\ee
where $\gamma_x$ denotes a curve ending at the point $x$. Correlation functions are invariant under deformations of the curve provided it does not encounter other charged operators. This background Wilson line is a non-dynamical topological operator that is trivial in the absence of background fields.

The introduction of a background flat connection is equivalent to inserting a network of symmetry defects $U_g(\Sigma_{D-1})$ representing constant transition functions on overlaps of patches~\cite{Frohlich:2006ch,Davydov:2010rm,Brunner:2014lua,Gaiotto:2014kfa}. Under this correspondence, the specification of a $G$-equivariant 1d TQFT, namely a 1d TQFT together with consistent intersections with the topological symmetry defects $U_g(\Sigma_{D-1})$, is an equivalent description of the background Wilson line $W_\rho$. 

Upon gauging the symmetry $G$ and summing over background flat connections, the background Wilson lines become dynamical topological Wilson lines generating the symmetry category $\mathsf{Rep}(G)$. Local operators transforming in an irreducible representation $\rho$ become twisted sector local operators attached to the topological line $W_\rho$.

\subsection{Example}

Let us conclude this section with a simple well-known example of local operators transforming in irreducible representation of a global symmetry group. Consider a gauge theory in $D=3$ with a connected simple gauge group $\mathbb{G}$. It has abelian magnetic 0-form symmetry $G = \pi_1(\mathbb{G})^{\vee}$ given by the Pontryagin dual of the fundamental group of $\mathbb{G}$.

The local operators charged under this symmetry are monopole operators. Consider a monopole operator at a fixed spacetime point $x$ linked by a small 2-sphere $S_x^2$. The magnetic symmetry measures the topological type of the $\mathbb{G}$-bundle generated by the monopole operator on $S_x^2$. This is classified by
\be
H^2(S^2_x,\pi_1(\mathbb{G})) \, \cong \, \pi_1(\mathbb{G})
\ee
and may be regarded as the transition function on the equator of $S^2_x$. The action of a group element $g \in \pi_1(\mathbb{G})^\vee$ on a monopole operator of topological type $p \in \pi_1(\mathbb{G})$ is then
\begin{equation}
    g \, \triangleright \, M_p(x) \; =  \; g(p) \, \cdot \, M_p(x) \, ,
\end{equation}
where we view the group element $g$ as a homomorphism $g: \pi_1(\mathbb{G}) \to U(1)$.

In this example, the auxiliary $G$-equivariant 1d TQFT in the categorical perspective, or equivalently the background Wilson line for the topological symmetry, is nothing but a Dirac string attached to the monopole operator.

\section{Lines and 2-representations of groups}
\label{sec:lines-groups}

In this section, we generalise the results from the previous section to show that line defects transform in 2-representations of a finite group symmetry. We emphasise that this conclusion applies to genuine non-topological line defects. We explain how it incorporates a number of known statements about line defects, such as the fact that local operators at their junctions transform in projective representations of the symmetry group as well as the phenomenon of symmetry fractionalisation.

We start with an elementary approach using properties of topological symmetry defects to construct the data of a 2-representation, before reformulating this in a more physical way that uncovers the mathematical notion of induction of 2-representations. We then rephrase the problem in a more categorical context that manifests the abstract definition of 2-representations of finite groups. Finally, we conclude with some examples. All of these considerations hold in any dimension $D$ and independently of 't Hooft anomalies for the symmetries.

\subsection{Preliminaries}
\label{subsec:lines-prelim}

\begin{figure}[h]
	\centering
	\includegraphics[height=2.9cm]{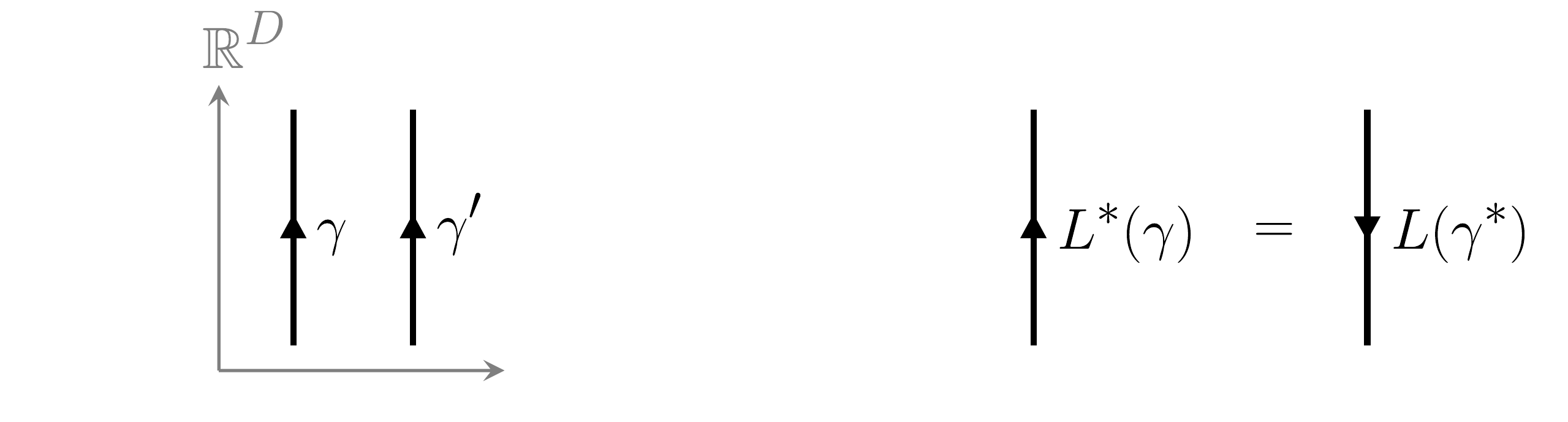}
	\vspace{-5pt}
	\caption{}
	\label{fig:line-preliminaries}
\end{figure}

We will consider line defects supported on oriented lines $\gamma,\gamma',\ldots$ aligned along a common axis in $D$-dimensional euclidean space-time $\mathbb{R}^D$. We draw this common axis vertically as illustrated on the left-hand side of figure~\ref{fig:line-preliminaries}. 

Given a line defect $L$, we may define a conjugate line $L^*$ by
\be
L^*(\gamma) \; := \; L(\gamma^*) \, ,
\ee
where $\gamma^*$ denotes the orientation reversal of $\gamma$ as illustrated on the right-hand side of figure \ref{fig:line-preliminaries}. We will typically omit the orientation from figures and implicitly assume an upward orientation of lines henceforth, unless stated otherwise.

A line defect $L(\gamma)$ may support topological local operators whose correlation functions are independent of their position on the line $\gamma$. The fusion of topological local operators generates an associative algebra $\cA_L$ associated to the line defect, which we will assume to be finite and semi-simple. As a consequence, it may be decomposed as a direct sum 
\be
\cA_L \; = \; \text{End}(V_1) \, \, \oplus \, \cdots \, \oplus \, \text{End}(V_s)
\ee
for some collection of finite-dimensional complex vector spaces $V_1,\ldots,V_s$. This motivates the introduction of notions of simple and reduced line defects:

\begin{itemize}
\item A line defect $L$ is \textit{simple} if it supports a simple algebra of topological local operators. This means $\cA_L = \text{End}(V)$ for some finite-dimensional vector space $V$. If a line defect $L$ is not simple, it admits a decomposition as a direct sum 
\be
L \; = \; L_1 \, \oplus \, \cdots \, \oplus \, L_s
\ee
of simple line defects with $\cA_{L_j} = \text{End}(V_j)$, where the direct sum of line defects is defined by the addition of their correlation functions. We may therefore restrict attention to simple line defects in what follows.

\item A simple line defect $L$ is \textit{reduced} if it only supports the identity operator as a topological operator, namely $\cA_{L} = \C$. A simple line defect $L'$ always admits a topological junction with a reduced simple line $L$, as illustrated in figure~\ref{fig:line-preliminaries-2}. The existence of this topological junction reflects the fact that the simple algebras $\cA_{L'} = \text{End}(V)$ and $\cA_{L} = \C$ are Morita-equivalent.

\begin{figure}[h]
	\centering
	\includegraphics[height=3cm]{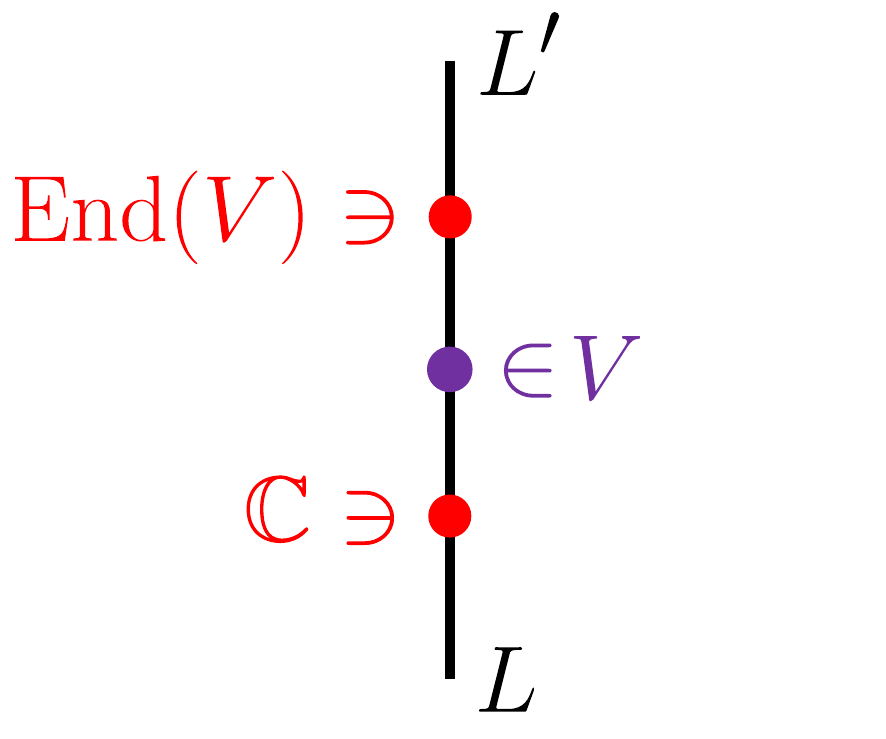}
	\vspace{-5pt}
	\caption{}
	\label{fig:line-preliminaries-2}
\end{figure}

As an example, consider stacking a reduced simple line defect $L$ with a 1d TQFT $\mathcal{T}_V$. The resulting simple line defect $L' = \mathcal{T}_V \otimes L$ supports a simple algebra $\mathcal{A}_{L'} = \text{End}(V)$ and admits topological junctions to $L$ labelled by elements of the vector space $V$, which arise from topological boundary conditions of the 1d TQFT.
\end{itemize}

Following these considerations, we may without loss of generality assume that, up to equivalence, all line operators are simple and reduced in the above sense. We will return to this issue from a more categorical perspective in section~\ref{ssec:lines-categorical-groups}.

\subsection{Elementary perspective}
\label{subsec:lines-groups-elementary}

\begin{figure}[h]
	\centering
	\includegraphics[height=3.9cm]{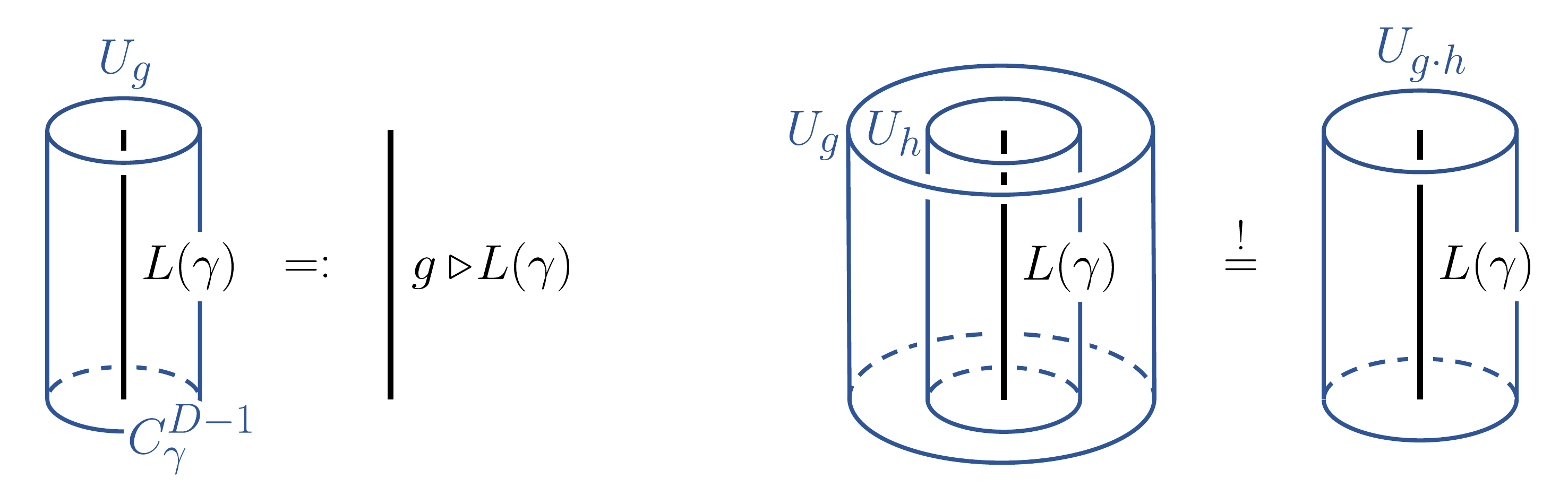}
	\vspace{-5pt}
	\caption{}
	\label{fig:line-action}
\end{figure}

Let us begin, as before, by considering a $D$-dimensional quantum field theory with a finite group symmetry $G$, generated by codimension-one topological defects $U_g(\Sigma_{D-1})$ labelled by group elements $g \in G$. We define an action of group elements $g \in G$ on line defects $L$ supported on a line $\gamma$ by
\begin{equation}\label{action-lines}
g \,\triangleright\, L(\gamma) \; := \; U_g(C_{\gamma}^{D-1}) \; L(\gamma) \, ,
\end{equation}
where $C_{\gamma}^{D-1} = \mathbb{R} \times S^{D-2}_{\gamma}$ denotes a small cylinder wrapping $\gamma$. This action satisfies the composition
\begin{equation}\label{action-compatibility-2}
g \triangleright (h \triangleright L(\gamma)) \; = \; (g \cdot h) \,\triangleright\, L(\gamma) 
\end{equation}
as a consequence of the fusion property of the topological defects. This definition and composition propeerty are illustrated in figure \ref{fig:line-action}.

The collection of line defects supported on a line $\gamma$ may form a discrete set of infinite cardinality or continuum or both. However, we may restrict ourselves to the study of finite subsets $\mathcal{S}$ by fixing a given line defect $L(\gamma)$ and defining
\begin{equation}
\mathcal{S} \; := \; \lbrace \, g \triangleright L(\gamma) \; | \; g \in G \, \rbrace \, ,
\end{equation}
which is a finite set due to the finiteness of $G$. The group action above then determines an irreducible permutation representation of $G$ on $\mathcal{S}$ or equivalently the structure of a transitive $G$-set on $\cS$. 

To be more concrete, we may choose to label line defects $L_i(\gamma) \in \mathcal{S}$ by a finite index $i=1,...,n$, which sets up a bijection $\mathcal{S} \cong \lbrace 1,...,n \rbrace$. The action (\ref{action-lines}) is then implemented by permutations $\sigma_g \in S_n$ such that
\begin{equation}
g \, \triangleright \, L_i(\gamma) \; = \; L_{\sigma_g(i)}(\gamma) \, .
\end{equation}
In particular, condition (\ref{action-compatibility-2}) ensures that these permutations satisfy
\begin{equation}
\sigma_g \, \circ \, \sigma_h \; = \; \sigma_{g \cdot h} \, ,
\end{equation}
so that $\sigma: G \to S_n$ defines a transitive permutation action of $G$ on the set of $n$ elements.  

\begin{figure}[h]
	\centering
	\includegraphics[height=4.2cm]{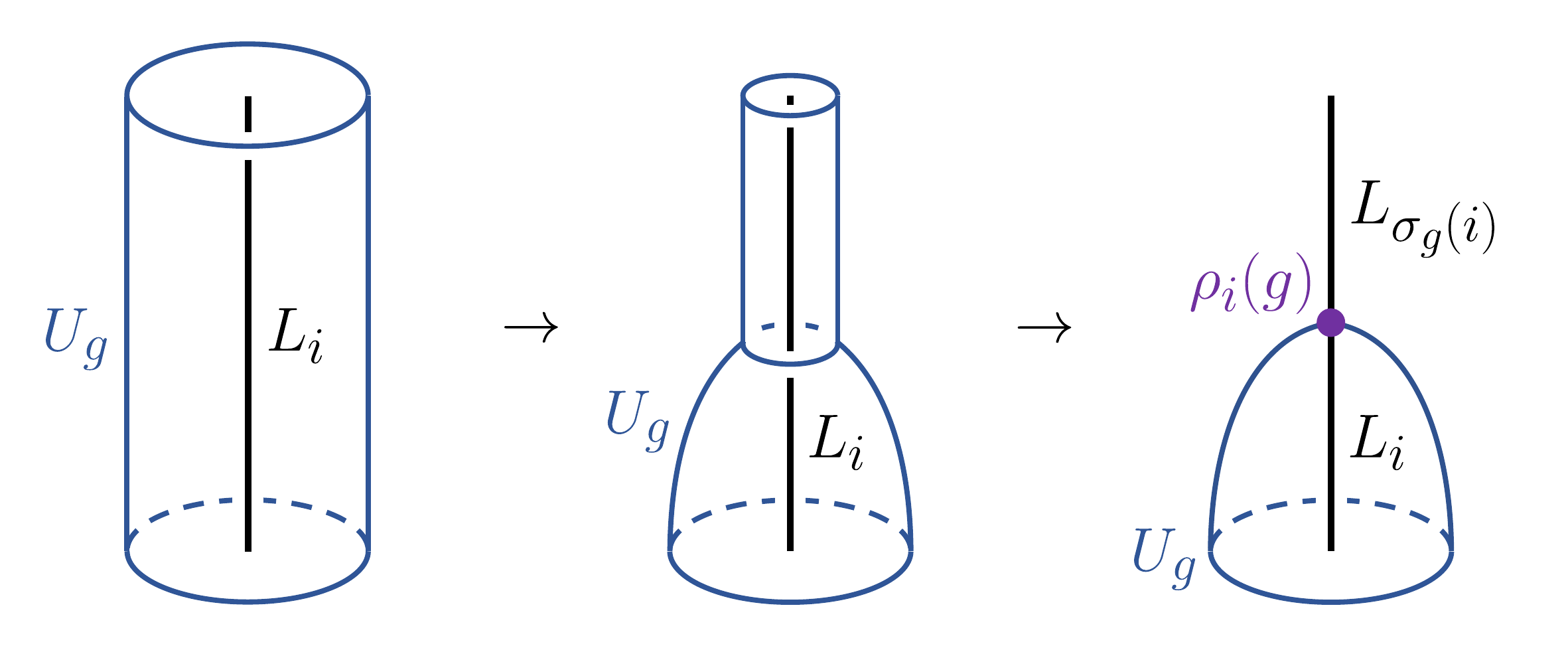}
	\vspace{-5pt}
	\caption{}
	\label{fig:intersection-operators}
\end{figure}

\begin{figure}[h]
	\centering
	\includegraphics[height=4.2cm]{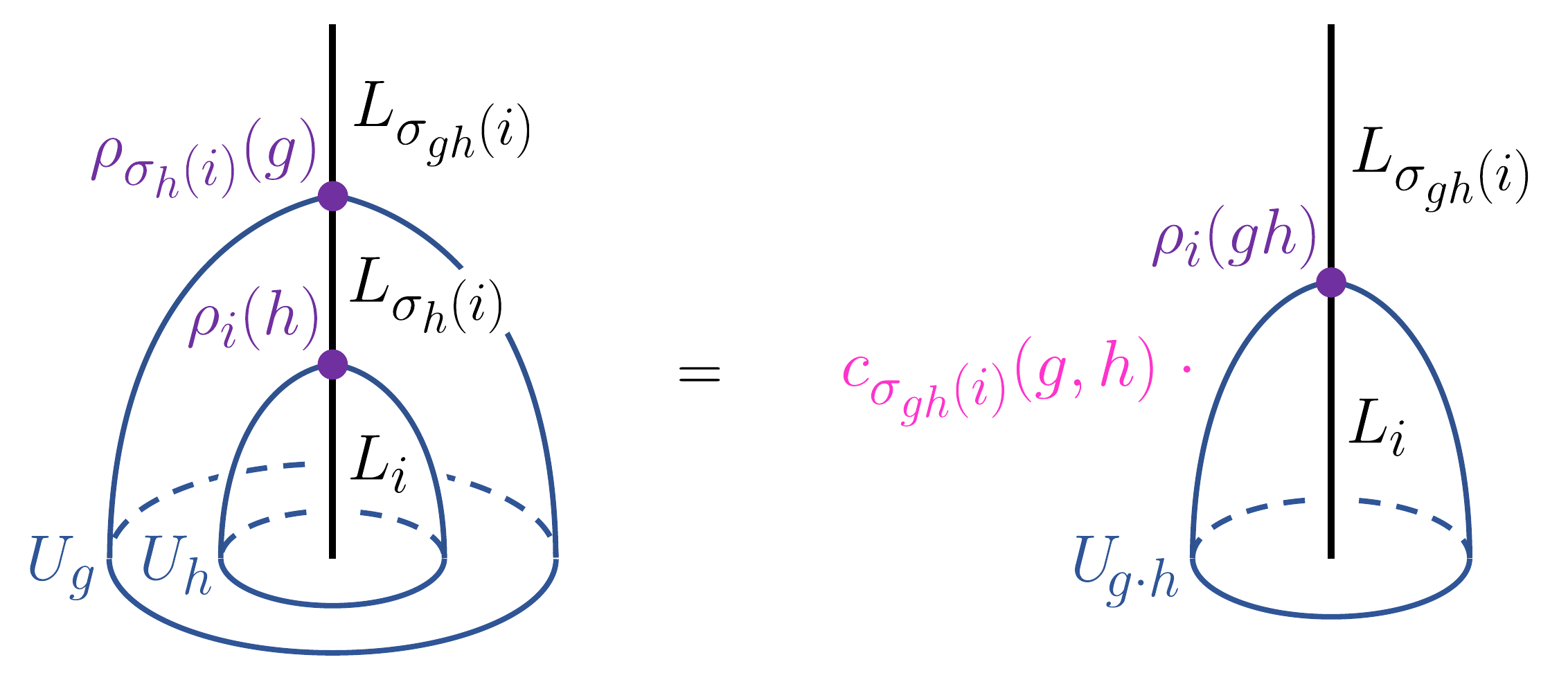}
	\vspace{-5pt}
	\caption{}
	\label{fig:projective-line-action}
\end{figure}

To proceed further, we consider the sequence of operations shown in figure \ref{fig:intersection-operators}. This illustrates that the topological defects $g \in G$ intersect each line $L_i$ at a unique topological junction operator $\rho_i(g)$ connecting it to the transformed line $L_{\sigma_g(i)}$.
While the topological defects in the bulk fuse according to the group law, the junction operators $\rho_i(g)$ may transform projectively, meaning
\begin{equation}
\rho_{\sigma_h(i)}(g) \, \circ \, \rho_i(h) \; = \; c_{\sigma_{gh}(i)}(g,h) \, \cdot \, \rho_i(g \cdot h)
\end{equation}
for some phases $c_i(g,h) \in U(1)$.
In other words, intersecting the line $L_i$ with two symmetry defects $g$ and $h$ consecutively is equivalent to intersecting it with their fusion $g \cdot h$ up to a multiplicative phase, as illustrated in figure \ref{fig:projective-line-action}.
This collection of phases may be viewed as a function
\begin{equation}
c: \; G \times G \, \to \, U(1)^n \, ,
\end{equation}
which, in order to be compatible with the fusion of three symmetry defects $g,h,k \in G$, must satisfy the twisted 2-cocycle condition
\begin{equation}\label{eq:twisted-2-cocycle-condition}
(d_{\sigma} c)_i\,(g,h,k) \;\; := \;\; \frac{c_{\sigma^{-1}_g(i)}(h,k) \cdot c_i(g,hk)}{c_i(gh,k) \cdot c_i(g,h)} \;\; \stackrel{!}{=} \;\; 1 \, .
\end{equation}
The function $c$ therefore defines a twisted 2-cocycle $c \in Z_{\sigma}^2(G,U(1)^n)$, where $U(1)^n$ is regarded as a $G$-module via the permutation action $\sigma$.

To summarise, the action of the symmetry group $G$ on a collection of line defects $L_i$ indexed by $i = 1,\ldots, n$ is specified by a pair $(\sigma,c)$ consisting of
\begin{enumerate}
\item A transitive permutation representation $\sigma : G \to S_n$.
\item A twisted 2-cocycle $c \in Z^2_\sigma(G,U(1)^n)$.
\end{enumerate}
This is precisely the data of an irreducible 2-representation of $G$ of dimension $n$. A general 2-representation is obtained by dropping the requirement of a transitive action. A more mathematical treatment of 2-representations of groups can be found in appendix \ref{app:2-representations}.

The pair $(\sigma,c)$ is subject to some redundancy corresponding to equivalent 2-represen-tations of $G$. First, we could have chosen an alternative labelling of the line defects
\begin{equation}
L_i' \; = \; L_{\tau^{-1}(i)}
\end{equation}
for some permutation $\tau \in S_n$, with associated permutation action
\begin{equation}
\sigma'_g \; = \; \tau \, \circ \, \sigma_g \, \circ \, \tau^{-1} \, .
\end{equation}
Similarly, the topological junctions describing the intersection of topological symmetry defects with the newly labelled line defects are given by
\begin{equation}
\rho_i'(g) \; = \; \rho_{\tau^{-1}(i)}(g) \, ,
\end{equation}
and their composition respects the group multiplication up to phases
\begin{equation}\label{eq:permutation-action-cocycle}
c'_i(g,h) \; = \; c_{\tau^{-1}(i)}(g,h) \, .
\end{equation}
The data $(\sigma',c')$ defines an equivalent 2-representation as expected since this is simply a relabelling of the line defects.

In addition to relabelling, we could also choose to redefine the junctions by
\begin{equation}
\rho'_i(g) \; \to \; b_{\sigma_g(i)}(g) \cdot \rho'_i(g)
\end{equation}
for some multiplicative phases $b_i(g) \in U(1)$, which we view as a function $b: G \to U(1)^n$. As a consequence, the projective phases are shifted by
\begin{equation}
c_i'(g,h) \; \to \; c_i'(g,h) \, \cdot \, (d_{\sigma'}b)_i(g,h) \, ,
\end{equation}
where we defined the twisted coboundary
\begin{equation}
(d_{\sigma'}b)_i\,(g,h) \;\; := \;\; \frac{b_{(\sigma'_g)^{-1}(i)}(h) \cdot b_i(g)}{b_i(gh)} \, .
\end{equation}
This corresponds to choosing another representative of the class $[c'] \in H_{\sigma'}^2(G,U(1)^n)$ and again is an equivalent 2-representation.

To summarise, pairs of 2-representations $(\sigma,c)$ and $(\sigma',c')$ are equivalent if there exists a permutation $\tau \in S_n$ such that
\begin{equation}\label{eq:equivalent-2reps}
\sigma' \; = \; \tau \circ \sigma \circ \tau^{-1} \qquad \text{and} \qquad [c'] \, = \, [\tau \triangleright c] \, ,
\end{equation}
where $\tau \triangleright c$ denotes the permutation action of $\tau$ on $c$ as in (\ref{eq:permutation-action-cocycle}) and $[.] : Z^2_{\sigma'}(G,U(1)^n) \to H^2_{\sigma'}(G,U(1)^n)$ denotes the projection onto twisted group cohomology.  This agrees with the mathematical notion of equivalence of 2-representations.

\subsubsection{Orientation reversal}

So far we implicitly assumed line operators to be oriented upwards with respect to a common vertical axis. Let us now examine how changing the orientation of a line affects its transformation behaviour under the finite group symmetry $G$.

Concretely, consider line operators $L_i(\gamma)$ supported on an upward-oriented line $\gamma$ and transforming in an irreducible 2-representation $(\sigma,c)$ of $G$. The conjugated line operators $L^{\ast}_i(\gamma) = L_i(\gamma^{\ast})$ are obtained by placing $L_i$ on the orientation-reversed line $\gamma^{\ast}$. As before, symmetry defects $g \in G$ intersect each line operator $L^{\ast}_i$ at a unique topological junction $\Bar{\rho}_i(g)$ connecting it to the transformed line $L^{\ast}_{\sigma_g(i)}$. 
\begin{figure}[h]
	\centering
	\includegraphics[height=8.4cm]{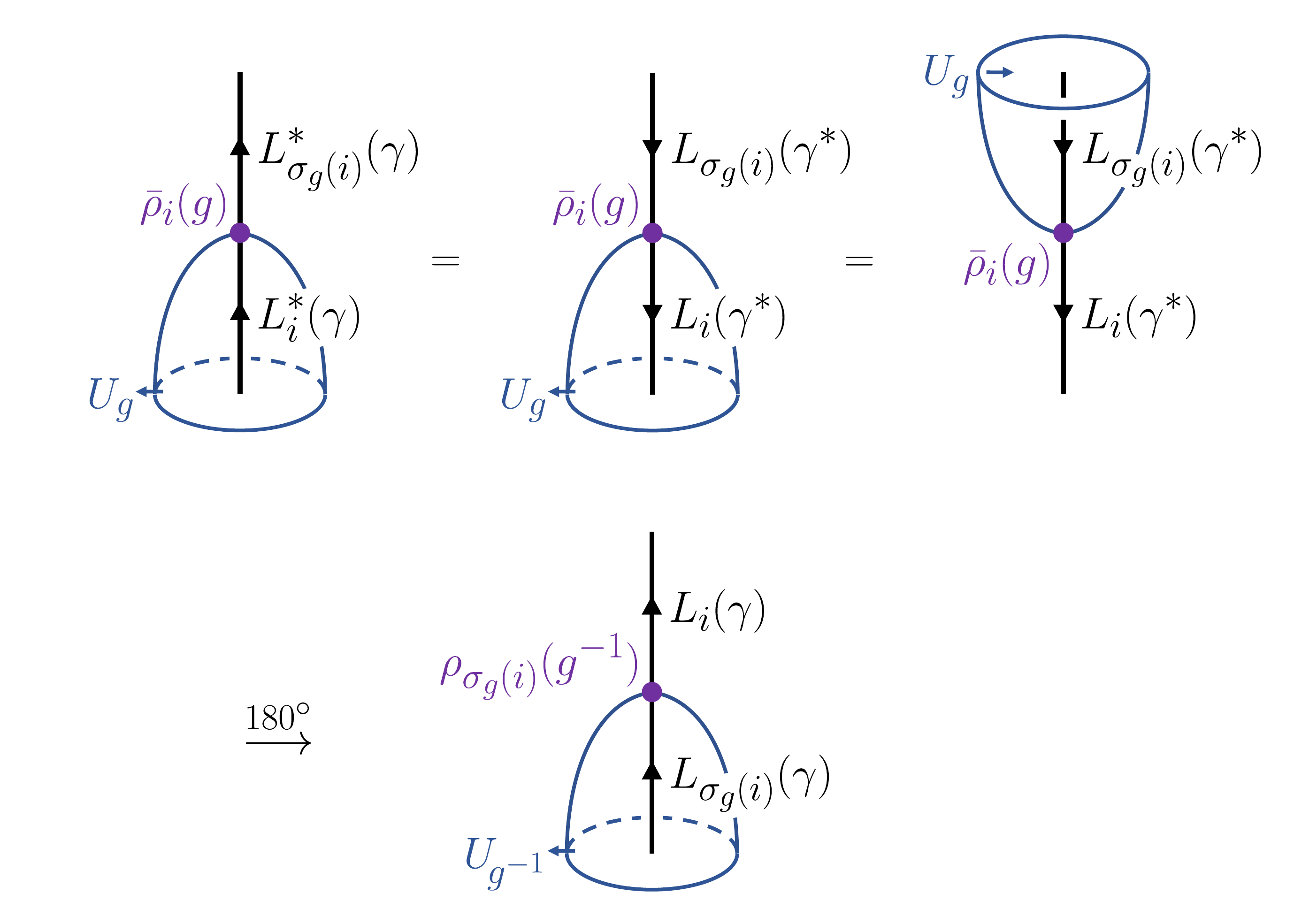}
	\vspace{-5pt}
	\caption{}
	\label{fig:orientation-reversal}
\end{figure}
The sequence of topological moves illustrated in figure \ref{fig:orientation-reversal} then illustrates that we can identify the junctions $\Bar{\rho}_i(g)$ with the junction operators $\rho_{\sigma_g(i)}(g^{-1})$ between the unconjugated lines $L_i$. Consequently, they compose projectively in the sense that
\begin{equation}
    \Bar{\rho}_{\sigma_h(i)}(g) \, \circ \, \Bar{\rho}_i(h) \; = \; \Bar{c}_{\sigma_{gh}(i)}(g,h) \, \cdot \, \Bar{\rho}_i(gh) \, ,
\end{equation}
where the projective phases are given by
\begin{equation}
    \Bar{c}_i(g,h) \; = \; c_{\sigma_{gh}^{-1}(i)}(h^{-1},g^{-1}) \, .
\end{equation}
Up to coboundaries, these can be identified with the complex conjugated phases $\overline{c_i(g,h)}$ and define a conjugate 2-cocycle $\bar c \in Z_{\sigma}^2(G,U(1)^n)$. As a consequence, the conjugated lines $L_i^{\ast}$ transform in the 2-representation $(\sigma,\Bar{c})$, which is the 2-representation conjugate to $(\sigma,c)$.

\subsubsection{Separated lines}
\label{sssec:naive-separated-lines}

We now consider the action of the symmetry group on collections of line defects supported on distinct lines in spacetime. The upshot is that they transform in an appropriate tensor product of 2-representations of $G$.

\begin{figure}[h]
	\centering
	\includegraphics[height=7cm]{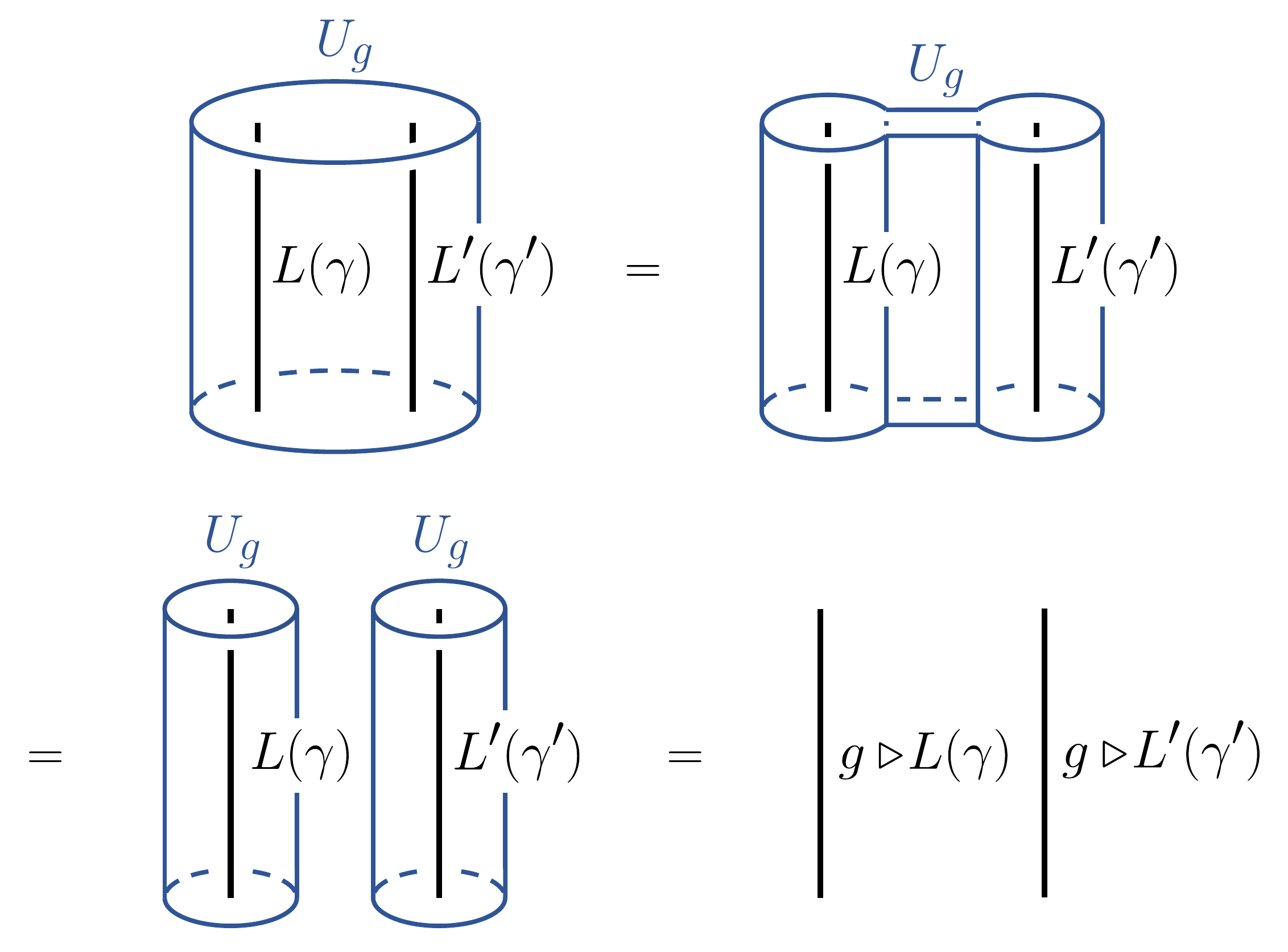}
	\vspace{-5pt}
	\caption{}
	\label{fig:tensor-product-line-action}
\end{figure}

Consider a pair of line defects $L(\gamma) \,L'(\gamma')$ supported on separated parallel lines $\gamma$, $\gamma'$. As before, group elements act on this pair by linking the operators with a sufficiently large $(D-1)$-cylinder. The sequence of topological operations illustrated in figure~\ref{fig:tensor-product-line-action} reveals this action is given by
\begin{equation}\label{tensor-action-2}
g \,\triangleright\, \big[ L(\gamma) \, L'(\gamma') \big] \; = \; \big[ g \triangleright L(\gamma) \big] \big[ g \triangleright L'(\gamma')\big]
\end{equation}
in terms of the individual actions of $g$ on $L(\gamma)$ and $L'(\gamma')$.

\begin{figure}[h]
	\centering
	\vspace{5pt}
	\includegraphics[height=13.5cm]{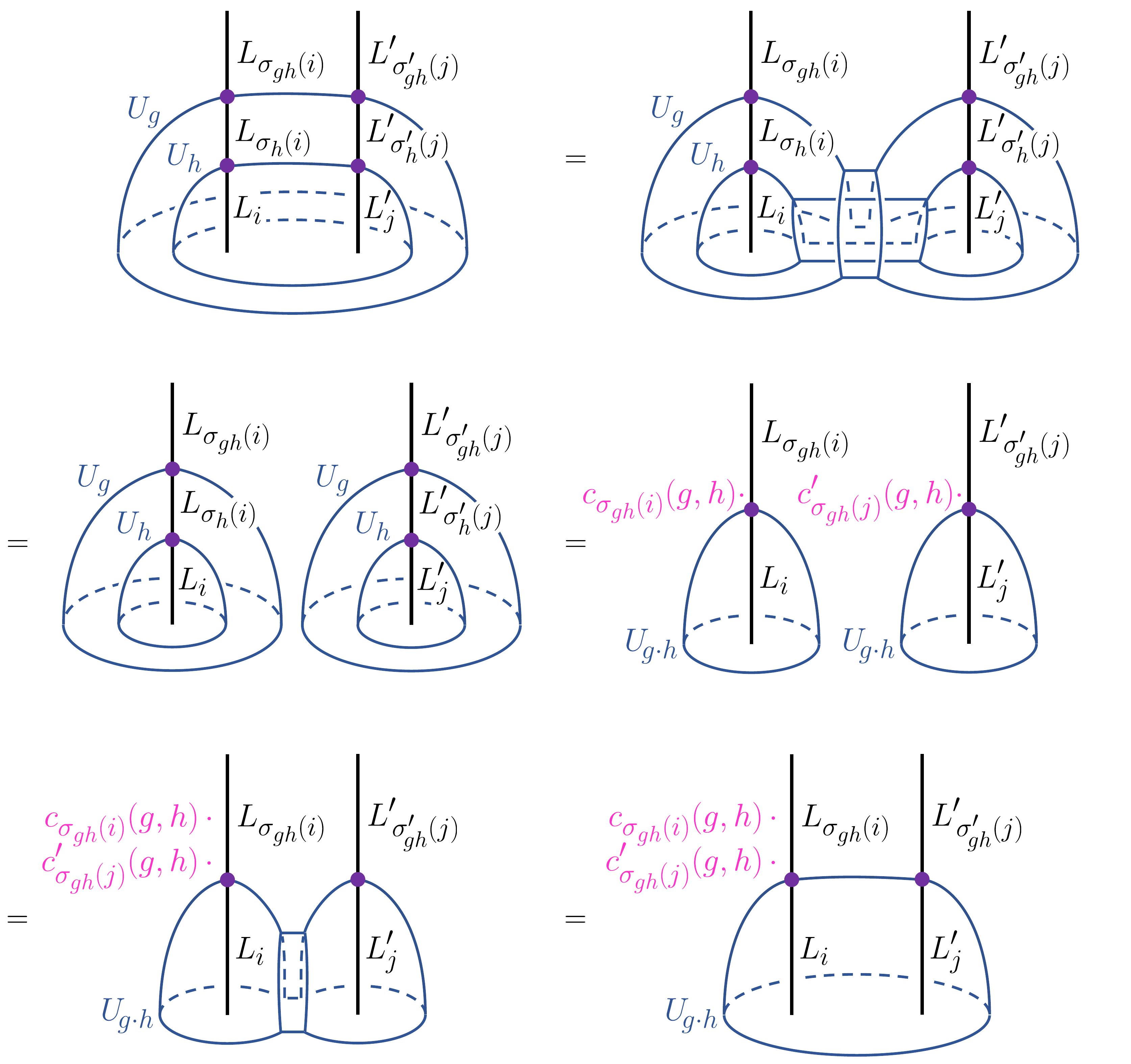}
	\vspace{-3pt}
	\caption{}
	\label{fig:tensor-product-projective-line-action}
\end{figure}

In order to interpret this result mathematically, suppose $L(\gamma)$ and $L'(\gamma')$ are elements of sets $\mathcal{S}$ and $\mathcal{S}'$ forming 2-representations $(\sigma,c)$ and $(\sigma',c')$ of $G$, respectively. The collection of pairs of line operators from these sets is then described by the Cartesian product set
\begin{equation}
\mathcal{S} \otimes \mathcal{S}' \; := \; \mathcal{S} \times \mathcal{S}' \, .
\end{equation}
The elements of this set are indexed by $L_i(\gamma) \otimes L'_j(\gamma')$, where $L_i(\gamma)$ and $L_j'(\gamma')$ denote indexations of $\mathcal{S}$ and $\mathcal{S}'$, respectively. According to (\ref{tensor-action-2}), the group action on such pairs of line operators is given by
\begin{equation}
g \, \triangleright \, \big[ L_i(\gamma) \otimes L_j'(\gamma') \big] \; = \; L_{\sigma_g(i)}(\gamma) \, \otimes  \, L'_{\sigma'_g(j)}(\gamma')  \, ,
\end{equation}
which can be identified with the tensor product permutation action $\sigma \otimes \sigma': G \to S_{nn'}$ sending $g \in G$ to the permutation
\begin{equation}\label{eq:tensor-product-permutation}
(\sigma \otimes \sigma')_g: \; (i,j) \, \mapsto \, (\sigma_g(i), \sigma_g'(j)) \, .
\end{equation}
Furthermore, the sequence of topological operations illustrated in figure \ref{fig:tensor-product-projective-line-action} reveals that the intersection of $L_i(\gamma)L'_j(\gamma')$ with symmetry defects $g,h \in G$ respects group multiplication up to phases
\begin{equation}\label{eq:tensor-product-2-cocycle}
(c \otimes c')_{(i,j)}(g,h) \;\, := \;\, c_i(g,h) \cdot c'_j(g,h) \; \in \; U(1) \, .
\end{equation}
This defines a tensor product 2-cocycle $c \otimes c' \in Z^2_{\sigma \otimes \sigma'}(G,U(1)^{nn'})$.

In summary, the collection of pairs of line defects $L(\gamma) \in \mathcal{S}$ and $L'(\gamma') \in \mathcal{S}'$ transforms in the $nn'$-dimensional 2-representation
\begin{equation}
(\sigma,c) \otimes (\sigma',c') \; := \; (\sigma \otimes \sigma', c \otimes c') \, ,
\end{equation}
where $\sigma \otimes \sigma'$ and $c \otimes c'$ are defined in (\ref{eq:tensor-product-permutation}) and (\ref{eq:tensor-product-2-cocycle}). This coincides with the natural tensor product of the 2-representations, mirroring the well-known property that products of local operators transform in tensor product representations.

\subsubsection{Junction operators}
\label{sssec:naive-junction-operators}

The fact that line defects transform in 2-representations of a symmetry group $G$ has important consequences for junction operators or non-genuine local operators on which they end. We emphasise that the following properties hold regardless of whether these junctions are topological or not.

Consider a collection of line defects $L_i$ transforming in an irreducible 2-representation $(\sigma,c)$ of $G$ and denote by $V_i$ the vector space of non-genuine local operators on which $L_i$ may end. This will typically be an infinite-dimensional complex vector space. However, as in section~\ref{sec:locals}, we can restrict ourselves to finite-dimensional subspaces generated by acting with group elements on a given local operator.

\begin{figure}[h]
	\centering
	\includegraphics[height=3.5cm]{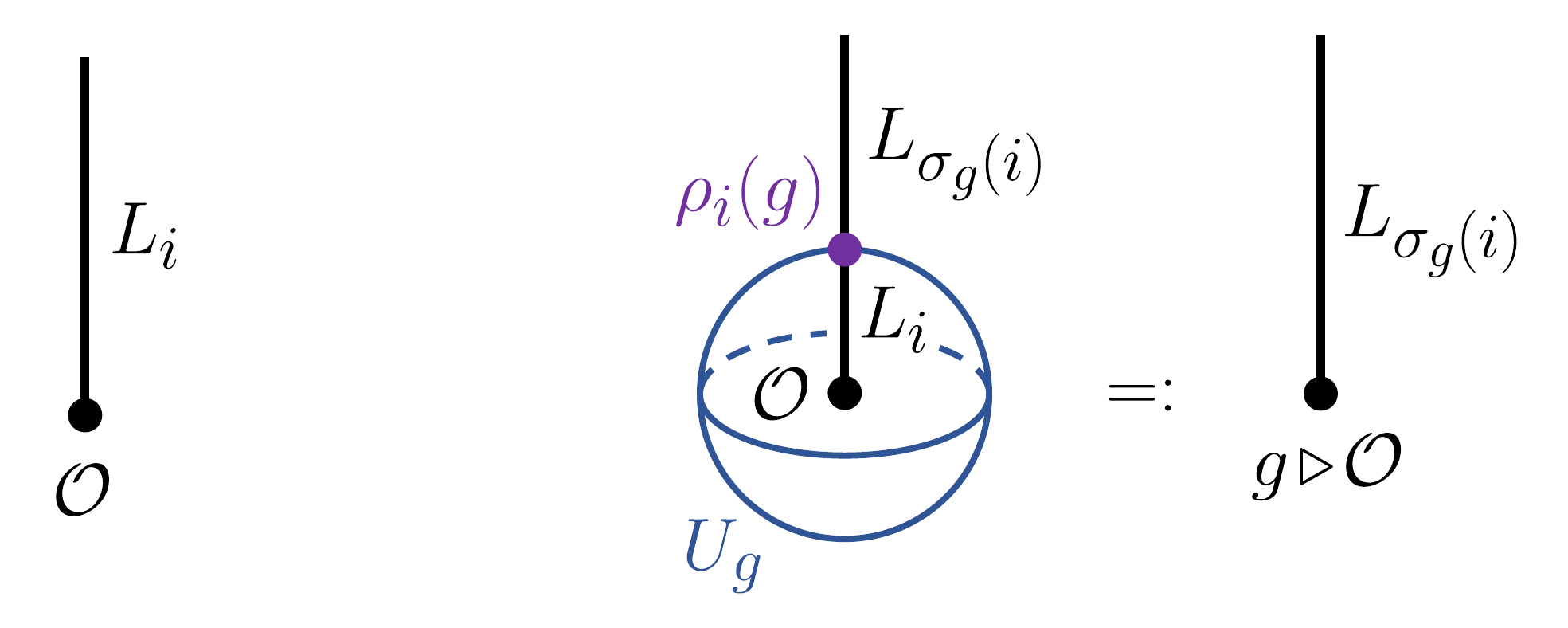}
	\vspace{-5pt}
	\caption{}
	\label{fig:projective-local-action}
\end{figure}

Upon linking an operator $\mathcal{O}$ ending the line $L_i$ with a symmetry defect $g \in G$, it is transformed into a new operator $g \triangleright \mathcal{O}$ ending the line $L_{\sigma_g(i)}$ as illustrated in figure \ref{fig:projective-local-action}. This determines a collection of linear maps
\begin{equation}
\Phi_i(g) : \; V_i \, \to \, V_{\sigma_g(i)} \, ,
\end{equation}
whose compatibility with the fusion of topological symmetry defects in the bulk is implemented by the condition
\begin{equation}
\Phi_{\sigma_h(i)}(g) \, \circ \, \Phi_i(h) \;\; \stackrel{!}{=} \;\; c_{\sigma_{gh}(i)}(g,h) \, \cdot \, \Phi_i(gh) \, .
\end{equation}
This structure has a geometric interpretation as a vector bundle
\be
\pi : \; V \, \rightarrow \, \lbrace 1,...,n \rbrace
\ee
with fibres $V_i = \pi^{-1}(i)$ together with bundle maps $\Phi(g): V \to V$ satisfying $\pi \circ \Phi(g) = \sigma_g$ as well as the composition rule
\begin{equation}
\Phi(g) \, \circ \, \Phi(h) \;\; = \;\; c(g,h) \, \cdot \, \Phi(gh) \, .
\end{equation}
More invariantly, this is a $G$-equivariant vector bundle on $\cS = \{1,\ldots,n\}$. 

Following~\cite{Bartsch:2022mpm,Bartsch:2022ytj}, we call the pair $(V,\Phi)$ a graded projective representation of $G$ of type $(\sigma,c)$. Abstractly, they correspond to intertwiners between the trivial 2-representation and the 2-representation $(\sigma,c)$. In summary, local operators ending line defects transforming in the 2-representation $(\sigma,c)$ transform in graded projective representations of $G$ of type $(\sigma,c)$.

\begin{figure}[h]
	\centering
	\includegraphics[height=4.4cm]{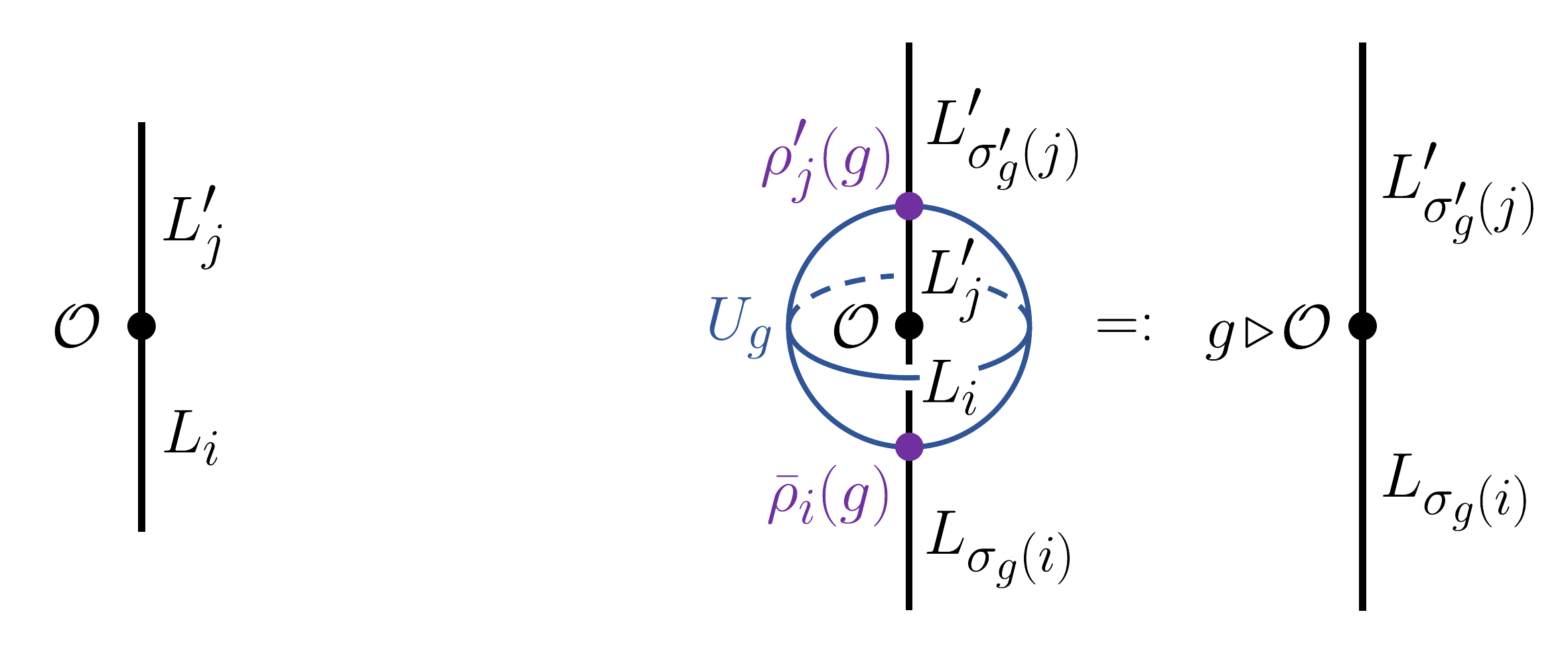}
	\vspace{-5pt}
	\caption{}
	\label{fig:projective-junction-action}
\end{figure}

More generally, we may consider local operators sitting at the junction between a pair of line operators $L$ and $L'$ transforming in 2-representations $(\sigma,c)$ and $(\sigma',c')$ as illustrated in figure \ref{fig:projective-junction-action}.
By arguments similar to the above, their transformation behaviour will be described by graded projective representations of type $(\sigma \otimes \sigma', \Bar{c} \otimes c')$, which correspond to 1-morphisms (or 1-intertwiners) between the 2-representations $(\sigma,c)$ and $(\sigma',c')$.

\subsection{Induction perspective}
\label{subsec:lines-groups-induction}

By construction, the 2-representations $(\sigma,c)$ considered in the previous subsection are irreducible in the sense that the associated permutation action $\sigma: G \to S_n$ is transitive\footnote{Concretely, this means that for each pair $i,j \in \lbrace 1,...,n \rbrace$ there exists a $g\in G$ such that $\sigma_g(i) = j$.}. As we will describe in this subsection, the irreducible 2-representations of $G$ allow for an alternative description that admits a more physical interpretation of the associated data and recovers the mathematical notion of induction of 2-representations~\cite{GANTER20082268,OSORNO2010369}. 

To see this, let us again consider a collection of line operators $L_i$ transforming in an $n$-dimensional irreducible 2-representation of $G$ labelled by $(\sigma,c)$. We fix the line operator $L = L_1 \in \mathcal{S}$ and consider its stabiliser
\begin{gather}
\begin{aligned}
H \; := \;\; &\lbrace \, h \in G \; | \; h \triangleright L = L \, \rbrace \\ 
\equiv \;\; &\lbrace \, h \in G \; | \; \sigma_h(1) = 1 \, \rbrace \, \subset \, G
\end{aligned}
\end{gather}
with respect to the $G$-action. Wrapping the line $L$ with symmetry defects $h \in H$ leaves it invariant, so we say $L$ preserves the subgroup $H \subset G$.

\begin{figure}[h]
	\centering
	\includegraphics[height=4.1cm]{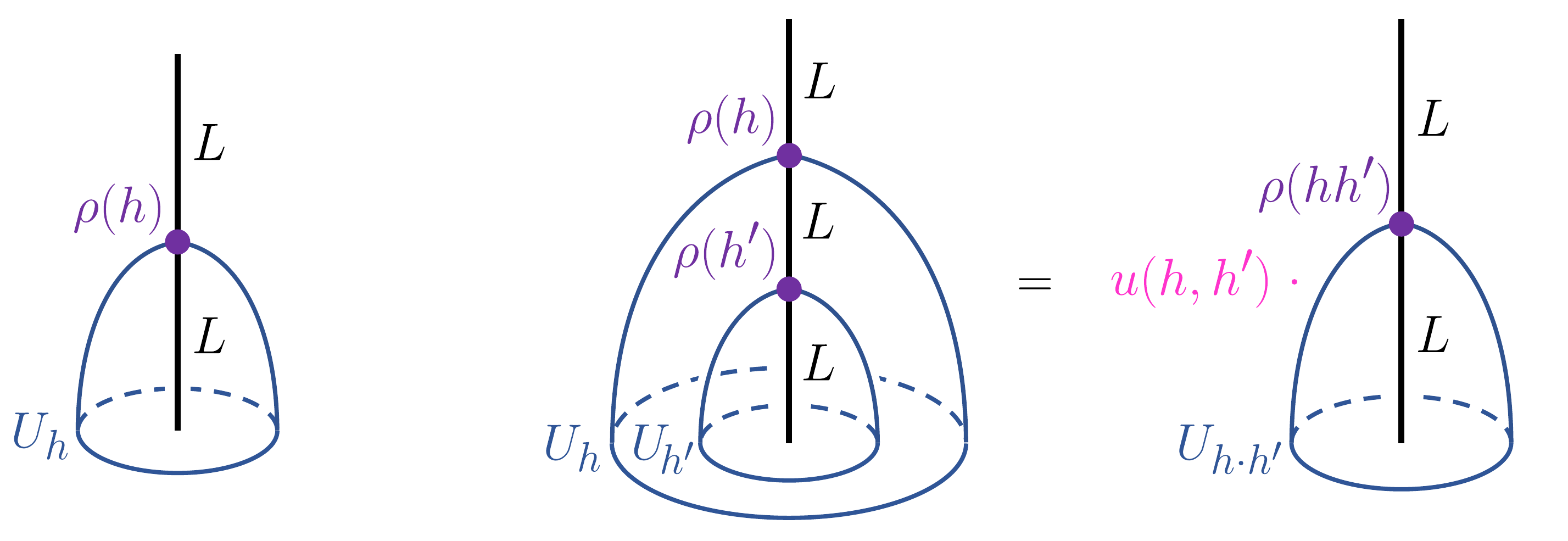}
	\vspace{-5pt}
	\caption{}
	\label{fig:simple-2-reps}
\end{figure}

As before, the line $L$ supports topological junctions $\rho(h) := \rho_1(h)$ that describe its intersection with symmetry defects $h \in H$ as illustrated on the left-hand side of figure \ref{fig:simple-2-reps}. These junctions are multiplicative up to phases
\begin{equation}
u(h,h') \; := \; c_1(h,h')
\end{equation}
as illustrated on the right-hand side of figure \ref{fig:simple-2-reps}. 
Associativity of the fusion of symmetry defects in the bulk constrains these phases to obey the ordinary 2-cocycle condition
\begin{equation}
(du)(h_1,h_2,h_3) \; = \; \frac{u(h_2,h_3) \cdot u(h_1,h_2h_3)}{u(h_1h_2,h_3) \cdot u(h_1,h_2)} \; \stackrel{!}{=} \; 1 \, .
\end{equation}
We may regard the 2-cocycle $u$ (or rather its group cohomology class $[u] \in H^2(H,U(1))$) as a defect 't Hooft anomaly for the symmetry $H \subset G$ preserved by $L$.

To summarise, we can label the irreducible 2-representations of $G$ by pairs $(H,u)$ consisting of
\begin{enumerate}
\item a subgroup $H \subset G$,
\item a 2-cocycle $u \in Z^2(H,U(1))$.
\end{enumerate}
Note that this corresponds to a 1-dimensional 2-representation $u$ of the subgroup $H$. Up to equivalence, the original 2-representation $(\sigma,c)$ can be reconstructed from this data as follows: First consider the space of left $H$-cosets 
\begin{equation}
G/H \; = \; \lbrace r_1H, \, ..., \, r_nH \rbrace
\end{equation}
in $G$ with fixed representatives $r_i \in G$ ($i=1,...,n$) such that $r_1 = 1$. Left multiplication by elements $g \in G$ then permutes the elements of the left coset space via
\begin{equation}
g \cdot (r_iH) \; = \; r_{\sigma_g(i)}H \, ,
\end{equation}
which induces a permutation action $\sigma: G \to S_n$. More concretely,
\begin{equation}
g \cdot r_i \; = \; r_{\sigma_g(i)} \cdot h_i(g)
\end{equation}
for some unique elements $h_i(g) \in H$, which, up to coboundaries, allow us to reconstruct the twisted 2-cocycle $c \in Z^2_{\sigma}(G,U(1)^n)$ from the ordinary 2-cocycle $u \in Z^2(H,U(1))$ via
\begin{equation}\label{eq:induction-cocycle}
c_i(g_1,g_2) \; = \; u\big(\, h_{\sigma^{-1}_{g_1}(i)}(g_1), \, h_{\sigma^{-1}_{g_1g_2}(i)}(g_2) \, \big) \, .
\end{equation}
Mathematically, this construction realises the 2-representation $(\sigma,c)$ as the induction of the 1-dimensional 2-representation $u$ of $H$, which we write as
\begin{equation}
(\sigma,c) \; = \; \text{Ind}_H^G(u) \, .
\end{equation}
It is known that, up to equivalence, all irreducible 2-representations of $G$ can be obtained via induction of a 1-dimensional 2-representation of a subgroup of $G$~\cite{GANTER20082268,OSORNO2010369}.

The data $(H,u)$ labelling an irreducible 2-representation again suffers from redundancies giving equivalent 2-representations. Namely, we could have chosen to fix a different line operator $L' = L_i \in \mathcal{S}$ with $i \neq 1$, whose stabiliser $H'$ would then be related to the stabiliser $H$ of $L=L_1$ by
\begin{equation}
H' \; = \; r_i H r_i^{-1} \, =: \, {}^{r_i\!}H \, .
\end{equation}
Similarly, using equation (\ref{eq:induction-cocycle}) we see that, up to coboundaries, the corresponding ordinary 2-cocycle $u' \in Z^2(H',U(1))$ would be given by
\begin{equation}
u'(h'_1,h'_2) \; \equiv \; c_i(h'_1,h'_2) \; = \; u\big(\, r_i^{-1}h_1'\, r_i, \; r_i^{-1}h_2'\, r_i  \, \big) \; =: \; ({}^{r_i}u)(h_1',h_2')
\end{equation}
for all $h_1',h_2' \in H'$. In summary, the pair $(H',u')$ is related to the pair $(H,u)$ by
\begin{equation}\label{eq:equivalent-simple-2reps}
H' \, = \, {}^{g\!}H \qquad \text{and} \qquad [u'] \, = \, [{}^gu] \, ,
\end{equation}
where $g \equiv r_i \in G$. Since the choice of fixed line $L \in \mathcal{S}$ is unphysical, we should consider the irreducible 2-representations labelled by $(H,u)$ and $(H',u')$ as equivalent. This agrees with the mathematical notion of equivalence of irreducible 2-representations: two irreducible induced 2-representations of $G$ labelled by $(H,u)$ and $(H',u')$ are equivalent if there exists a $g \in G$ such that (\ref{eq:equivalent-simple-2reps}) holds.

\subsubsection{Separated lines}

Now consider again pairs of line defects $L(\gamma)$, $L'(\gamma')$ supported on parallel lines $\gamma$, $\gamma'$ and transforming in irreducible 2-representations $(\sigma,c)$, $(\sigma',c')$ of $G$. As above, we can alternatively label the latter by their stabiliser subgroups $H = \text{Stab}_{\sigma}(1)$ and $H' = \text{Stab}_{\sigma'}(1)$ and 2-cocycles $u = c_1|_H$ and $u' = c'_1|_{H'}$. By the orbit-stabilizer theorem, this sets up a correspondence of $G$-orbits
\begin{equation}
G/H \; \cong \; \lbrace 1,...,n \rbrace \qquad \text{and} \qquad G/H' \; \cong \; \lbrace 1,...,n' \rbrace \, ,
\end{equation}
where $G$ acts on the left-hand sides by left multiplication and on the right-hand sides via the permutation actions $\sigma$ and $\sigma'$, respectively. 

As described in subsection \ref{sssec:naive-separated-lines}, the pair $(L,L')$ transforms in the tensor product 2-representation $(\sigma,c) \otimes (\sigma',c')$ of $G$. In general, the tensor product of two irreducible 2-representations is not irreducible but decomposes as a direct sum of irreducible 2-represen-tations. This decomposition is induced by the decomposition of the product $G$-set
\begin{equation}
(G/H) \, \times \, (G/H') \;\; \cong \bigsqcup_{[g] \, \in \, H \backslash G / H'} G/(H \cap {}^{g\!}H')
\end{equation}
into a disjoint union of $G$-orbits indexed by double cosets $[g] \in H \backslash G / H'$ and with stabilisers $H \cap {}^{g\!}H'$. Correspondingly, the tensor product of the irreducible 2-representations labelled by $(H,u)$ and $(H',u')$ decomposes as
\begin{equation}\label{eq:tensor-product-2-reps-of-groups}
(H,u) \, \otimes \, (H',u') \;\; = \bigoplus_{[g] \, \in \, H \backslash G / H'} \big( H \cap {}^{g\!}H', \; u \cdot {}^gu' \big) \, ,
\end{equation}
where $({}^gu')(k,k') := u'(g^{-1}kg, \, g^{-1}k'g)$ as before.

\subsubsection{Junction operators}

Let us again consider the vector spaces $V_i$ of local operators that end lines $L_i$ transforming in a irreducible $n$-dimensional 2-representation $(\sigma,c)$ of $G$. As described in subsection \ref{sssec:naive-junction-operators}, their collection forms a graded projective representation $(V,\Phi)$ of $G$ of type $(\sigma,c)$, where $\Phi_i$ captures the action of $G$ on $V_i$ that comes from linking the corresponding local operators with symmetry defects $g \in G$.

\begin{figure}[h]
	\centering
	\includegraphics[height=3.5cm]{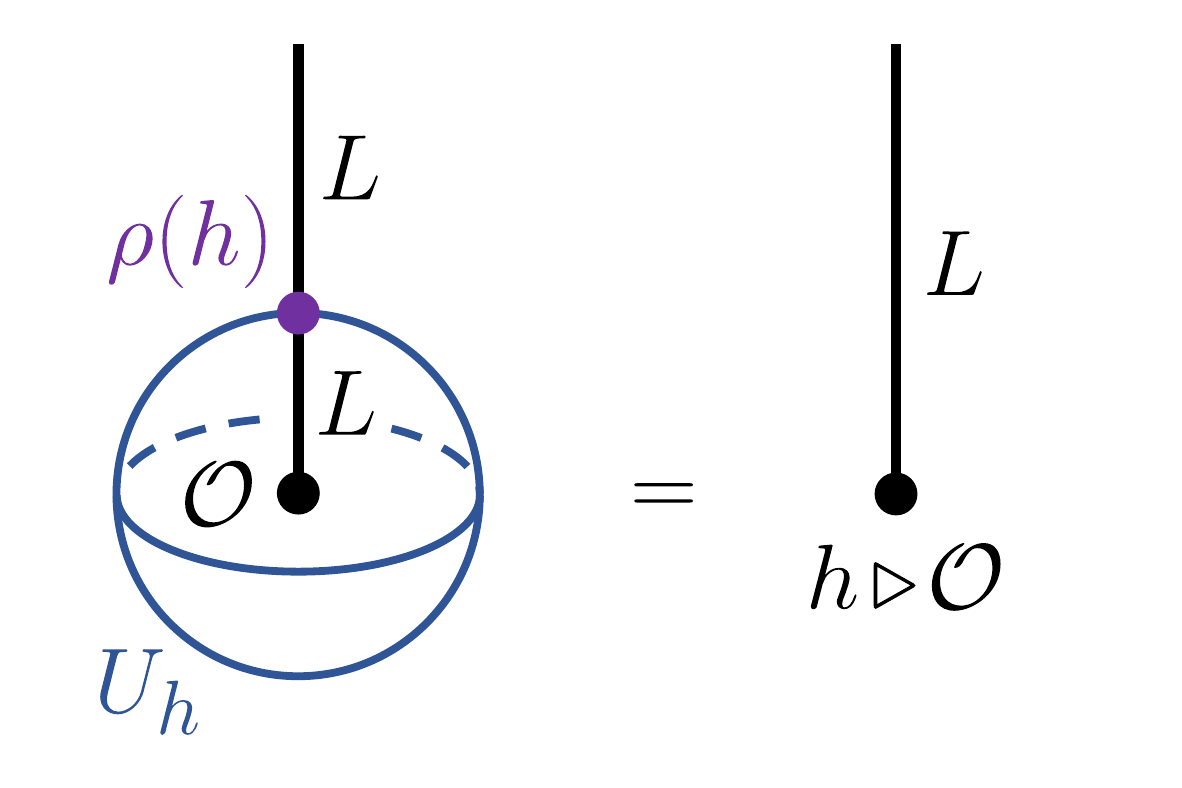}
	\vspace{-5pt}
	\caption{}
	\label{fig:induction-local-action}
\end{figure}

The action of the stabiliser $H = \text{Stab}_{\sigma}(1) \subset G$ on local operators ending the line $L=L_1$  is described by the group homomorphism
\begin{equation}
\phi \, := \, \Phi_1|_H : \;\; H \; \to \; \text{End}(W) \, ,
\end{equation}
where $W := V_1$ denotes the vector spaces of local operators ending the line $L$. This is illustrated in figure \ref{fig:induction-local-action}. Compatibility with the fusion of symmetry defects in the bulk is implemented by the condition
\begin{equation}
\phi(h) \, \circ \, \phi(h') \; \stackrel{!}{=} \; u(h,h') \, \cdot \, \phi(hh') \, ,
\end{equation}
where the multiplicative phase is given by $u(h,h') = c_1(h,h')$ as before. 

In other words, the vector space $W$ of local operators ending the line defect $L$ forms a projective representation of $H$ with 2-cocycle $u \in Z^2(H,U(1))$. In the language of subsection \ref{sssec:naive-junction-operators}, $(W,\phi)$ is a graded projective representation of $H$ of type $(1,u)$. Up to equivalence, the original graded projective representation $(V,\Phi)$ of type $(\sigma,c)$ can be reconstructed from this data as follows: First consider the space of left $H$-cosets 
\begin{equation}
G/H \; = \; \lbrace r_1H, \, ..., \, r_nH \rbrace
\end{equation}
in $G$ with fixed representatives $r_i \in G$ ($i=1,...,n$) such that $r_1 = 1$. As before, elements $g \in G$ act on the latter via
\begin{equation}
g \cdot r_i \; = \; r_{\sigma_g(i)} \cdot h_i(g)
\end{equation}
for some unique elements $h_i(g) \in H$. Using this, we can define a collection of vector spaces
\begin{equation}
V_i \; := \; r_i \otimes W
\end{equation}
on which $g \in G$ acts via the linear maps
\begin{equation}
\Phi_i(g) : \;\; r_i \otimes \mathcal{O} \,\; \mapsto \,\; r_{\sigma_g(i)} \otimes \big[ \phi\big(h_i(g)\big) \cdot \mathcal{O} \big] \, .
\end{equation}
Using the projectivity of $\phi$, one can then check that the maps $\Phi_i(g)$ satisfy
\begin{equation}
\Phi_{\sigma_{h}(i)}(g) \, \circ \, \Phi_i(h) \;\; = \;\; c_{\sigma_{gh}(i)}(g,h) \, \cdot \, \Phi_i(gh)
\end{equation}
with phases $c_i(g,h)$ as in (\ref{eq:induction-cocycle}). Up to equivalence, we therefore recover the graded projective representation $(V,\Phi)$. This realises $(V,\Phi)$ as the induction of the graded projective representation $(W,\phi)$ of $H \subset G$, which we write as
\begin{equation}
(V,\Phi) \; = \; \text{Ind}_H^G(W,\phi) \, .
\end{equation}

In summary, upon labelling an irreducible 2-representations of $G$ by pairs $(H,u)$ as above, local operators at the end of line defects transforming in this 2-representation transform in projective representations of $H$ with 2-cocycle $u$. The fact that local operators ending line defects may transform in projective representations of a symmetry group has played in important role in many previous works~\cite{Bhardwaj:2022dyt,Delmastro:2022pfo,Brennan:2022tyl,Bhardwaj:2023zix}.

\subsection{Categorical perspective}
\label{ssec:lines-categorical-groups}

\begin{figure}[h]
	\centering
	\includegraphics[height=4.6cm]{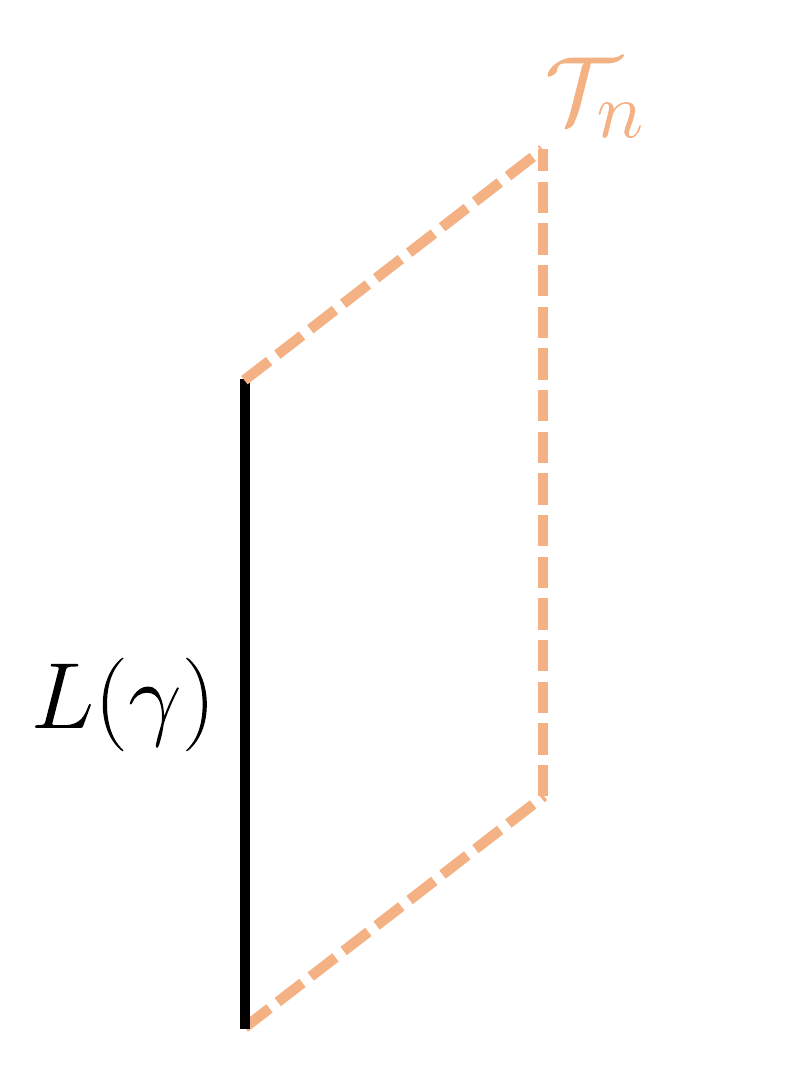}
	\vspace{-5pt}
	\caption{}
	\label{fig:2d-tqft-1}
\end{figure}

It is convenient to reformulate the findings of the previous subsection in a more categorical and therefore more invariant manner. To do this, we consider the collection of $n$ line operators $L_i(\gamma)$ as gapped boundary conditions for an attached auxiliary fully-extended framed 2d TQFT $\mathcal{T}_n$, as illustrated in figure \ref{fig:2d-tqft-1}.

Fully-extended framed 2d TQFTs are in 1-1 correspondence with (equivalence classes of) objects in the fusion 2-category $2\mathsf{Vec}$ of finite-dimensional 2-vector spaces. This has a number of different constructions in the literature, see for instance~\cite{71,bartlett2015modular,Gaiotto:2019xmp}. Each of these constructions provides useful perspectives on the problem at hand.

\subsubsection{2-vector spaces}

Following the description of line operators via their algebras of topological operators, a first formulation of the 2-category $\mathsf{2Vec}$ can be given as follows:
\begin{itemize}
\item Objects are finite semi-simple associative algebras $\cA$.
\item 1-morphisms between objects $\mathcal{A}$ and $\mathcal{B}$ are $\cA$-$\cB$-bimodules.
\item 2-morphisms are bimodule maps.
\end{itemize}
In this description, objects are 2d TQFTs with a choice of gapped boundary condition that supports an associative algebra of topological local operators. This is natural if we want to set up a correspondence between gapped boundary conditions and line defects $L$: the latter support an associative algebra $\cA_L$ of topological local operators as discussed previously in section~\ref{subsec:lines-prelim}.

Intrinsically, 2d TQFTs are in 1-1 correspondence with equivalence classes of objects in $2\mathsf{Vec}$, which forgets the choice of gapped boundary condition. It is clear from the definition of 1-morphisms that equivalence in $2\mathsf{Vec}$ is Morita-equivalence of associative algebras. Any finite semi-simple associative algebra $\mathcal{A}'$ may be decomposed as a sum of matrix algebras
\be
\cA' \; \cong \; \bigoplus_{i=1}^n \; \text{End}(V_i) 
\ee
and is therefore Morita equivalent to as associative algebra
\be
\cA \; = \;  \bigoplus_{i=1}^n \, \C \, .
\ee
The 2d TQFTs are therefore in 1-1 correspondence with natural numbers $n \in \mathbb{N}$, corresponding to the number of vacua. We denote this 2d TQFT by $\cT_n$ in what follows.

\begin{figure}[h]
	\centering
	\includegraphics[height=4.6cm]{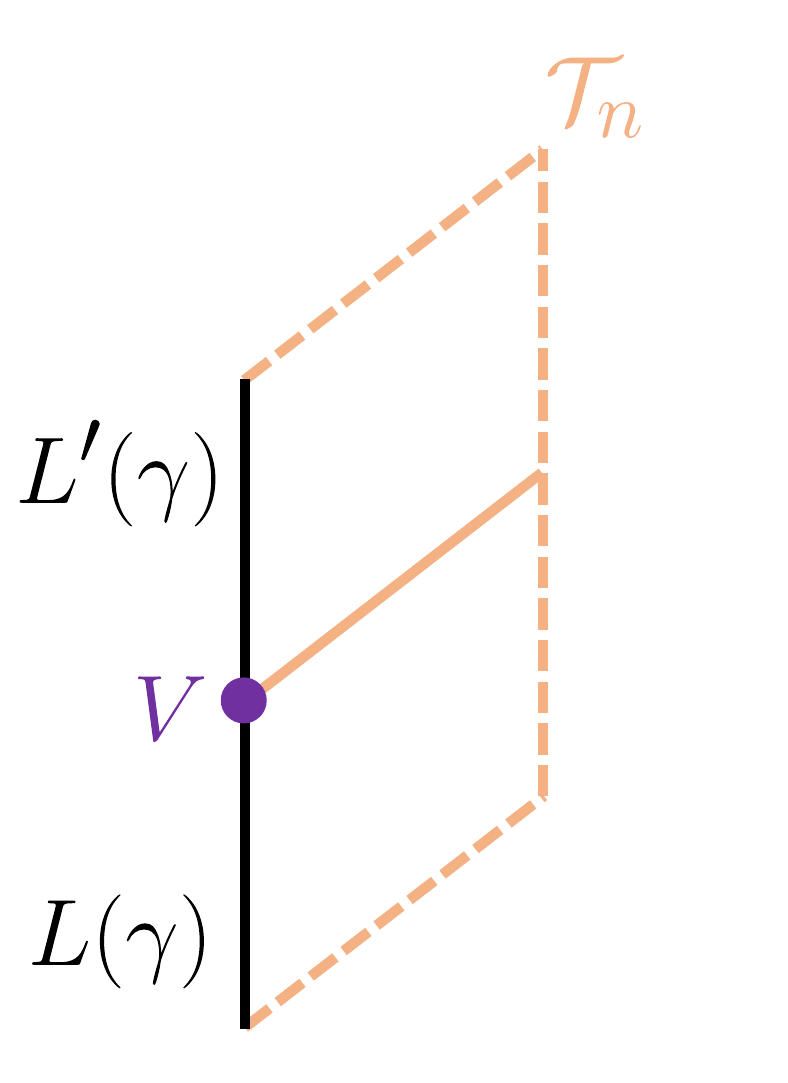}
	\vspace{-5pt}
	\caption{}
	\label{fig:2d-tqft-1-2}
\end{figure}

We can now set up a more concrete identification between a collection of simple line defects $L'_i(\gamma)$ indexed by $i = 1,\ldots,n$ supporting associative algebras $\cA_{L'_i} = \text{End}(V_i)$ and gapped boundary conditions for an auxiliary 2d TQFT $\cT_n$. Note that invertible bi-modules implementing Morita-equivalence arise from topological line defects in the 2d TQFT that impinge on the gapped boundary, as illustrated in figure~\ref{fig:2d-tqft-1-2}. 
Under the correspondence between gapped boundary conditions for $\cT_n$ and line defects, this implies the existence of a topological interface between a simple line $L'(\gamma)$ with $\cA_{L'} = \text{End}(V)$ and a reduced line $L(\gamma)$ with $\cA_{L} = \C$, as claimed previously in section~\ref{subsec:lines-prelim}.

In this spirit, it convenient (but not necessary) to restrict attention to simple line defects and work with formulations of $2\mathsf{Vec}$ where equivalent objects are identified. One such formulation is that of Kapranov-Voevodsky~\cite{71}, which we summarise as follows:
\begin{itemize}
\item Objects are natural numbers $n \in \mathbb{N}$, or finite semi-simple categories $\mathsf{Vec}^n$. This is the category of boundary conditions in the 2d TQFT $\cT_n$.
\item 1-morphisms between objects $n$ and $m$ are functors $A: \mathsf{Vec}^n \to \mathsf{Vec}^m$ and correspond to topological interfaces between 2d TQFTs $\cT_n$ and $\cT_m$.
\item 2-morphisms are natural transformations $\Phi: A \Rightarrow B$ and correspond to topological  junctions between topological interfaces.
\end{itemize}
The fusion of objects corresponds to the stacking of 2d TQFTs: $\cT_n \otimes \cT_m \, \cong \, \cT_{n \cdot m}$.

\subsubsection{2-representations}

Using the correspondence between collections of simple line defects $L_i(\gamma)$ and gapped boundary conditions for $\cT_n$, the action of a symmetry group $G$ on line defects may be translated into a $G$-equivariant structure on $\mathcal{T}_n$ and provides a natural construction of the abstract categorical definition of 2-representations.

\begin{figure}[h]
	\centering
	\includegraphics[height=6cm]{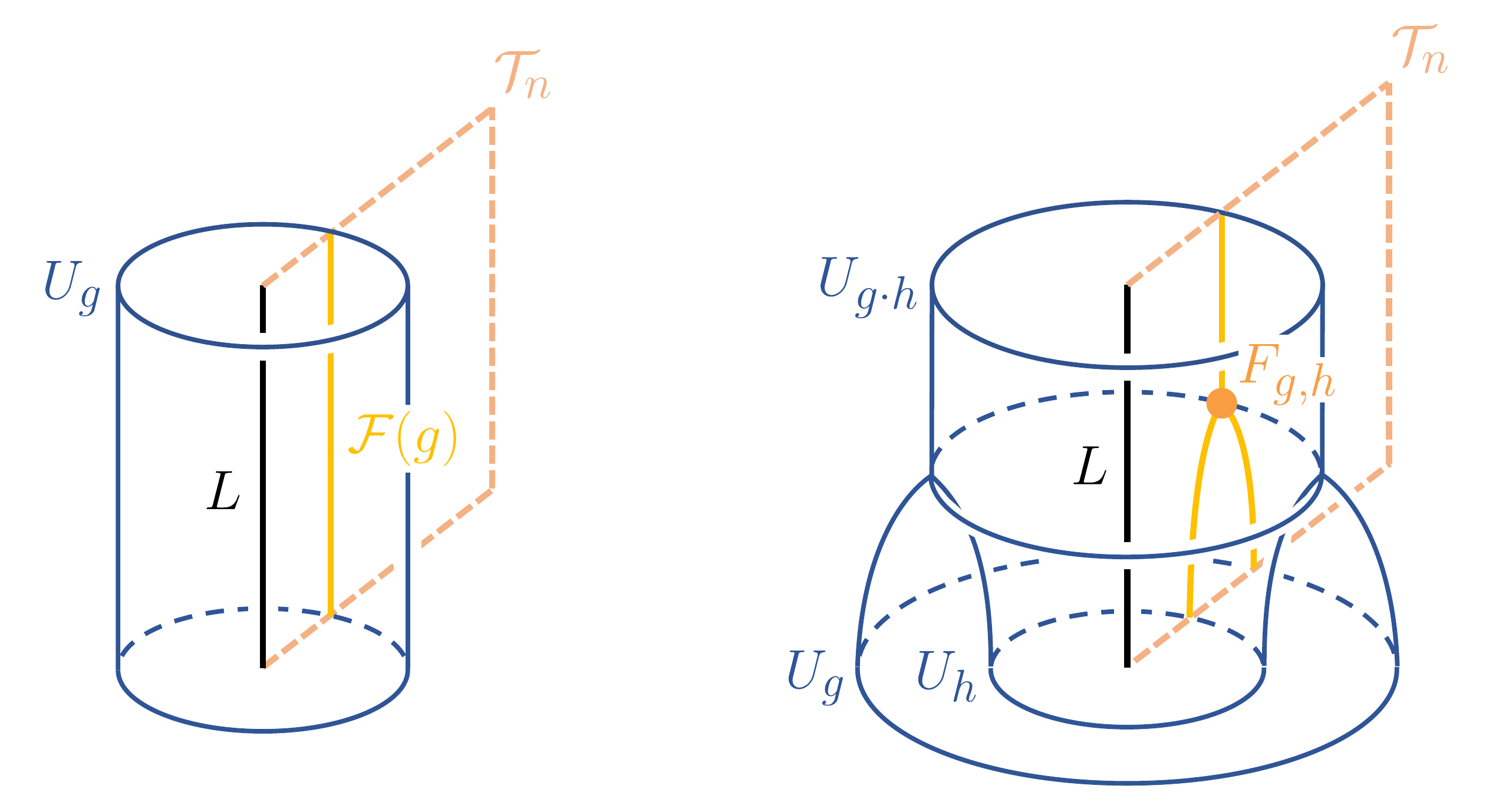}
	\vspace{-5pt}
	\caption{}
	\label{fig:2d-tqft-2}
\end{figure}

Concretely, consider wrapping a line defect $L(\gamma)$ with a topological symmetry defect $U_g(C_\gamma)$ on a cylinder $C^{D-1}_\gamma = \R \times S_\gamma^{D-2}$. Due to the attached auxiliary TQFT $\mathcal{T}_n$, this requires choosing a topological intersection $\mathcal{F}(g)$ between the symmetry defect and $\mathcal{T}_n$, as illustrated on the left of figure \ref{fig:2d-tqft-2}.

This intersection may be viewed as a topological line defect on $\mathcal{T}_n$ and therefore determines a functor $\mathcal{F}(g): \mathsf{Vec}^n \to \mathsf{Vec}^n$. The symmetry defects act on the simple boundary conditions $L \in \mathsf{Vec}^n$ via
\begin{equation}
L \;\, \mapsto \;\, \mathcal{F}(g)(L) \, .
\end{equation}
This corresponds to shrinking down the cylinder $C_\gamma^{D-1}$ onto $L$ and thus implements the action of $g$ on line defects.

The compatibility of the functors $\mathcal{F}(g)$ with the fusion of symmetry defects $g,h \in G$ in the bulk is implemented by natural isomorphisms
\begin{equation}\label{eq:pseudo-2-morphisms}
F_{g,h} \, : \;\; \mathcal{F}(g) \, \circ \, \mathcal{F}(h) \;\; \Rightarrow \;\; \mathcal{F}(g \cdot h) \, ,
\end{equation}
which correspond to topological junctions arising from the intersection of $\mathcal{T}_n$ with the fusion of two symmetry defects $g,h \in G$ in the bulk, as illustrated on the right of figure \ref{fig:2d-tqft-2}. 

\begin{figure}[h]
	\centering
	\includegraphics[height=10.6cm]{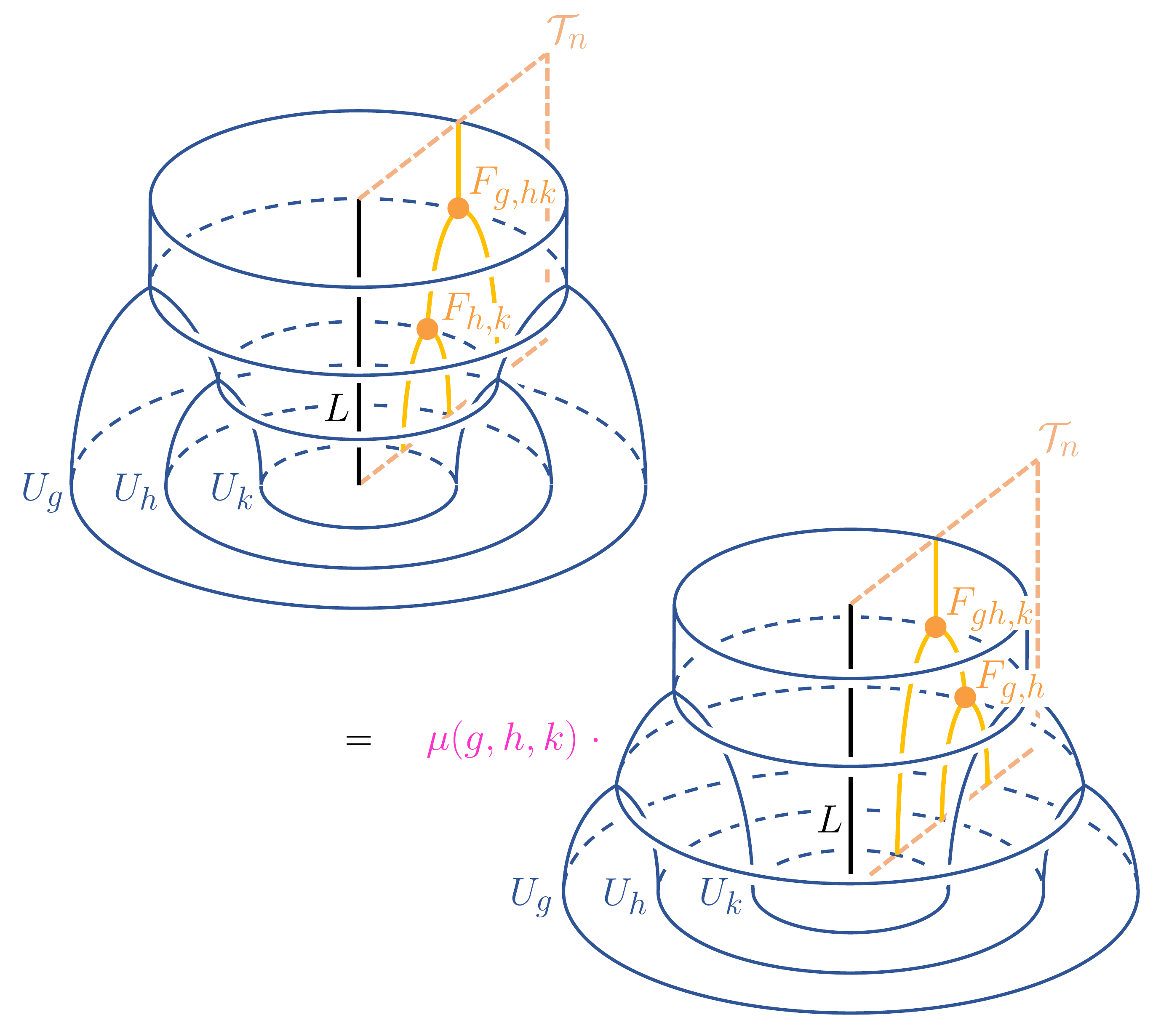}
	\vspace{-5pt}
	\caption{}
	\label{fig:2d-tqft-8}
\end{figure}

The junctions may respect the group law of $G$ only projectively in the sense that intersecting $\mathcal{T}_n$ with the the fusion of three symmetry defects $g,h,k \in G$ in two possible manners yields equivalent results up to a multiplicative phase $\mu(g,h,k)$, as illustrated in figure \ref{fig:2d-tqft-8}. The collection of phases may be regarded as a map
\begin{equation}
\mu: \; G \times G \times G \; \to \; U(1) \, ,
\end{equation}
which must satisfy the 3-cocycle condition
\begin{equation}
(d\mu)(g,h,k,l) \; := \; \frac{\mu(h,k,l) \cdot \mu(g,hk,l) \cdot \mu(g,h,k)}{\mu(gh,k,l) \cdot \mu(g,h,kl)} \; \stackrel{!}{=} \; 1 
\end{equation}
in order to be compatible with the intersection of four symmetry defects $g,h,k,l \in G$.
If the associated group cohomology class is trivial, $[\mu] = 1$, the junctions can be renormalised to satisfy the required compatibility condition\footnote{Here, we denote by $\circ$ and $\star$ the vertical and horizontal composition of 2-morphisms in $\mathsf{2Vec}$, respectively.}
\begin{equation}\label{eq:pseudo-2-hom-compatibility}
F_{g,hk} \, \circ \, \big[ \text{Id}_{\mathcal{F}(g)} \, \star \, F_{h,k} \big]  \;\; = \;\; F_{gh,k} \, \circ \, \big[ F_{g,h} \, \star \, \text{Id}_{\mathcal{F}(k)} \big] \, .
\end{equation}
The associated group cohomology class $[\mu] \in H^3(G,U(1))$ is therefore obstruction for the natural transformations $F_{g,h}$ to define consistent junctions that are compatible with the fusion of symmetry defects in the bulk.

In summary, the action of the symmetry group $G$ on a collection of line defects $L_i(\gamma)$ indexed by $i =1,\ldots,n$ in this framework can be described by the following data:
\begin{enumerate}
\item A 2d TQFT $\mathcal{T}_n$ (an object in $\mathsf{2Vec}$).
\item A topological line $\mathcal{F}(g)$ in $\mathcal{T}_n$ (a 1-endomorphism in $\mathsf{2Vec}$) for each $g \in G$.
\item A topological junction $F_{g,h}$ (a 2-morphism in $\mathsf{2Vec}$) for each pair $g,h \in G$, satisfying the compatibility condition (\ref{eq:pseudo-2-hom-compatibility}).
\end{enumerate}
This collection of data can be recognised as a (pseudo-)2-functor
\begin{equation}
\mathcal{F}: \; \widehat{G} \, \to \, \mathsf{2Vec} \, ,
\end{equation}
where the group $G$ is regarded as a 2-category $\widehat{G}$ with a single object $\ast$, whose 1-endomor-phisms are $\text{End}_{\widehat{G}}(\ast) = G$ and whose 2-morphisms are trivial. The collection of such 2-functors itself forms a 2-category whose
\begin{itemize}
\item objects are 2-functors $\mathcal{F}: \widehat{G} \to \mathsf{2Vec}$,
\item 1-morphisms are natural transformations $\eta: \mathcal{F} \Rightarrow \mathcal{F}'$,
\item 2-morphisms are modifications $\Xi: \eta \Rrightarrow \eta'$.
\end{itemize}
We denote this category by $[\widehat{G},\mathsf{2Vec}]$ and recognise it as the 2-category of finite-dimensional 2-representations of $G$, 
\begin{equation}
[\widehat{G}, \mathsf{2Vec}] \, = \, \mathsf{2Rep}(G) \, .
\end{equation}
Equivalently, this may be regarded as the 2-category of $G$-equivariant fully-extended framed 2d TQFTs. 

This reproduces from a more abstract perspective that line defects transform in 2-representations of the symmetry group $G$. It enables us to abstract the construction of 2-representations of $G$ away from specific line defects and to reframe it into the existence of $G$-equivariant structures on 2d TQFTs. Well known arguments show that irreducible 2-representations are classified by the concrete data in subsections~\ref{subsec:lines-groups-elementary} and~\ref{subsec:lines-groups-induction}, as described for instance in~\cite{ELGUETA200753,OSORNO2010369} and summarised in appendix \ref{app:2-representations}.

\subsubsection{Separated lines}

\begin{figure}[h]
	\centering
	\includegraphics[height=5.85cm]{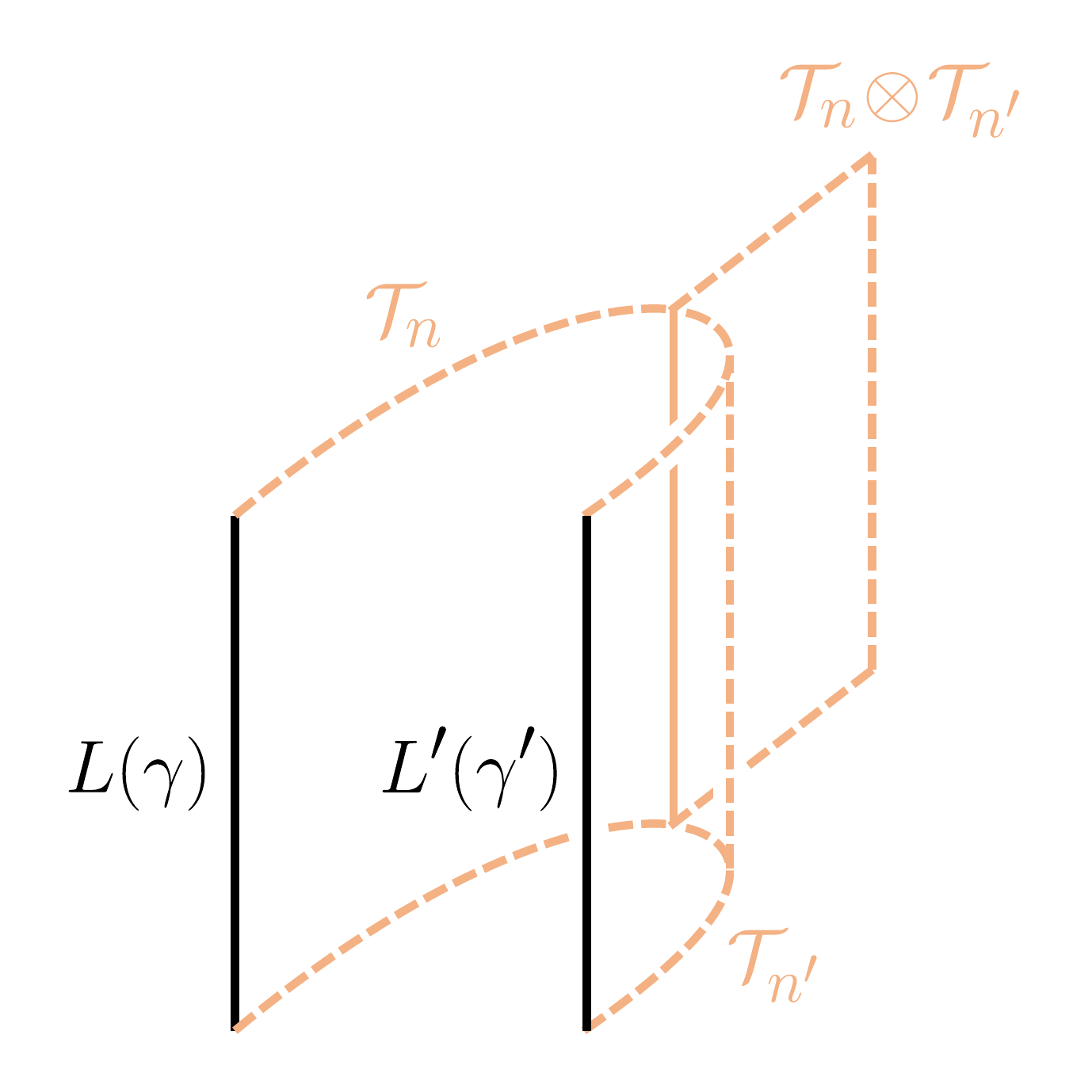}
	\vspace{-5pt}
	\caption{}
	\label{fig:2d-tqft-4}
\end{figure}

The transformation of line defects $L(\gamma) \,L'(\gamma')$ supported on separated parallel lines $\gamma, \gamma'$ can now be reframed into the stacking of the associated TQFTs $\mathcal{T}_n$, $\mathcal{T}_{n'}$, which is determined by the fusion structure on $2\mathsf{Vec}$. This is illustrated in figure \ref{fig:2d-tqft-4}. 

As a result, we obtain a fusion structure on the 2-category $2\mathsf{Rep}(G) = [\widehat{G},\mathsf{Vec}]$ of $G$-equivariant 2d TQFTs, which corresponds to the tensor product of 2-representations and their 1- and 2-intertwiners. This reproduces the results from subsections~\ref{subsec:lines-groups-elementary} and~\ref{subsec:lines-groups-induction} that line defects at separated spacetime loci transform in tensor product 2-representations of the symmetry group $G$.

\subsubsection{Junction operators}

Similarly to the identification of line defects as boundary conditions for an attached 2d TQFT, we can regard local operators at the junction between two line defects as a boundary condition for a topological interface between the corresponding 2d TQFTs. 

\begin{figure}[h]
	\centering
	\includegraphics[height=5.85cm]{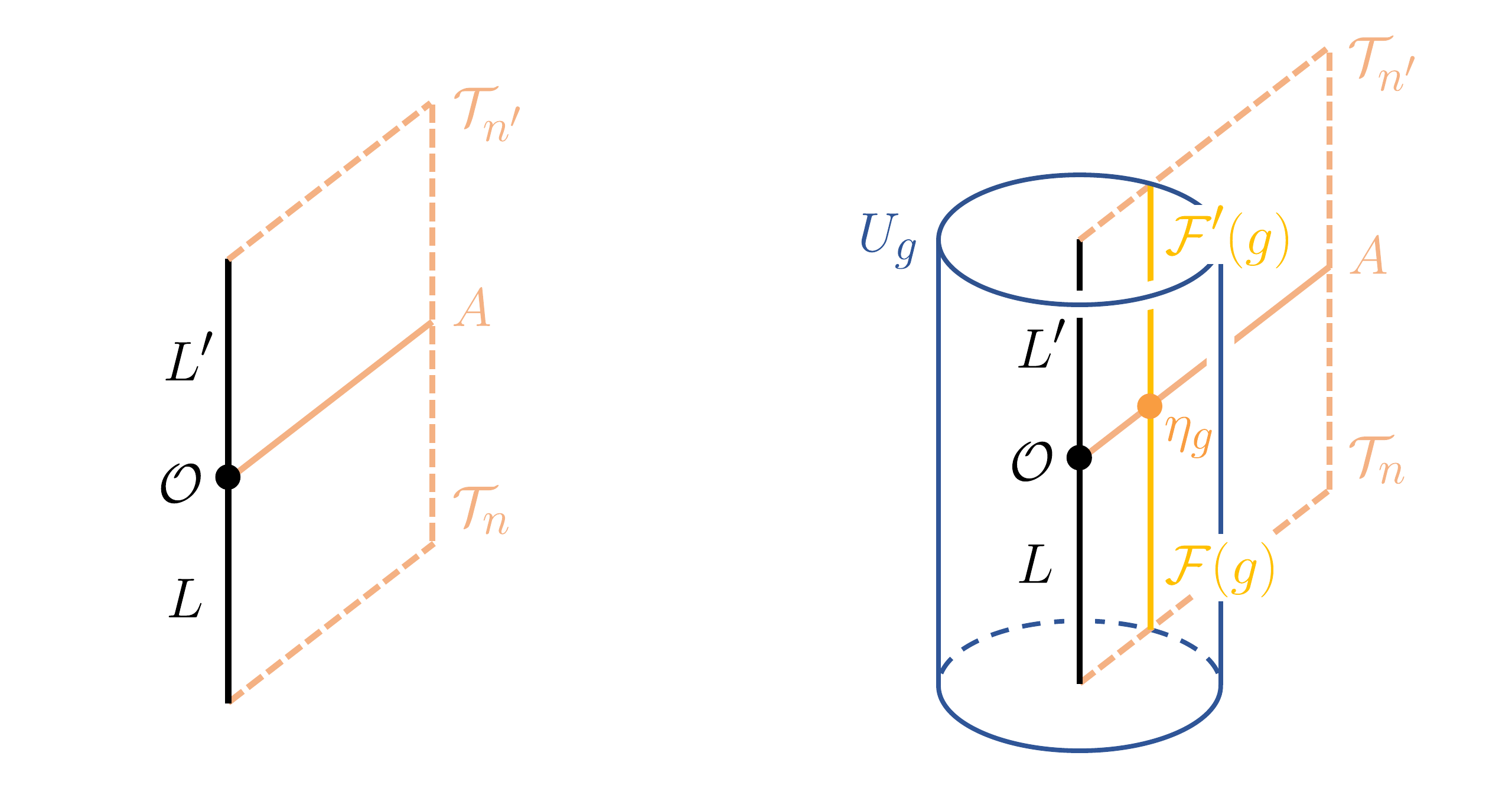}
	\vspace{-5pt}
	\caption{}
	\label{fig:2d-tqft-5}
\end{figure}

Concretely, consider two line defects $L$, $L'$ viewed as objects in the categories $\mathsf{Vec}^n$, $\mathsf{Vec}^{n'}$ of boundary conditions for $\mathcal{T}_n$, $\mathcal{T}_{n'}$. Let $A$ be a topological interface between $\mathcal{T}_n$ and $\mathcal{T}_{n'}$ corresponding to a functor $A: \mathsf{Vec}^n \to \mathsf{Vec}^{n'}$. Then, local junction operators between $L$ and $L'$ are elements 
\begin{equation}
\mathcal{O} \, \in \, \text{Hom}(A(L),L')
\end{equation}
of the vector space of boundary conditions for the interface $A$, as illustrated on the left of figure \ref{fig:2d-tqft-5}.

\begin{figure}[h]
	\centering
	\includegraphics[height=6.5cm]{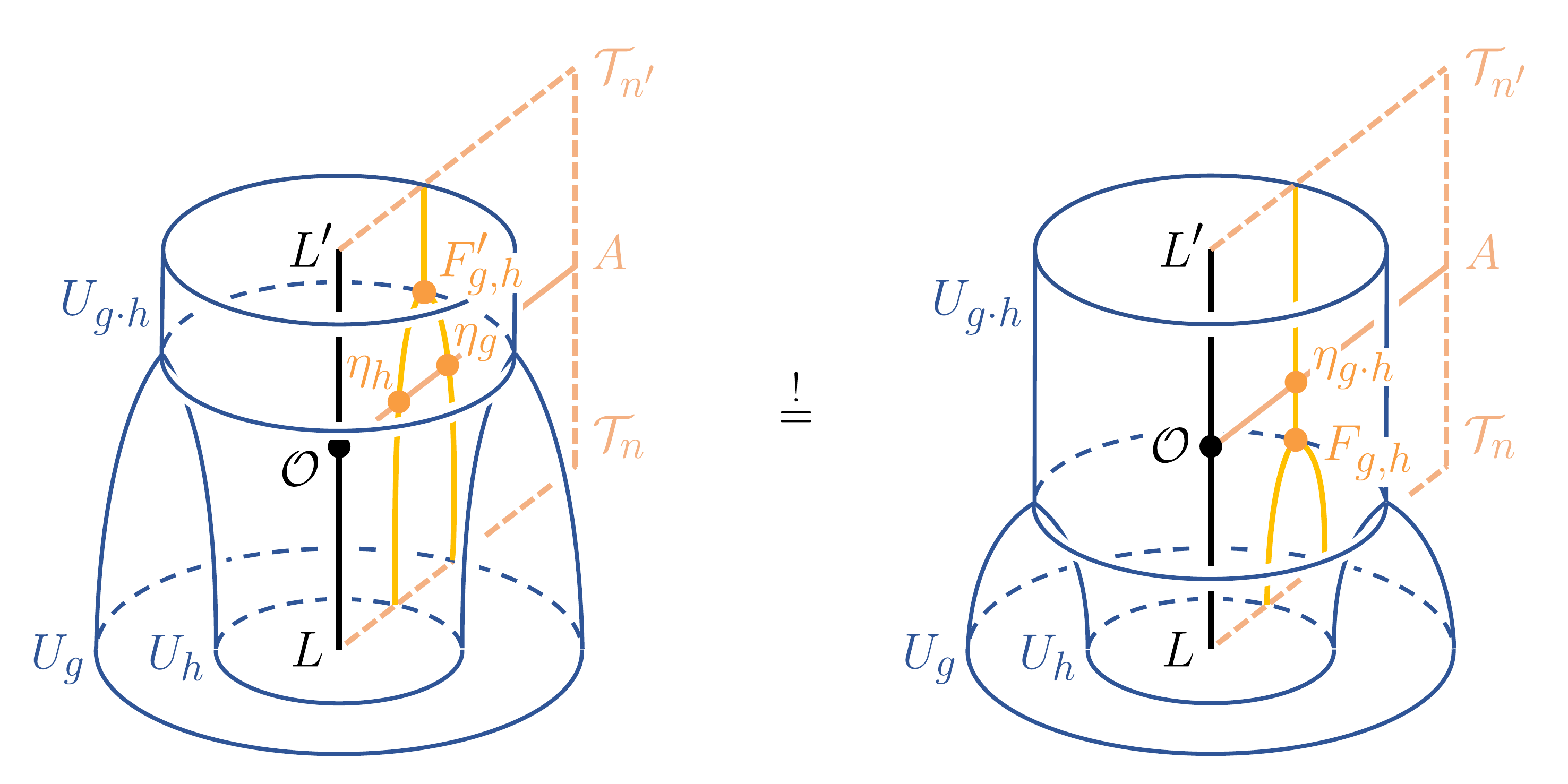}
	\vspace{-5pt}
	\caption{}
	\label{fig:2d-tqft-6}
\end{figure}

Group elements $g \in G$ act on the lines $L$, $L'$ by wrapping together with intersections $\mathcal{F}(g)$, $\mathcal{F}'(g)$ of the corresponding symmetry defect with $\mathcal{T}_n$, $\mathcal{T}_{n'}$. However, due to the presence of the topological interface $A$, we additionally need to specify a local junction operator $\eta_g$ as shown on the right of figure \ref{fig:2d-tqft-5}. This is a 2-morphism
\begin{equation}\label{eq:natural-transformation-2-morphisms}
\eta_g : \;\; A \, \circ \, \mathcal{F}(g) \;\; \Rightarrow \;\; \mathcal{F}'(g) \, \circ \, A
\end{equation}
in $\mathsf{2Vec}$, which must be compatible with the consecutive wrapping action of two symmetry defects $g,h \in G$ in the sense that
\begin{equation}\label{eq:nat-transformation-compatibility}
\big[ F'_{g,h} \star \text{Id}_A \big] \, \circ \, \big[ \text{Id}_{\mathcal{F}'(g)} \star \eta_h \big] \, \circ \, \big[ \eta_g \star \text{Id}_{\mathcal{F}(h)} \big] \;\; \stackrel{!}{=} \;\; \eta_{g \cdot h} \, \circ \, \big[ \text{Id}_A \star F_{g,h} \big] \, .
\end{equation}
This captures the fact that intersecting the topological interface $A$ with two symmetry defects $g,h \in G$ consecutively is equivalent to intersecting it with their fusion $g \cdot h \in G$, as illustrated in figure \ref{fig:2d-tqft-6}.

Using the junction operators $\eta_g$, the topological symmetry defects $g \in G$ then act on the boundary junctions $\mathcal{O} \in \text{Hom}(A(L),L')$ via
\begin{equation}
\mathcal{O} \;\; \mapsto \;\; \mathcal{F}'(g)(\mathcal{O}) \, \circ \, \eta_g(L)  \, ,
\end{equation}
where $\eta_g(L)$ denotes the component of the natural transformation at $L \in \mathsf{Vec}^n$. Physically, this corresponds to shrinking the cylinder onto $L$ and $L'$ and thus implements the action of the group element $g$ on the corresponding lines and their junctions.

In summary, the action of the symmetry group $G$ on local operators at the junction between two line operators can be described by the following data:
\begin{enumerate}
\item A topological interface $A$ between $\mathcal{T}_n$ and $\mathcal{T}_{n'}$ (a 1-morphism in $\mathsf{2Vec}$),
\item A topological junction $\eta_g$ (2-morphisms in $\mathsf{2Vec}$) for each $g \in G$, satisfying the compatibility condition (\ref{eq:nat-transformation-compatibility}).
\end{enumerate}
This collection of data can be recognized as a natural transformation
\begin{equation}
\eta : \,\; \mathcal{F} \; \Rightarrow \; \mathcal{F}'
\end{equation}
between the functors $\mathcal{F},\mathcal{F}' \in [\widehat{G},\mathsf{2Vec}]$ describing the transformation behaviour of the lines $L$ and $L'$. As such, they correspond to 1-morphisms between the corresponding 2-representations of $G$. This reproduces from a more abstract perspective the result from subsections~\ref{subsec:lines-groups-elementary} and~\ref{subsec:lines-groups-induction} that local junction operators between two line operators transform as 1-intertwiners between the corresponding 2-representations of $G$. 

\subsection{Examples}

Let us conclude this section with a few examples of line operators transforming in 2-representations of a finite group symmetry. We will first consider classes of line operators that arise naturally in the context of gauge theory and continue by describing 2-representations with non-trivial 2-cocycle that are induced by certain types of mixed anomalies.

\subsubsection{Gauge theory}
\label{ssec:gauge-theory-charge-conjugation}

We consider a pure gauge theory in $D$ dimensions with a simple connected gauge group $\mathbb{G}$. The theory has a symmetry group $G = \text{Out}(\mathbb{G})$ of outer automorphisms of $\mathbb{G}$, often called charge conjugation symmetry. We denote elements of this group by equivalence classes $[f] \in \text{Out}(\mathbb{G})$ of automorphisms $f \in \text{Aut}(\mathbb{G})$ modulo inner automorphisms that act by conjugation. Thus $[f] = [f']$ iff $f' = f \circ \text{conj}_g$ for some group element $g \in \mathbb{G}$.

A natural class of line operators in any gauge theory are dynamical Wilson lines $W_R(\gamma)$ labelled by (equivalence classes of) representations $R$ of $\mathbb{G}$. Note that the conjugate of a Wilson line is the Wilson line in the conjugate representation $R^*$, namely
\begin{equation}
W_R(\gamma^{\ast}) \; = \; W_{R^*}(\gamma) \, .
\end{equation}
Outer automorphisms $[f] \in G$ act on Wilson lines via
\begin{equation}\label{eq:Wilson-line-action}
    [f] \, \triangleright \, W_R(\gamma) \; = \; W_{R \, \circ  f^{-1}}(\gamma) \, ,
\end{equation}
where $R \circ f^{-1}$ denotes the representation of $\mathbb{G}$ that is obtained by precomposing $R$ with the inverse of a representative $f$ of the outer automorphism $[f]$\footnote{The action on Wilson lines is independent of the representative $f$ since choosing a different representative $f' = f \circ \text{conj}_g$ for some $g \in \mathbb{G}$ leads to the equivalent representation $R \circ (f')^{-1} = R(g)^{-1} \circ [R \circ f^{-1}] \circ R(g)$ and Wilson lines depend only on the equivalence class of the representation.}. This induces a permutation action on the set of isomorphism classes of irreducible representations of $\mathbb{G}$, whereupon they decompose into irreducible 2-representations of $G = \text{Out}(\mathbb{G})$.

Let us look at this in a few concrete examples:

\begin{itemize}
    \item Consider $\mathbb{G} = SU(N)$ with $N > 2$ so that $\text{Out}(\mathbb{G}) = \mathbb{Z}_2$. Charge conjugation exchanges the fundamental representation $\mathbf{N}$ and the anti-fundamental representation $\overline{\textbf{N}}$ of $SU(N)$. The fundamental and anti-fundamental Wilson lines $W_{\mathbf{N}}$, $W_{\overline{\mathbf{N}}}$ thus transform in the 2-dimensional 2-representation of $\mathbb{Z}_2$. This extends naturally to anti-symmetric powers of the fundamental representation.

    \item Consider $\mathbb{G} = \text{Spin}(2N)$ with $N > 4$ so that $\text{Out}(\mathbb{G}) = \mathbb{Z}_2$. Charge conjugation exchanges the spinor and conjugate spinor representations $S^{\pm}$ of $\text{Spin}(2N)$, which therefore transform in the 2-dimensional 2-representation of $\mathbb{Z}_2$.

    \item If $\mathbb{G} = \text{Spin}(8)$, the outer automorphism 0-form symmetry enhances to the non-abelian symmetric group
    \begin{equation}
        \text{Out}(\mathbb{G}) \, = \, S_3 \, = \, \mathbb{Z}_3 \rtimes \mathbb{Z}_2 \, =: \, \braket{r,c} \, ,
    \end{equation}
    which permutes the two spinor representations $S^{\pm}$ and the vector representation $V$ amongst each other according to
    \vspace{-1pt}
    \begin{equation}\label{eq:triality}
        \begin{gathered}
        \includegraphics[height=2.1cm]{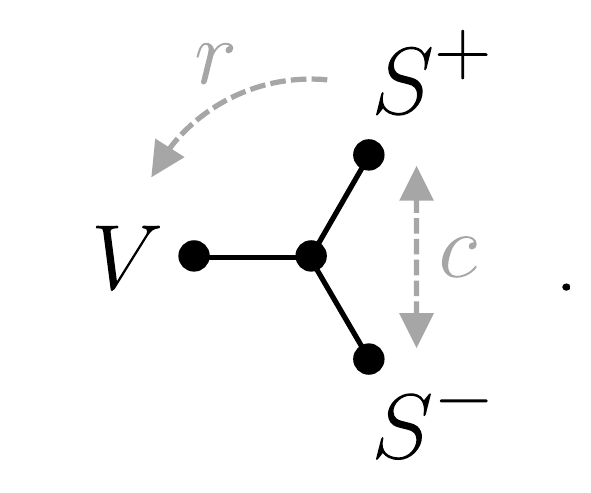}
        \end{gathered}
    \vspace{-4pt}
    \end{equation}
    They therefore transform in the 3-dimensional irreducible 2-representation of $S_3$ labelled by the $\mathbb{Z}_2$-subgroup $\braket{rc} \subset S_3$.
\end{itemize}

Another interesting class of line defects in three dimensions $D=3$ are Gukov-Witten defects $Y_{[g]}(\gamma)$ labelled by conjugacy classes $[g]$ in $\mathbb{G}$. They restrict the holonomy of the gauge connection around a small circle linking the line $\gamma$ to lie in the conjugacy class $[g]$. Outer automorphisms $[f] \in \text{Out}(\mathbb{G})$ act on Gukov-Witten lines via
\begin{equation}
    [f] \, \triangleright \, Y_{[g]}(\gamma) \; = \; Y_{[f(g)]}(\gamma) \, ,
    \label{eq:Gukov-Witten}
\end{equation}
which induces a permutation action of outer automorphisms on the set of conjugacy classes in $\mathbb{G}$. The latter forms a continuous family parameterised by $\mathbb{T}/\mathcal{W}$, where $\mathbb{T} \subset \mathbb{G}$ denotes a maximal torus and $\mathcal{W}$ the associated Weyl group. In this way, we obtain continuous families of irreducible 2-representations of the outer automorphism 0-form symmetry group. 

Let us look at a few concrete examples:
\begin{itemize}
	\item Consider $\mathbb{G} = U(1)$ with $\text{Out}(\mathbb{G}) = \mathbb{Z}_2$. Gukov-Witten defects are labelled by a phase $\alpha = e^{i\theta}$ with charge conjugation acting by complex conjugation $\alpha \to \overline{\alpha}$. There is therefore a continuous family of 2-dimensional 2-representations parametrised by $\theta \in (0,\pi)$ and two trivial 2-representations corresponding to $\theta = 0,\pi$.
    \item  Consider $\mathbb{G} = SU(N)$ with $N > 2$, so that $\text{Out}(\mathbb{G}) = \mathbb{Z}_2$. A maximal torus can be parametrised by $N$ phases $\alpha_i$ satisfying $\prod_{i=1}^N \alpha_i = 1$ and the associated Weyl group $\cW = S_N$ permutes the phases. Charge conjugation acts by complex conjugation, $\alpha_i \mapsto \overline{\alpha_i}$, thus providing continuous families of 2-dimensional 2-representations away from fixed points.

    \item Consider $\mathbb{G} = \text{Spin}(2N)$ with $N > 4$, so that $\text{Out}(\mathbb{G}) = \mathbb{Z}_2$. A maximal torus can be parameterised by $N$ phases $\alpha_i$. These phases are acted upon by $(\mathbb{Z}_2)^N \! \rtimes S_N$, where the generator of each $\mathbb{Z}_2$ factor complex conjugates the corresponding phase $\alpha_i \mapsto \overline{\alpha_i}$ and elements of the symmetric group permute the phases $\alpha_i$ among themselves. The Weyl group is the kernel of the group homomorphism
    \begin{equation}
        (\mathbb{Z}_2)^N \! \rtimes S_N \; \to \; \lbrace \pm 1 \rbrace \, , \qquad \big( (x_1,...,x_N), s \big) \; \mapsto \; (-1)^{(\Sigma_i x_i)} \, .
    \end{equation}
    Outer automorphisms act by complex conjugating a single phase, say $\alpha_1 \mapsto \overline{\alpha_1}$, again providing continuous families of 2-representations.
\end{itemize}

\subsubsection{Mixed anomalies}

The examples presented above all correspond to 2-representations $(\sigma,c)$ where $\sigma : G \to S_n$ is a non-trivial permutation representation but the projective phase is trivial, $[c] = 1$. We now provide a class of examples with non-trivial projective phases.

Consider a theory in $D$ dimensions which, in addition to a finite 0-form symmetry $G$, supports a finite abelian $(D-2)$-form symmetry $A$. The latter is generated by topological line defects and for our purposes is viewed as an auxiliary device to construct non-trivial 2-representations of $G$. Let us furthermore assume that $A$ and $G$ have a mixed 't Hooft anomaly represented by a $(D+1)$-dimensional SPT phase of the form
\begin{equation}\label{eq:mixed-anomaly}
    \int_{X^{D+1}} \! \braket{ \mathsf{a} \cup \mathsf{g}^{\ast}(e)} \, ,
\end{equation}
where $\mathsf{a} \in Z^{D-1}(X,A)$ and $\mathsf{g}: X \to BG$ denote the background fields for $A$ and $G$, respectively, and $e$ represents a class $[e] \in H^2(G,A^{\vee})$.

\begin{figure}[h]
	\centering
	\includegraphics[height=2.4cm]{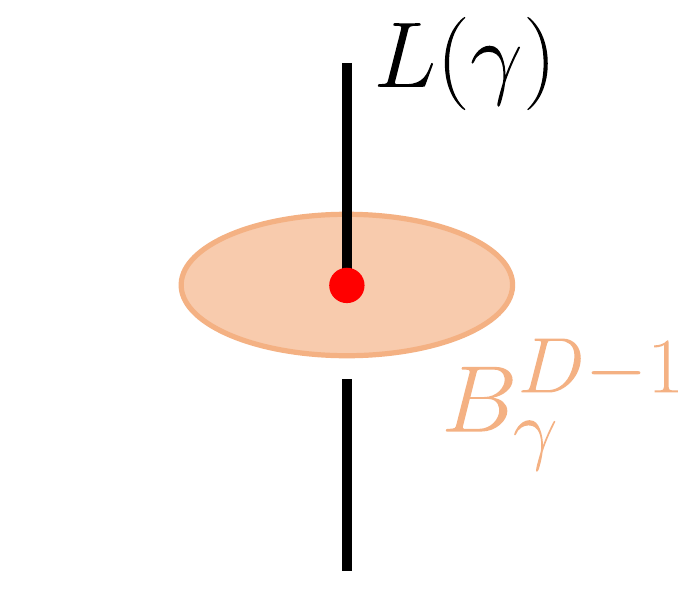}
	\vspace{-5pt}
	\caption{}
	\label{fig:line-intersection}
\end{figure}

Now consider a line defect $L(\gamma)$ that is invariant under the wrapping action of $G$ and induces a background field for $A$ such that
\begin{equation}
    \int_{B^{D-1}_\gamma} \mathsf{a} \; = \; a \, \in \, A
\end{equation}
for any small $(D-1)$-ball $B^{D-1}_{\gamma}$ intersecting the line $\gamma$ at the origin, as illustrated in figure \ref{fig:line-intersection}. Examples include the topological line defects generating $A$ and their compositions with other non-topological line defect. In the language of~\cite{Bhardwaj:2022dyt}, They are solitonic defects for the $(D-2)$-form symmetry $A$. 

Following the arguments of~\cite{Bhardwaj:2022dyt}, the 't Hooft anomaly~\eqref{eq:mixed-anomaly} comes from solitonic line operators $L(\gamma)$ ending on local operators transforming in projective representations of $G$ with 2-cocycle $\braket{a,e} \in Z^2(G,U(1))$. These line defects will therefore transform in 1-dimensional  2-representations of $G$ with $c = \braket{a,e}$. 

An example is pure $\mathbb{G} = SO(4N)$ gauge theory in dimension $D=3$. This has a 0-form symmetry $G = \pi_1(\mathbb{G})^{\vee} \times \text{Out}(\mathbb{G}) = \mathbb{Z}_2 \times \mathbb{Z}_2$ given by the combination of the magnetic and charge conjugation symmetries, as well as an electric 1-form symmetry $A = Z(\mathbb{G}) = \mathbb{Z}_2$. These symmetries exhibit a cubic mixed 't Hooft anomaly represented by the four-dimensional SPT phase
\begin{equation}
     \int_{X^4} \braket{\mathsf{a} \cup \mathsf{g}^{\ast}(e)}
\end{equation}
in terms of the background fields $\mathsf{a} \in Z^2(X,\mathbb{Z}_2)$ and $\mathsf{g} : X \to B(\mathbb{Z}_2 \times \mathbb{Z}_2)$ and the representative $e$ of the non-trivial class $[e] \in H^2(\mathbb{Z}_2 \times \mathbb{Z}_2, \mathbb{Z}_2) \cong \mathbb{Z}_2$. The topological Gukov-Witten defect line defect generating the electric 1-form symmetry thus transforms in a 1-dimensional 2-representation of $\mathbb{Z}_2 \times \mathbb{Z}_2$ characterised by the non-trivial 2-cocycle $\braket{z,e} \in Z^2(\mathbb{Z}_2 \times \mathbb{Z}_2, U(1))$.

\section{Lines and 2-representations of 2-groups}
\label{sec:lines-2groups}

In the previous section we considered the transformation behaviour of line operators under an ordinary finite group symmetry. We now generalise these considerations to a finite 2-group symmetry. We are briefer here and focus on the new features precipitated by the additional data of a 2-group.

\subsection{Preliminaries}

\begin{figure}[h]
	\centering
	\includegraphics[height=4cm]{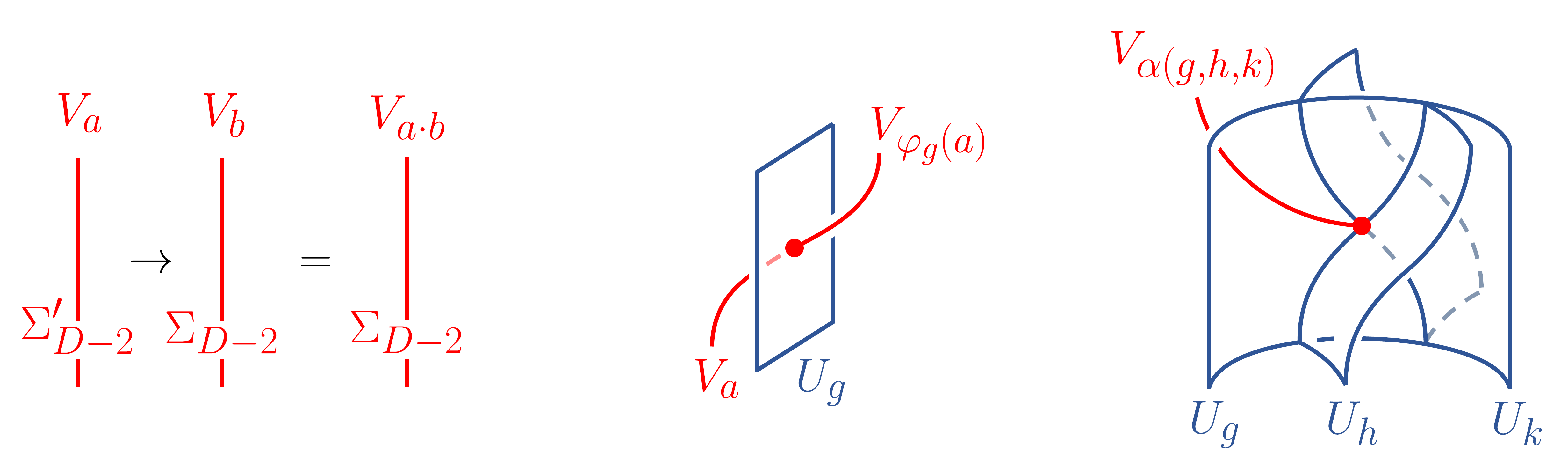}
	\vspace{-5pt}
	\caption{}
	\label{fig:2-group-symmetry}
\end{figure}

In addition to the 0-form symmetry group $G$, there is now an abelian 1-form symmetry group $A$. This is implemented by topological defects $V_a(\Sigma_{D-2})$ labelled by group elements $a \in A$ and supported on codimension-two submanifolds $\Sigma_{D-2}$ that fuse according to the group law of $A$, as illustrated on the left-hand side of figure \ref{fig:2-group-symmetry}.

The 0- and 1-form symmetries $G$ and $A$ may interact with one another in a non-trivial manner, which can be divided into two parts:
\begin{enumerate}
\item When a codimension-two defect $V_a$ pierces a codimension-one defect $U_g$, it may get transformed into a new defect $V_{\varphi_g(a)}$ determined by a homomorphism
\begin{equation}
\varphi: \; G \, \to \, \text{Aut}(A) \, ,
\end{equation}
as illustrated in the middle of figure \ref{fig:2-group-symmetry}.

\item The fusion of three codimension-one defects $U_g$, $U_h$ and $U_k$ is associative up to an emergent codimension-two defect labelled by $\alpha(g,h,k) \in A$. This is illustrated on the right-hand side of figure \ref{fig:2-group-symmetry}. Compatibility with the fusion of four codimension-one defects requires $\alpha$ to obey the twisted 3-cocycle condition,
\begin{equation}
\alpha \in Z^3_{\varphi}(G,A) \, .
\end{equation}
The associated class $[\alpha] \in H^3_{\varphi}(G,A)$ is called the \textit{Postnikov class}. 
\end{enumerate}
The collection of data $\cG = (G,A,\varphi,\alpha)$ determines a 2-group symmetry.

\begin{figure}[h]
	\centering
	\includegraphics[height=3.8cm]{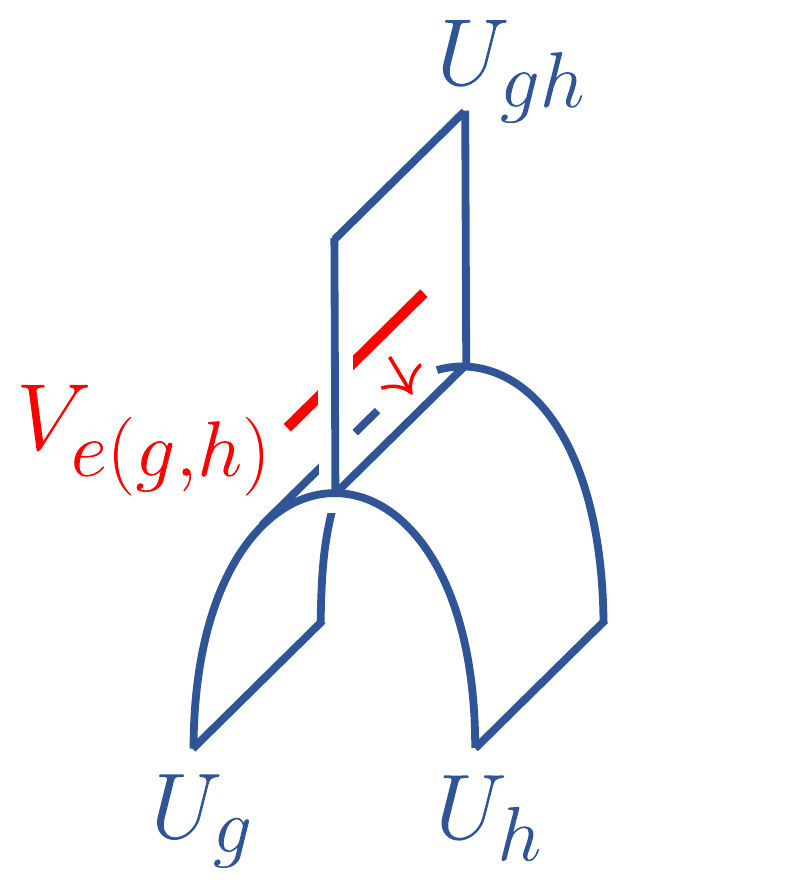}
	\vspace{-5pt}
	\caption{}
	\label{fig:symmetry-fractionalisation-1}
\end{figure}

If the Postnikov class vanishes, $[\alpha] = 1$, the 2-group is said to be split. In this case, there is additional data due to the possibility to shift the trivialization of the Postnikov class by a 2-cocycle
\be
e \, \in \, Z_\varphi^2(G,A) \, .
\ee
The associated class $[e] \, \in \, H^2_{\varphi}(G,A)$ is known as the symmetry fractionalisation~\cite{Benini:2018reh,Delmastro:2022pfo}.
Physically, it captures the dressing of the junction of two defects $g,h \in G$ with a specified 1-form defect $e(g,h) \in A$, as illustrated in figure~\ref{fig:symmetry-fractionalisation-1}. 

\subsection{Elementary perspective}
\label{subsec:lines-2groups-elementary}

Now consider again a finite collection of line defects $L_i(\gamma)$ indexed by $i = 1,\ldots,n$ and transforming under the 0-form symmetry group according to
\begin{equation}
g \, \triangleright \, L_i(\gamma) \; = \; L_{\sigma_g(i)}(\gamma)
\end{equation}
for some transitive permutation $\sigma : G \to S_n$.

\begin{figure}[h]
	\centering
	\includegraphics[height=2.95cm]{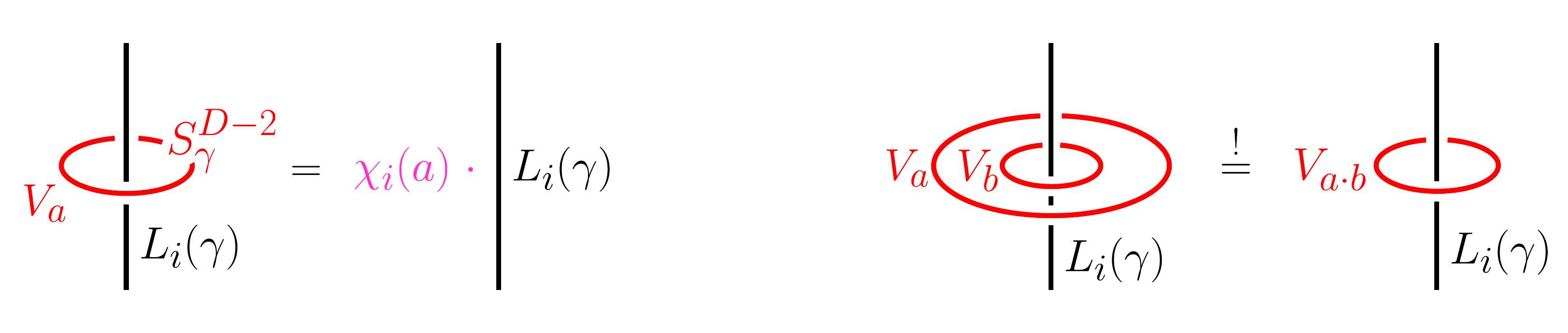}
	\vspace{-5pt}
	\caption{}
	\label{fig:linking-action-1}
\end{figure}

In addition, elements $a \in A$ of the 1-form symmetry group can now act on line defects by linking with the corresponding topological symmetry defect supported on a small $S^{D-2}_{\gamma}$. Shrinking $S_{\gamma}^{D-2}$ results in a topological local operator supported on $L_i(\gamma)$. Assuming the line defects are reduced, this is a multiple of the identity operator and therefore generates a multiplicative phase
\begin{equation}
a \, \triangleright \, L_i(\gamma) \; = \; \chi_i(a) \cdot L_{i}(\gamma) \, ,
\end{equation}
with $\chi_i(a) \in U(1)$\footnote{The same conclusion holds if $L_i(\gamma)$ is not reduced but a further short argument is needed. Namely, compatibility with the existence of a topological junction with a reduced line, as illustrated in figure~\ref{fig:line-preliminaries-2}, requires the 1-form generator act by a multiple of the identity operator on $L_i(\gamma)$.}. This illustrated on the left-hand side of figure \ref{fig:linking-action-1}.

The collection of phases $\chi_i(a)$ indexed by $i=1,\ldots,n$ can be regarded as a map
\begin{equation}
\chi: \; A \, \to \, U(1)^n \, .
\end{equation}
Compatibility with the consecutive action of two symmetry defects $a,b \in A$ requires that this map is a group homomorphism,
\begin{equation}
\chi(a) \cdot \chi(b) \; \stackrel{!}{=} \, \chi(a \cdot b) \, .
\end{equation}
This is the statement that linking a line $L_i(\gamma)$ with two topological defects $a,b \in A$ consecutively is equivalent to linking with their fusion $a \cdot b \in A$, as illustrated on the right-hand side of figure \ref{fig:linking-action-1}. We therefore have a collection of characters $\chi \in (A^{\vee})^n$ specifying the charges of the lines $L_i(\gamma)$ under the 1-form symmetry $A$.

\begin{figure}[h]
	\centering
	\includegraphics[height=3.7cm]{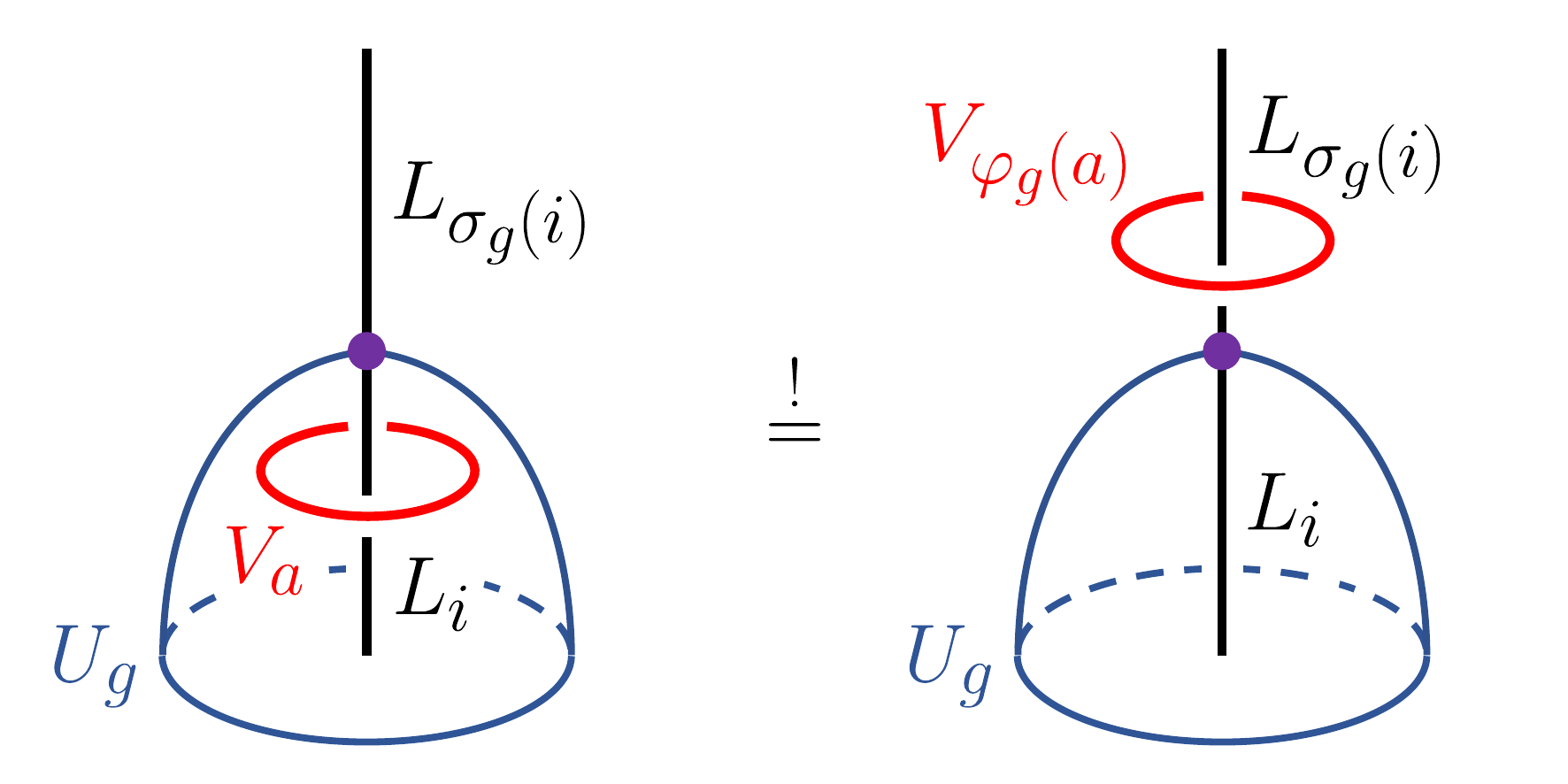}
	\vspace{-5pt}
	\caption{}
	\label{fig:linking-action-2}
\end{figure}

The characters must be compatible with the 2-group structure.
First, linking $L_i(\gamma)$ with a 1-form symmetry defect $a \in A$ and subsequently wrapping with a 0-form symmetry defect $g \in G$ must be equivalent to first wrapping with $g$ and subsequently linking the transformed line $L_{\sigma_g(i)}$ with the transformed 1-form defect $\varphi_g(a) \in A$.
This is illustrated in figure \ref{fig:linking-action-2} and gives the condition
\begin{equation}
\chi_i(a) \; \stackrel{!}{=} \; \chi_{\sigma_g(i)}\big(\varphi_g(a)\big) \, .
\end{equation}

Second, the intersections of the lines $L_i(\gamma)$ of 0-form symmetry defects $g,h \in G$ may again multiply with projective phases $c_i(g,h)$ forming a function
\be
c : \; G \times G \, \to \, U(1)^n \, .
\ee
However, since the fusion of 0-form symmetry defects is now only associative up to 1-form symmetry defects determined by the representative $\alpha$ of the Postnikov class, the twisted 2-cocycle condition (\ref{eq:twisted-2-cocycle-condition}) is shifted to
\begin{equation}
(d_{\sigma}c)_i(g,h,k) \; \stackrel{!}{=} \; \chi_i\big( \alpha(g,h,k) \big) \, .
\label{eq:dc-2group}
\end{equation}
If the Postnikov class vanishes and the 2-group is split, the projective phases can be renormalised to satisfy the twisted 2-cocyle condition $d_\sigma c = 1$.

In summary, we can label the action of the finite 2-group on the collection of line defects $L_i(\gamma)$ by triples $(\sigma,c,\chi)$ consisting of
\begin{enumerate}
\item a permutation action $\sigma: G \to S_n$,
\item a 2-cochain $c \in C_\sigma^2(G,U(1)^n)$,
\item a collection of characters $\chi \in (A^{\vee})^n$,
\end{enumerate}
satisfying the conditions
\begin{equation}
g \triangleright_{\sigma} \chi(a) \; = \; \chi(g \triangleright_{\varphi} a) \qquad \text{and} \qquad d_{\sigma}c \; = \; \braket{\chi,\alpha}
\end{equation}
for all $g \in G$ and $a \in A$, where we denoted $(g \triangleright_{\sigma} \chi)_i := \chi_{\sigma^{-1}_g(i)}$. This is the data of an $n$-dimensional 2-representation of the 2-group. When $A$ is trivial, this reduces to the labelling of 2-representations of an ordinary group $G$ we encountered in section~\ref{sec:lines-groups}. A more mathematical treatment of 2-representations of 2-groups can be found in appendix \ref{app:2-representations}.

Similarly to before, two 2-representations $(\sigma,c,\chi)$ and $(\sigma',c',\chi')$ are considered equivalent if there exists a permutation $\tau \in S_n$ such that
\begin{equation}
\sigma' \; = \; \tau \circ \sigma \circ \tau^{-1} \, , \qquad  [c'] \, = \, [\tau \triangleright c] \, , \qquad  \chi' \, = \, \tau \triangleright \chi \, .
\end{equation}

The tensor product of two 2-representations $(\sigma,c,\chi)$ and $(\sigma',c',\chi')$ is given by
\begin{equation}
(\sigma,c,\chi) \, \otimes \,(\sigma',c',\chi') \; = \; \big( \sigma \otimes \sigma', \, c \otimes c' , \, \chi \otimes \chi' \big)
\end{equation}
with $\sigma \otimes \sigma'$ and $c \otimes c'$ as in (\ref{eq:tensor-product-permutation}) and (\ref{eq:tensor-product-2-cocycle}) and $\chi \otimes \chi' \in (A^{\vee})^{n \cdot n'}$ given by
\begin{equation}
(\chi \otimes \chi')_{(i,j)} \; = \; \chi_i \cdot \chi_j \, .
\end{equation}

Lastly, the conjugate of a 2-representation $(\sigma,c,\chi)$ is the 2-representation $(\sigma,\Bar{c},\Bar{\chi})$ obtained by complex conjugating $c$ and $\chi$.

\subsubsection{Symmetry fractionalization}

If the 2-group is split, one may shift the trivialisation of the Postnikov class by a 2-cocycle $e \in Z_{\varphi}^2(G,A)$, which corresponds to dressing the junction of 0-form symmetry defects $g,h \in G$ by a 1-form symmetry defect $e(g,h) \in A$, as was illustrated in figure~\ref{fig:symmetry-fractionalisation-1}.

\begin{figure}[h]
	\centering
	\includegraphics[height=3.8cm]{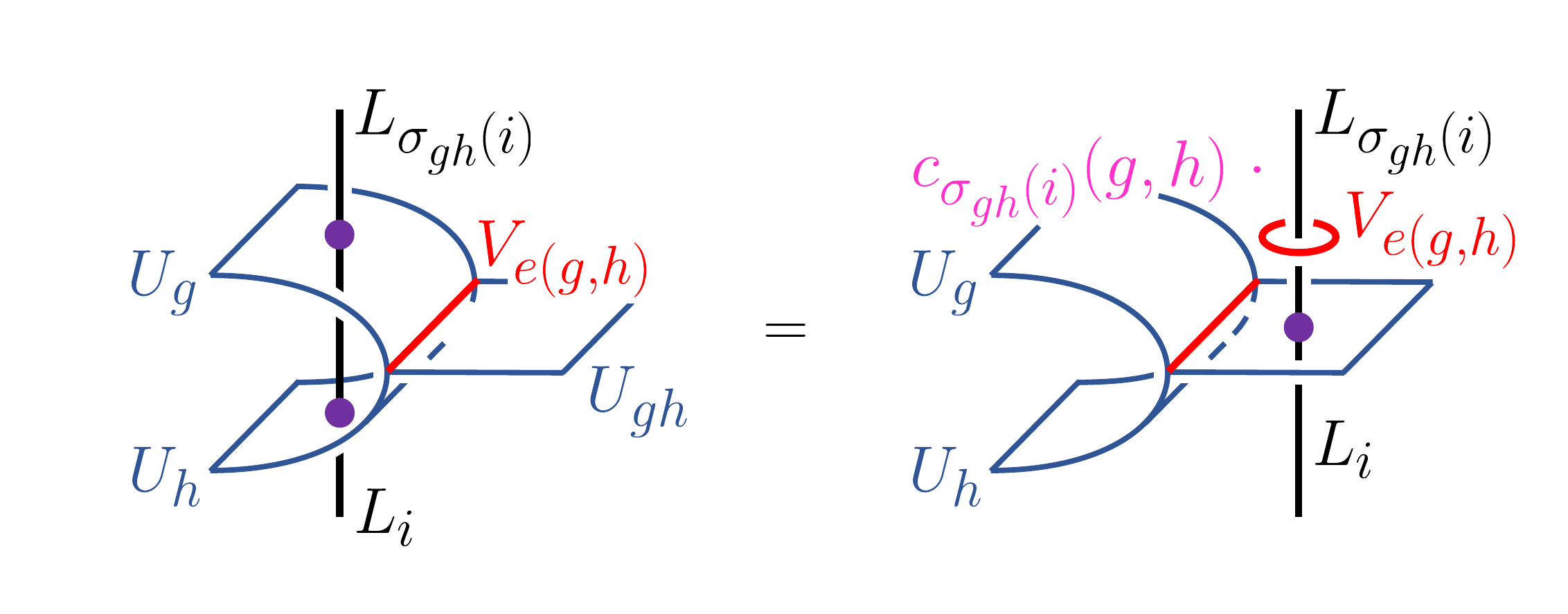}
	\vspace{-5pt}
	\caption{}
	\label{fig:symmetry-fractionalisation-2}
\end{figure}

It is straightforward to see from~\eqref{eq:dc-2group} that shifting the trivialisation results in a shift of the projective phases of 2-representations involving line defects $L_i(\gamma)$ charged under the 1-form symmetry,
\begin{equation}
c \; \to \; c \cdot \langle \chi , e \rangle \, .
\label{eq:symm-frac-shift}
\end{equation}
This may be understood graphically as illustrated in figure~\ref{fig:symmetry-fractionalisation-2}. 
Namely, dragging the fusion of $g,h \in G$ through a line defect $L_i(\gamma)$ generates an additional linking by the 1-form symmetry defect $e(g,h) \in A$ and therefore a phase
\begin{equation}
\chi_{\sigma_{gh}(i)}(e(g,h)) \; \in \; U(1)
\end{equation}
in addition to the projective phase $c_{\sigma_{gh}(i)}(g,h)$ from before. This reproduces the shift in equation~\eqref{eq:symm-frac-shift}.

\subsubsection{Junction Operators}

\begin{figure}[h]
	\centering
	\includegraphics[height=3.4cm]{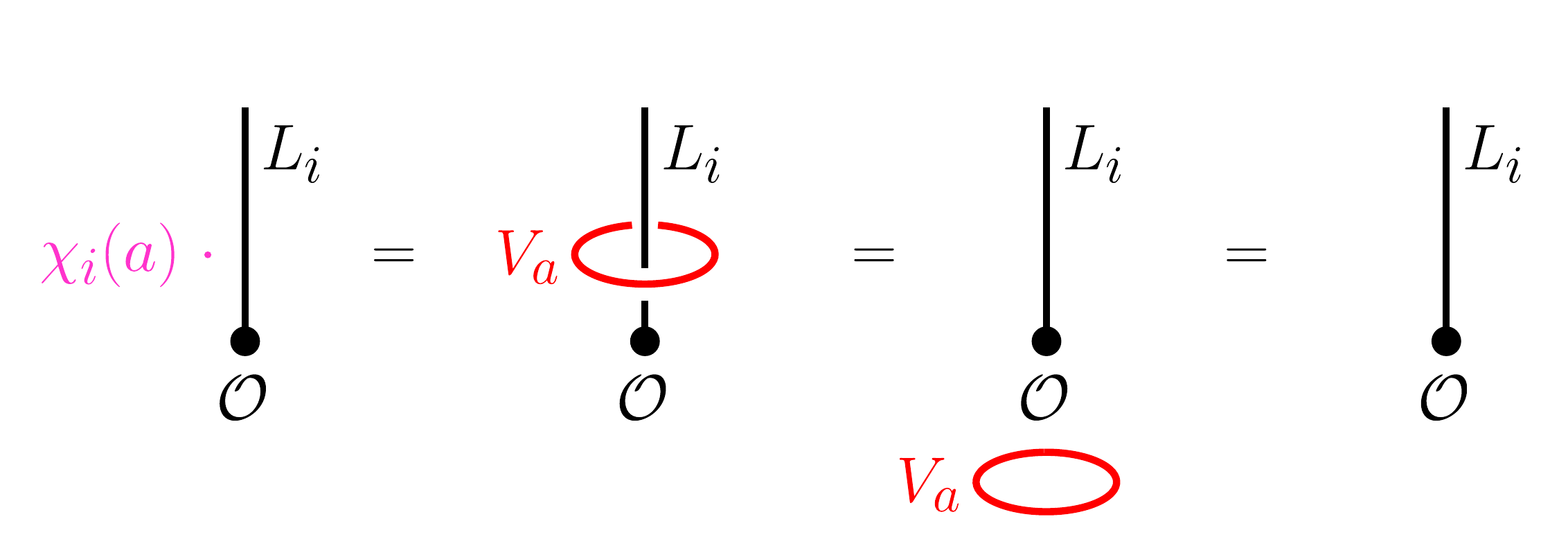}
	\vspace{-5pt}
	\caption{}
	\label{fig:linking-action-3}
\end{figure}

Let us now consider finite-dimensional vector spaces $V_i$ of local operators on which $L_i$ end. In the presence of a 1-form symmetry $A$, they are trivial unless the corresponding character $\chi_i \in A^{\vee}$ is trivial. This captures the fact that if $L_i$ can end on a local operator $\mathcal{O}$, it cannot be charged under a 1-form symmetry due to the possibility to unlink any codimension-two defect  as illustrated in figure \ref{fig:linking-action-3}.

From the perspective of higher representation theory, this reproduces the fact that 1-intertwiners between the trivial 2-representation and $(\sigma,c,\chi)$ are given by graded projective representations of $G$ of type $(\sigma,c)$ whose support is restricted to those $i \in \lbrace 1,...,n\rbrace$ for which $\chi_i=1$~\cite{ELGUETA200753}.

More generally, line defects $L_i$, $L'_j$ that admit local operator junctions $\mathcal{O}$ must have the same 1-form charge due to similar arguments as above. This reproduces the fact that 1-intertwiners between 2-representations $(\sigma,c,\chi)$ and $(\sigma',c',\chi')$ of of a 2-group $\mathcal{G}$ are given by graded projective representations $G$ of type $(\sigma \otimes \sigma', \Bar{c} \otimes c')$ whose support is restricted to those $(i,j) \in \lbrace 1,...,n \rbrace \times \lbrace1,...,n' \rbrace$ for which $\chi_i = \chi'_j$~\cite{ELGUETA200753}.

\subsection{Induction perspective}
\label{subsec:lines-2groups-induction}

\begin{figure}[h]
	\centering
	\includegraphics[height=3.7cm]{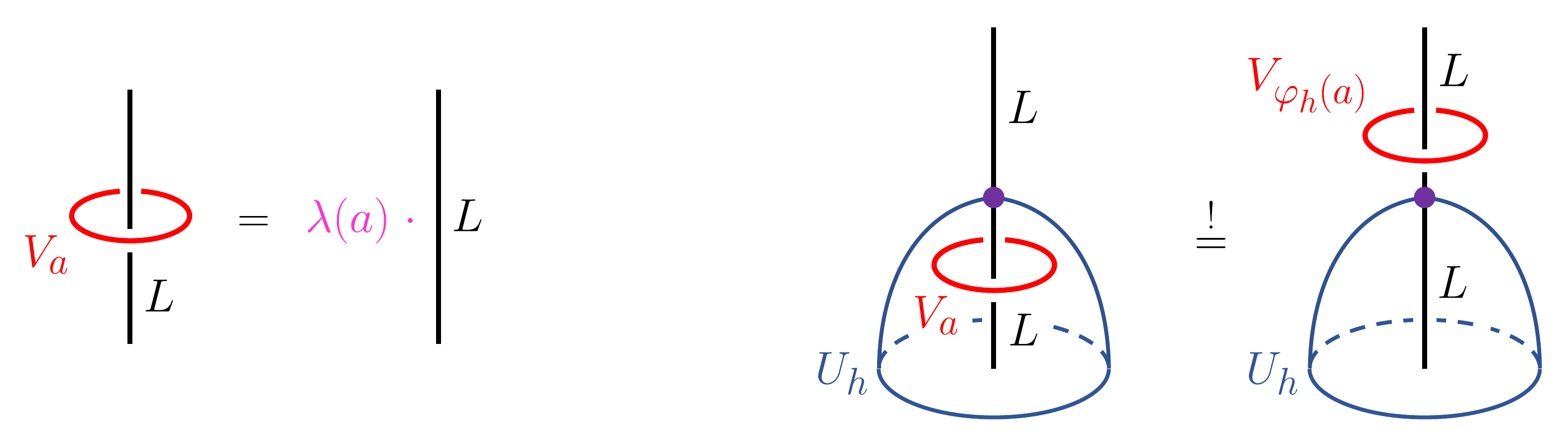}
	\vspace{-5pt}
	\caption{}
	\label{fig:induction-linking}
\end{figure}

Let us now consider the generalisation of the induction approach introduced in subsection~\ref{subsec:lines-groups-induction}. We again fix a line $L = L_1$ transforming in an irreducible 2-representation $(\sigma,c,\chi)$ with stabiliser subgroup $H := \text{Stab}_{\sigma}(1) \subset G$. This line may now be charged under the 1-form symmetry $A$, described by a character $\lambda \, := \, \chi_1 \, \in \, A^{\vee}$, as illustrated on the left of figure \ref{fig:induction-linking}.

This data must again be compatible with the 2-group structure. First, linking $L$ with a 1-form symmetry defect $a \in A$ and subsequently wrapping with a 0-form symmetry defect $h \in H$ must be equivalent to linking $L$ with the transformed 1-form symmetry defect $\varphi_h(a) \in A$, as illustrated on the right of figure \ref{fig:induction-linking}. This gives the condition
\begin{equation}
\lambda(a) \; \stackrel{!}{=} \; \lambda(\varphi_h(a)) \, ,
\end{equation} 
meaning the character $\lambda \in A^\vee$ must be $H$-invariant.

Furthermore, the collection of phases $u(h,h') := c_1(h,h')$ arising from multiplication of intersections of $H$-defects with $L$ are required to satisfy
\begin{equation}
(du)(h_1,h_2,h_3) \; \stackrel{!}{=} \; \lambda \big(\alpha(h_1,h_2,h_3) \big) \, ,
\end{equation}
where $\alpha \in Z^3_{\varphi}(G,A)$ is a representative of the Postnikov class describing the non-associati-vity of the fusion of 0-form symmetry defects in the bulk. 

%Omitting the arguments, the condition is $d u = \langle \lambda , \alpha|_H \rangle$. Note that if the character $\lambda$ is trivial, $L$ can end on local operators transforming in a projective representation of $H$ with 2-cocycle $u$.

In summary, the action of the 2-group $\mathcal{G} = (G,A,\varphi,\alpha)$ on a line defect $L$ is determined by a triple $(H,u,\lambda)$ consisting of
\begin{enumerate}
\item a subgroup $H \subset G$,
\item a 2-cochain $u \in C^2(H,U(1))$,
\item a $H$-invariant character $\lambda \in A^{\vee}$,
\end{enumerate}
such that $du = \langle \lambda, \alpha|_H \rangle$. Note that this data corresponds to a 1-dimensional 2-representa-tion of the 2-subgroup $\mathcal{H} = (H,A,\varphi|_H, \alpha|_H) \subset \mathcal{G}$. Similarly to the discussion in section~\ref{subsec:lines-groups-induction}, the original irreducible 2-representation $(\sigma,c,\chi)$ can be reconstructed from this data by induction,
\begin{equation}
(\sigma,c,\chi) \; = \; \text{Ind}_{\mathcal{H}}^{\mathcal{G}}(u,\lambda) \, .
\end{equation}
It is known that all irreducible 2-representations arise in this way from induction of 1-dimensional 2-representations of subgroups~\cite{ELGUETA200753,OSORNO2010369}. 

Similarly to before, two irreducible 2-representations $(H,u,\lambda)$ and $(H',u',\lambda')$ are considered equivalent if there exists a $g \in G$ such that
\begin{equation}
H' \, = \, {}^{g\!}H \, , \qquad [u'] \, = \, [{}^{g}u] \, , \qquad \lambda' \, = \, {}^{g\!}\lambda \, , 
\end{equation}
where we denoted $({}^{g\!}\lambda)(.) := \lambda(\varphi^{-1}_g(.) )$.

The tensor product of two irreducible 2-representations $(H,u,\lambda)$ and $(H',u',\lambda')$ is in general not irreducible but decomposes as a direct sum of irreducible 2-representations according to
\begin{equation}
(H,u,\lambda) \, \otimes \, (H',u',\lambda') \;\; = \bigoplus_{[g] \, \in \, H \backslash G / K} \big( H \cap {}^{g\!}H', \; u \cdot {}^gu', \; \lambda \cdot {}^g\lambda' \big) \, ,
\end{equation}
which generalises the result from (\ref{eq:tensor-product-2-reps-of-groups}).

The space of 1-intertwiners between the trivial 2-representation and $(H,u,\lambda)$ is given by projective representations of $H$ with 2-cocycle $u$ provided that $\lambda = 1$, otherwise it is trivial. This again reflects the fact that if the line $L$ can end on non-genuine local operators, its 1-form charge $\lambda$ must be trivial.

Lastly, the conjugate of an irreducible 2-representation $(H,u,\lambda)$ is the irreducible 2-representation $(H,\Bar{u},\Bar{\lambda})$ obtained by complex conjugating $u$ and $\lambda$.

\subsection{Categorical perspective}
\label{ssec:lines-categorical-2groups}

Let us now extend the categorical perspective on 2-representations presented in subsection~\ref{ssec:lines-categorical-groups} to incorporate 1-form symmetries and 2-groups. 

\begin{figure}[h]
	\centering
	\includegraphics[height=5.5cm]{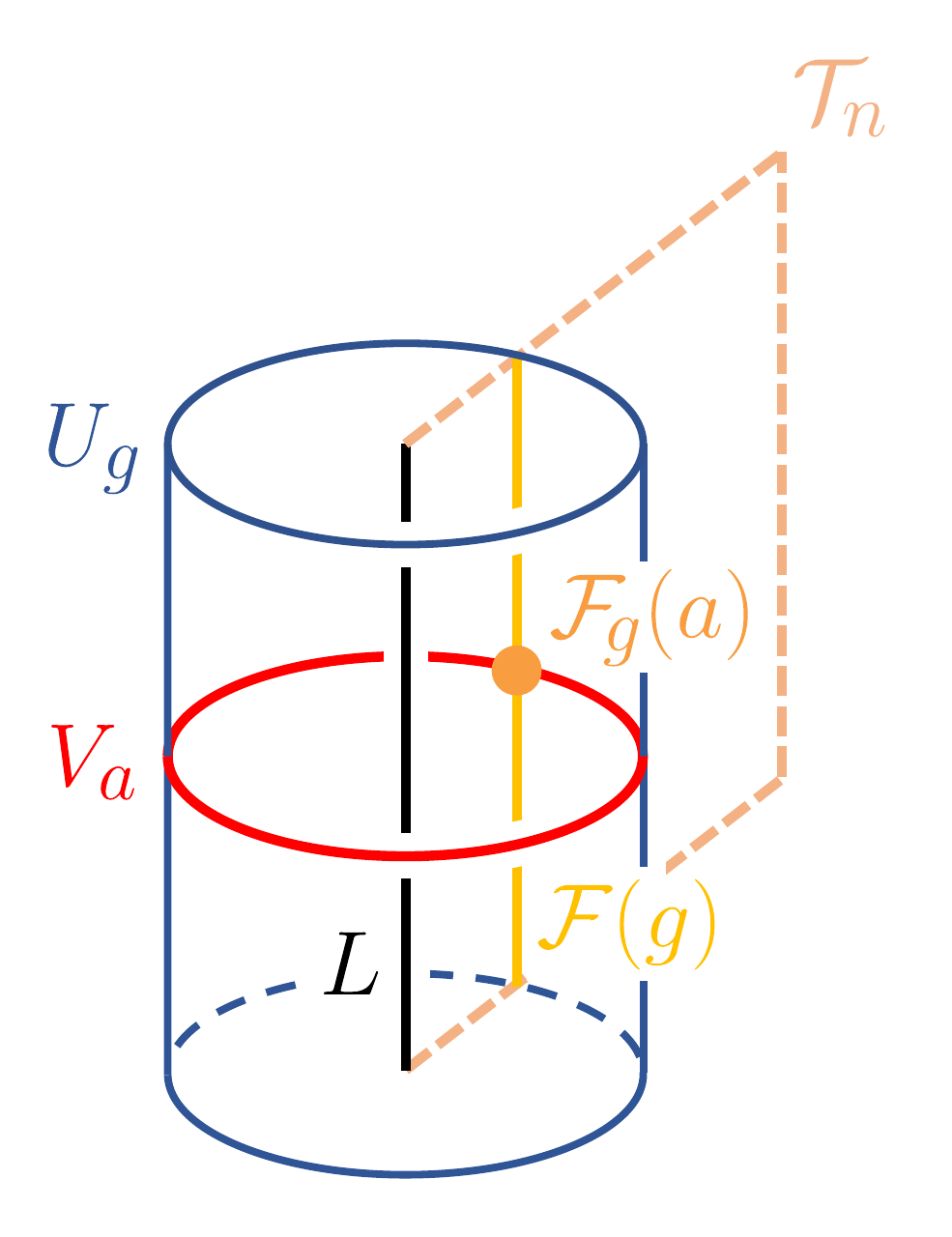}
	\vspace{-5pt}
	\caption{}
	\label{fig:2d-tqft-3}
\end{figure}

\begin{figure}[h]
	\centering
	\includegraphics[height=11.7cm]{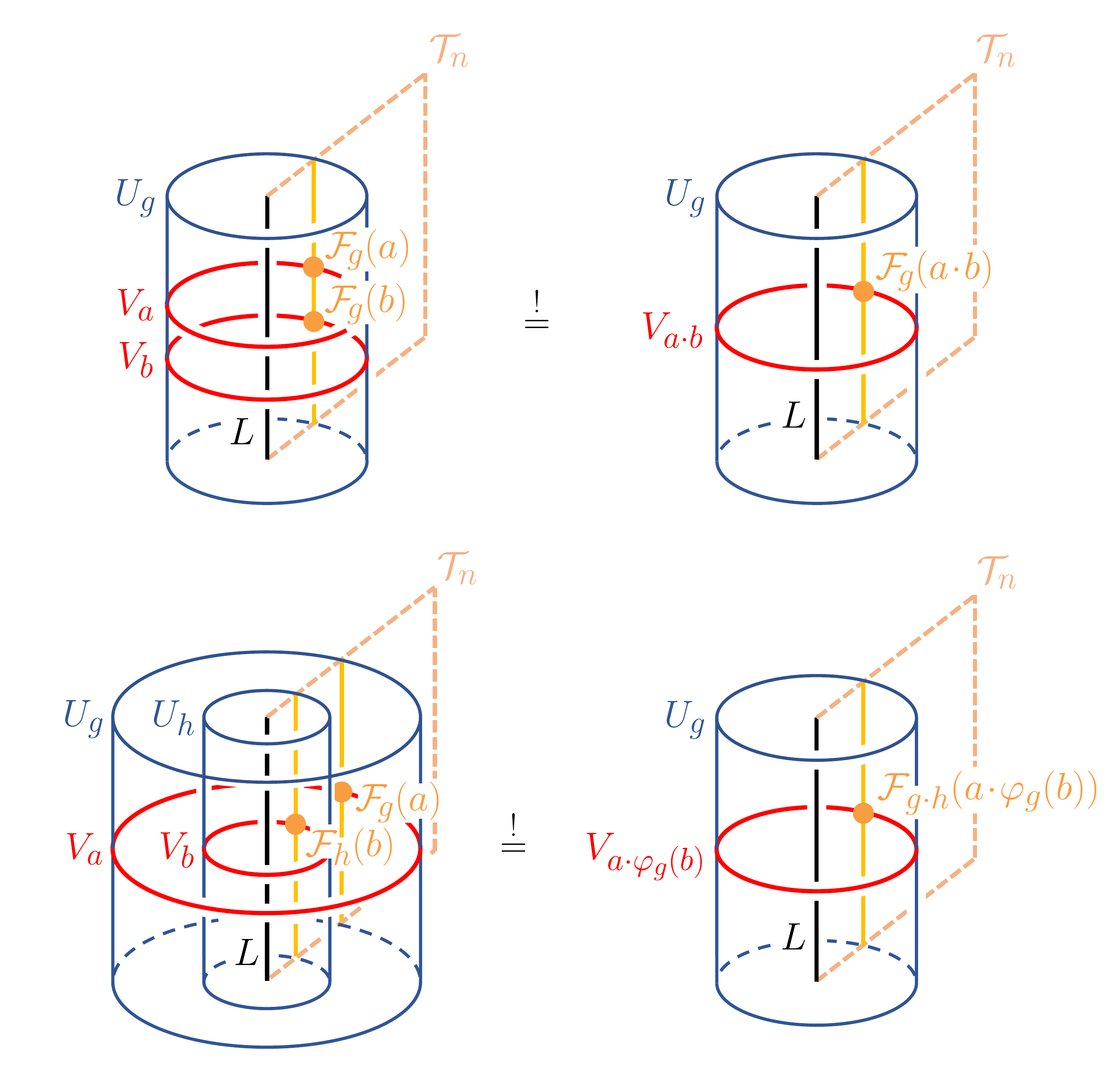}
	\vspace{-5pt}
	\caption{}
	\label{fig:2d-tqft-7}
\end{figure}

Recall that we identify lines defects $L_i(\gamma)$ indexed by $i = 1,\ldots,n$ with boundary conditions for an auxiliary 2d TQFT $\mathcal{T}_n$. In order to incorporate the 1-form symmetry $A$, we now consider linking line defects $L$ with configurations of 0-form and 1-form symmetry defects illustrated in figure \ref{fig:2d-tqft-3}.
In addition to the topological intersection $\cF(g)$, this now requires a choice of a point-like intersection $\mathcal{F}_g(a)$ between the 1-form defect $V_a$ and $\mathcal{T}_n$. This may be regarded as a topological local operator on the topological line $\mathcal{F}(g)$ and can therefore be identified with a natural transformation 
\be
\mathcal{F}_g(a): \; \mathcal{F}(g) \, \Rightarrow \, \mathcal{F}(g)
\ee
or, equivalently, a 2-morphism in $2\mathsf{Vec}$. These natural transformations must be compatible with the intersection of two 1-form symmetry defects $a,b \in A$ in the sense that
\begin{equation}\label{eq:1-form-2-homs-compatibility}
\mathcal{F}_g(a) \, \circ \, \mathcal{F}_g(b) \; \stackrel{!}{=} \; \mathcal{F}_g(a \cdot b) \qquad \text{and} \qquad \mathcal{F}_g(a) \, \star \, \mathcal{F}_h(b) \; \stackrel{!}{=} \; \mathcal{F}_{g \cdot h}(a \cdot \varphi_g(b)) \, .
\end{equation}
These conditions reflect the fact that linking with two 1-form symmetry defects $a,b \in A$ either vertically or horizontally is equivalent to linking it with their fusion, as illustrated in figure \ref{fig:2d-tqft-7}.

In summary, the additional data arising from the 1-form symmetry $A$ is a collection of natural transformations $\mathcal{F}_g(a)$ (2-morphisms in $\mathsf{2Vec}$) satisfying the compatibility conditions (\ref{eq:1-form-2-homs-compatibility}). Together with the 1-morphisms $\mathcal{F}(g)$ and 2-morphisms $F_{g,h}$ from subsection~\ref{ssec:lines-categorical-groups}, this data can be recognised as a (pseudo-)2-functor 
\begin{equation}
\mathcal{F} : \; \widehat{\mathcal{G}} \, \to \, \mathsf{2Vec} \, ,
\end{equation}
where the 2-group $\mathcal{G} = (G,A,\varphi,\alpha)$ is regarded as a 2-category $\widehat{\mathcal{G}}$ with a single object $\ast$ with 1-endomorphisms $\text{1-End}_{\widehat{\mathcal{G}}}(\ast) = G$ and 2-morphisms $\text{2-Hom}_{\widehat{\mathcal{G}}}(g,h) = \delta_{g,h} \cdot A$. 

The collection of such 2-functors again forms a 2-category $[\widehat{\mathcal{G}},\mathsf{2Vec}]$ which can be recognised as the 2-category of finite-dimensional 2-representations of $\mathcal{G}$,
\begin{equation}
[\widehat{\mathcal{G}},\mathsf{2Vec}] \; = \; \mathsf{2Rep}(\mathcal{G}) \, .
\end{equation}
Equivalently, this may be regarded as the 2-category of $\mathcal{G}$-equivariant framed fully-extended 2d TQFTs.

This reproduces from a more abstract perspective the fact that line defects transform in 2-representations of a finite 2-group symmetry $\cG$. It again enables us to abstract the construction of 2-representations away from specific line defects and to reframe it into the existence of $\cG$-equivariant structures on 2d TQFTs. It is known that irreducible 2-representations of finite 2-groups are classified by the concrete data in subsections~\ref{subsec:lines-2groups-elementary} and~\ref{subsec:lines-2groups-induction}, as described in~\cite{ELGUETA200753} and summarised in appendix \ref{app:2-representations}.

\subsection{Examples}

Let us conclude this section with examples of line defects transforming in 2-representations of a finite 2-group symmetry. We will first consider classes of line operators that arise naturally in gauge theory and continue with 2-representations that are induced by certain types of mixed anomalies.

\subsubsection{Gauge theory}

Consider again a pure gauge theory in $D$ dimensions with a simple connected gauge group $\mathbb{G}$. In addition to the charge conjugation symmetry $G = \text{Out}(\mathbb{G})$ discussed in subsection \ref{ssec:gauge-theory-charge-conjugation}, this theory has an electric 1-form symmetry $A = Z(\mathbb{G})$, which together with $G$ forms a split 2-group $\mathcal{G} =(G,A,\varphi)$ called the automorphism 2-group of $\mathbb{G}$. Here, the action $\varphi: G \to \text{Aut}(A)$ of outer automorphisms $[f] \in \text{Out}(\mathbb{G})$ on central group elements $z \in Z(\mathbb{G})$ defined by $\varphi_{[f]}(z) = f(z)$.

We consider the action of $\mathcal{G}$ on Wilson lines $W_R(\gamma)$ labelled by a representation $R$ of $\mathbb{G}$. As discussed in subsection \ref{ssec:gauge-theory-charge-conjugation}, outer automorphisms $[f] \in G$ act by
\begin{equation}\label{eq:Wilson-line-action}
    [f] \, \triangleright \, W_R(\gamma) \; = \; W_{R \, \circ  f^{-1}}(\gamma) \, ,
\end{equation}
and induces a permutation action $\sigma$ of $G$ on the set of isomorphism classes of irreducible representations of $\mathbb{G}$.

In addition, central elements $z \in A$ now act by
\begin{equation}
    z \, \triangleright \, W_R(\gamma) \; = \; \chi_R(z) \cdot W_R(\gamma) \, ,
    \label{eq:center-Wilson-loops}
\end{equation}
where $\chi_R \in A^{\vee}$ is the central character associated to the irreducible representation $R$. The central characters are compatible with the permutation action $\sigma$ in the sense that
\begin{equation}
\begin{aligned}
    \chi_{\,\sigma_{[f]}(R)}(z) \; &= \; \frac{1}{\text{dim}(\sigma_{[f]}(R))} \, \cdot \, \text{Tr}\big[ \, \sigma_{[f]}(R)(z) \, \big] \\
    &= \; \frac{1}{\text{dim}(R)} \, \cdot \, \text{Tr}\big[\,(R \circ f^{-1})(z)\,\big] \; = \; \chi_R\big(\,\varphi_{[f]^{-1}}(z)\,\big)
\end{aligned}
\end{equation}
for all outer automorphisms $[f] \in G$ and central elements $z \in A$.

The collection of central characters together with the permutation action thus defines a 2-representation of the automorphism split 2-group on the set of Wilson lines. Let us look at this in a few concrete examples:

\begin{itemize}
    \item Consider $\mathbb{G} = SU(N)$ with $N > 2$ so that $\text{Out}(\mathbb{G}) = \mathbb{Z}_2 =: \braket{c}$ and $Z(\mathbb{G}) = \mathbb{Z}_N =: \braket{z}$. We denote by $\Lambda^{\!k}(\mathbf{N})$ the $k$-th antisymmetric power  of the fundamental representation, on which charge conjugation acts by $c \triangleright \Lambda^{\!k}(\mathbf{N}) = \Lambda^{\!N-k}(\mathbf{N})$ for $k =1,...,N$ (note that $\Lambda^{\!N-1}(\mathbf{N}) = \overline{\mathbf{N}}$). The associated central characters are given by
    \begin{equation}
        \chi_{\Lambda^{\!k}(\mathbf{N})}(z) \, = \, e^{\frac{2\pi i k}{N}} \, ,
    \end{equation}
    which are compatible with the action of $\mathbb{Z}_2$ on $\mathbb{Z}_N$ given by $c \triangleright z = z^{-1}$. The Wilson lines $W_{\Lambda^{\!k}(\mathbf{N})}$ and $W_{\Lambda^{\!N-k}(\mathbf{N})}$ thus transform in a two-dimensional irreducible 2-representation labelled by the trivial subgroup of $\mathbb{Z}_2$ and the $k$-th power of the generator of $\mathbb{Z}_N^{\vee} \cong \mathbb{Z}_N$. An exception is $N$ even and $k = \frac{N}{2}$, in which case $W_{\Lambda^{\!N/2}(\mathbf{N})}$ transforms in a one-dimensional irreducible 2-representation labelled by the full 0-form group $\mathbb{Z}_2$ and the invariant character $\chi_{\Lambda^{\!N/2}(\mathbf{N})}(z) = -1$.

    \item Consider $\mathbb{G} = \text{Spin}(2N)$ with $N > 4$ so that $\text{Out}(\mathbb{G}) = \mathbb{Z}_2 =: \braket{c}$. Charge conjugation $c$ exchanges the spinor and conjugate spinor representations $S^{\pm}$ of $\text{Spin}(2N)$. The associated central characters depend on whether $N$ is even or odd:

    \begin{itemize}
        \item[$\circ$] If $N$ is even, $Z(\mathbb{G}) = \mathbb{Z}_2 \times \mathbb{Z}_2 =: \braket{z_1,z_2}$ and the central characters associated to the two spinor representations are
        \begin{equation}
        \begin{aligned}[c]
            \chi_{S^+}(z_1) \, &= \, 1 \\
            \chi_{S^+}(z_2) \, &= \, -1
        \end{aligned}
        \qquad\text{and}\qquad
        \begin{aligned}[c]
            \chi_{S^-}(z_1) \, &= \, -1 \\
            \chi_{S^-}(z_2) \, &= \, 1
        \end{aligned} 
        \quad .
        \end{equation}
        These are compatible with the action of $\mathbb{Z}_2$ on $\mathbb{Z}_2 \times \mathbb{Z}_2$ given by $c \triangleright z_1 = z_2$. The Wilson lines $W_{S^{\pm}}$ thus transform in the two-dimensional irreducible 2-representation labelled by the trivial subgroup of $\mathbb{Z}_2$ and the generator of the second factor of $(\mathbb{Z}_2 \times \mathbb{Z}_2)^{\vee} \cong \mathbb{Z}_2 \times \mathbb{Z}_2$.

        \item[$\circ$] If $N$ is odd, $Z(\mathbb{G}) = \mathbb{Z}_4 =: \braket{z}$ and the central characters associated to the two spinor representations are
        \begin{equation}
        \chi_{S^+}(z) \, = \, i \qquad \text{and} \qquad \chi_{S^-}(z) \, = \, -i \; .
        \end{equation}
        These are compatible with the action of $\mathbb{Z}_2$ on $\mathbb{Z}_4$ given by $c \triangleright z = z^3$. The Wilson lines $W_{S^{\pm}}$ thus transform in the two-dimensional irreducible 2-representation labelled by the trivial subgroup of $\mathbb{Z}_2$ and the generator of $\mathbb{Z}_4^{\vee} \cong \mathbb{Z}_4$.
    \end{itemize}

    \item If $\mathbb{G} = \text{Spin}(8)$, charge conjugation enhances to the non-abelian symmetric group
    \begin{equation}
        \text{Out}(\mathbb{G}) \, = \, S_3 \, = \, \mathbb{Z}_3 \rtimes \mathbb{Z}_2 \, =: \, \braket{r,c} \, ,
    \end{equation}
    which permutes $S^{\pm}$ and the vector representation $V$ as in (\ref{eq:triality}). The associated central characters of $Z(\mathbb{G}) = \mathbb{Z}_2 \times \mathbb{Z}_2 =: \braket{z_1,z_2}$ are
    \begin{equation}
    \begin{aligned}[c]
        \chi_{S^+}(z_1) \, &= \, 1 \\
        \chi_{S^+}(z_2) \, &= \, -1
    \end{aligned}
    \;\; , \qquad
    \begin{aligned}[c]
        \chi_{S^-}(z_1) \, &= \, -1 \\
        \chi_{S^-}(z_2) \, &= \, 1
    \end{aligned}
    \qquad\text{and}\qquad
    \begin{aligned}[c]
        \chi_V(z_1) \, &= \, -1 \\
        \chi_V(z_2) \, &= \, -1
    \end{aligned} 
    \;\; .
    \end{equation}
    These are compatible with the action of $S_3$ on $\mathbb{Z}_2 \times \mathbb{Z}_2$ given by
    \begin{equation}
        r \,\triangleright\, z_1 \; = \; z_1 \cdot z_2 \, , \qquad r \, \triangleright \, z_2 \; = \; z_1 \, , \qquad c \, \triangleright \, z_1 \; = \; z_2 \, .
    \end{equation}
    The Wilson lines $W_{S^{\pm}}$, $W_V$ thus transform in the three-dimensional irreducible 2-representation labelled by the $\mathbb{Z}_2$-subgroup $\braket{rc} \subset S_3$ and the $\braket{rc}$-invariant character corresponding to the generator of the second factor of $(\mathbb{Z}_2 \times \mathbb{Z}_2)^{\vee} \cong \mathbb{Z}_2 \times \mathbb{Z}_2$.
\end{itemize}

\subsubsection{U(1) gauge theory with charged matter}

Consider now a $\mathbb{G}=U(1)$ gauge theory with two charged complex scalar fields $\Phi_1$ and $\Phi_2$, both of which with charge $q=2$. 
If we neglect the charge conjugation and the magnetic $U(1)$ $0$-form symmetries, we can focus on the split $2$-group symmetry $\mathcal{G}=(SO(3),\bZ_2,1,w_3)$ that the theory enjoys. 

In particular, $SO(3)\cong SU(2)/\bZ_2$ is the flavour symmetry of the theory, where the $\bZ_2$ quotient is given by the diagonal global transformation $\Phi_I \rightarrow -\Phi_I$, which can be reabsorbed via a gauge transformation $e^{i\pi/q}\in U(1)$. 
Instead, the $1$-form component is given by the standard center $Z(\mathbb{G})=U(1)$ $1$-form symmetry generated by Gukov-Witten defects $Y_{[g]}(\gamma)$ broken down to $\mathbb{Z}_q$, which corresponds to the defects that are still topological in the presence of charged matter. 
Since the flavour symmetry acts trivially on these Gukov-Witten operators, the $2$-group must be split. However, $\mathcal{G}$ is characterised by a non-trivial Postnikov class $w_3=\text{Sq}^1(w_2)\in H^3(BSO(3),\bZ_2).$
To see this, we can turn on a non-trivial background for $SO(3)$ which does not lift to a valid one for $SU(2)$, i.e. corresponding to a non-trivial Stiefel-Whitney class $w_2\in H^2(BSO(3),\bZ_2)$.
Note then that there is a subgroup of gauge transformations $\bZ_4\subset U(1)$ that satisfies
\be 
\begin{tikzcd}
    1 \ar[r]& \bZ_2 \ar[r,hook]& \bZ_{4} \ar[r,"\pi"]& \bZ_2 \ar[r] &1,
\end{tikzcd}
\label{eq:short-exact-sequence-z2z4z2}
\ee 
where the first $\bZ_2$ subgroup describes gauge transformations which act trivially on the local fields $\Phi_I$, but not on Wilson lines. The resulting $\pi(\bZ_{4})\cong \bZ_2$ corresponds instead to the $\bZ_2$ quotient that affects the $0$-form symmetry structure, namely $SO(3)\cong SU(2)/\pi(\bZ_4)$. From \eqref{eq:short-exact-sequence-z2z4z2} it is a standard result \cite{Benini:2018reh} how the associated Bockstein homomorphism $\text{Bock}:H^*(-,\bZ_2)\rightarrow H^{*+1}(-,\bZ_2)$ induces a non-trivial Postnikov class $\text{Bock}(w_2)\equiv\text{Sq}^1(w_2)=w_3$.

Therefore, we expect extended operators to transform under $2$-representations of $\mathcal{G}$, in particular Wilson lines $W_n(\gamma)$, each of which must provide a unique irreducible $2$-representation $\mathcal{R}_n$, since the flavour symmetry acts trivially on them. 
While we do not provide a full classification of $2$-representations for continuous $2$-groups, we can restrict ourselves to study the pullback $\mathcal{F}_n=\kappa^* \mathcal{R}_n$ with respect to discrete $2$-subgroups $\mathcal{K}\overset{\kappa}{\hookrightarrow}\mathcal{G}$. In particular, we focus our attention on the case $\mathcal{K}=(\bZ_2\times \bZ_2,\bZ_q,1,\kappa^*w_3)$.

Let us start by evaluating $\kappa^*w_3$. Being the Bockstein map a natural cohomology operation, it follows that
\be
\kappa^* \text{Sq}^1(w_2) = \text{Sq}^1(\kappa^* w_2)\,,
\ee
where $\kappa^*w_2$ is the characteristic class that defines the extension of $\bZ_2\times\bZ_2$ by $\bZ_2$ in $SU(2)$. In other words, $\kappa^*w_2$ is the characteristic class that makes the following diagram commute
\be
\begin{tikzcd}
    1 \ar[r] &\bZ_2 \ar[r] & SU(2) \ar[r] & SO(3) \ar[r] & 1\\
    1 \ar[r] &\bZ_2 \ar[r] \ar[u] & \mathbb{Q}_8 \ar[r] \ar[u, hook] & \bZ_2 \times \bZ_2 \ar[r] \ar[u, hook, "\kappa"] & 1,
\end{tikzcd}
\label{eq:Q8}
\ee
where $\mathbb{Q}_8=\{\pm 1,\pm \mathbf{i},\pm \mathbf{j}, \pm \mathbf{k}\}$ is the group of quaternions. We can also interpret $\kappa^* w_2\equiv w_2(\rho)$ as the characteristic class associated to the $3$-dimensional representation $\rho=\rho_{1,0}\oplus\rho_{1,1}\oplus\rho_{0,1}$ of $\bZ_2\times\bZ_2$ in $SO(3)$ \cite{GUNARWARDENA1989327}, where $\rho_{n,m}$ denotes the representation of charges $(n,m)$. It is then easy to find that \be
w_2(\rho)= w_1(\rho_{1,0}) \cup w_1(\rho_{1,1}) + w_1(\rho_{1,1})\cup w_1(\rho_{0,1}) + w_1(\rho_{1,0})\cup w_1(\rho_{0,1})
\ee 
is non-trivial, where $w_1(\rho_{n,m})=\text{det}(\rho_{n,m})$. Moreover, it follows that 
\be
\text{Sq}^1(w_2(\rho)) = w_3(\rho) = w_1(\rho_{1,0})\cup w_{1}(\rho_{1,1})\cup w_1(\rho_{0,1})
\ee
describes the non-trivial generator of $H^3(\bZ_2\times \bZ_2,\bZ_2)$.

To determine the representations $\mathcal{F}_n$, we can first notice that by overlapping Wilson lines with various charges, say $n$ and $m$, we must have 
\be 
\mathcal{F}_{n+m} = \mathcal{F}_n\oplus \mathcal{F}_m.
\label{eq:2-rep-decomposition}
\ee
The first piece of data that classify $\mathcal{F}_n$ is $\chi_n:\bZ_2\rightarrow U(1)$ describing the linking phase between $W_n(\gamma)$ and the $\bZ_2$-generating Gukov-Witten defects. Based on \eqref{eq:center-Wilson-loops}, $\chi_n$ are simply the characters of the representation carried by $W_n(\gamma)$ restricted on the subgroup $\bZ_2\subset U(1)$. Note that this is also in agreement with \eqref{eq:2-rep-decomposition} since $\chi_n = \chi^n_1$. 
The second data left to determine $\mathcal{F}_n$ is then a cochain $c_n\in C^2 (\bZ_2\times\bZ_2,U(1))$ satisfying 
\be
dc_n = \langle \chi_n,w_3(\rho)\rangle,
\label{eq:Cn-2-rep}
\ee
where, because of \eqref{eq:2-rep-decomposition}, we must have again $c_n = c_1^n$. In particular, solutions $c_n$ of \eqref{eq:Cn-2-rep} for $n$ even identify cohomology classes in $H^2(\bZ_2\times\bZ_2,U(1))\cong \bZ_2$, while for $n$ odd define only a torsor over it. 

We can start by analyzing the case $n=2$, which physical meaning is more transparent. Since $W_2(\gamma)$ is endable on local operators $\Phi_I$, $c_2$ is interpreted as the $2$-coycle determining the projective action of the flavor subgroup $\bZ_2\times\bZ_2$ on them. In fact, this is indeed projective up to a $\bZ_2$ phase, determined by the extension \eqref{eq:Q8} and equivalently identified by $w_2(\rho) \in H^2(\bZ_2\times\bZ_2,\bZ_2)$. By regarding $\bZ_2$ subgroup of $U(1)$ via the inclusion $s:\bZ_2\hookrightarrow U(1)$ we can verify that $w_2(\rho)$ is mapped to the non-trivial generating cohomology class\footnote{Here  $x_i=s_*w_1(\rho_{i\text{ mod }2,i+1\text{ mod }2})$ represent the non-trivial generator of $H^1(\bZ_2, U(1))$ for each of the two $\bZ_2$ factors in $\bZ_2\times\bZ_2$. Therefore, in terms of group cohomology $(x_1\cup x_2)(n,m):=\exp(i \pi n_1 m_2)$, where $n_i,\,m_i$ denotes the $i$-esimal $\bZ_2$ components of $n,\,m\in\bZ_2\times\bZ_2$.}
\be 
s_* w_2(\rho) = s_* \left(w_1(\rho_{1,0})\cup w_1(\rho_{0,1})\right)=x_1 \cup x_2 \in H^2(\bZ_2\times \bZ_2,U(1)).
\ee
Based on this argument, we can identify 
\be
c_2=x_1\cup x_2.
\ee
This determine the $2$-representation $\mathcal{F}_{2n}$ for any $W_{2n}(\gamma)$ of even charge: for example, $\mathcal{F}_4= \mathcal{F}_2\oplus\mathcal{F}_2$. Accordingly, $c_4=c_2^2$ describes a trivial cohomology class, in agreement with the fact that $W_{4}(\gamma)$ can end of polynomials of degree $2$ of $\Phi_I$, for which $\bZ_2\times\bZ_2$ is not projective anymore. 
Finally, the $2$-representation associated to Wilson lines of odd charge follows from the two possible solutions $c_{1,3}$ of
\be 
dc_{1,3} = s_* w_3(\rho),
\ee
valued in $H^3(\bZ_2\times\bZ_2,U(1))$ and which will correspond to the cochains associated to $W_{1,3}(\gamma)$. Again, these differ by the non-trivial class $c_2=x_1\cup x_2 \in H^2(\bZ_2\times\bZ_2,U(1))$, representing the constraint $c_3= c_1 \cdot c_2$. Note that the solutions $c_{1,3}$ defining a torsion over $H^2(\bZ_2\times\bZ_2,U(1))$ are a direct consequence of the fact that these Wilson lines can not end on any local operators.

\subsubsection{Anomalous 2-group symmetry}
\label{sec:anomalous-2-group}

Consider a theory in $D=3$ with a finite 2-group symmetry $\mathcal{G} = (G,A,\varphi, \alpha)$. 't Hooft anomalies of 2-groups have been studied in~\cite{Benini:2018reh}.

\begin{figure}[h]
	\centering
	\includegraphics[height=1.3cm]{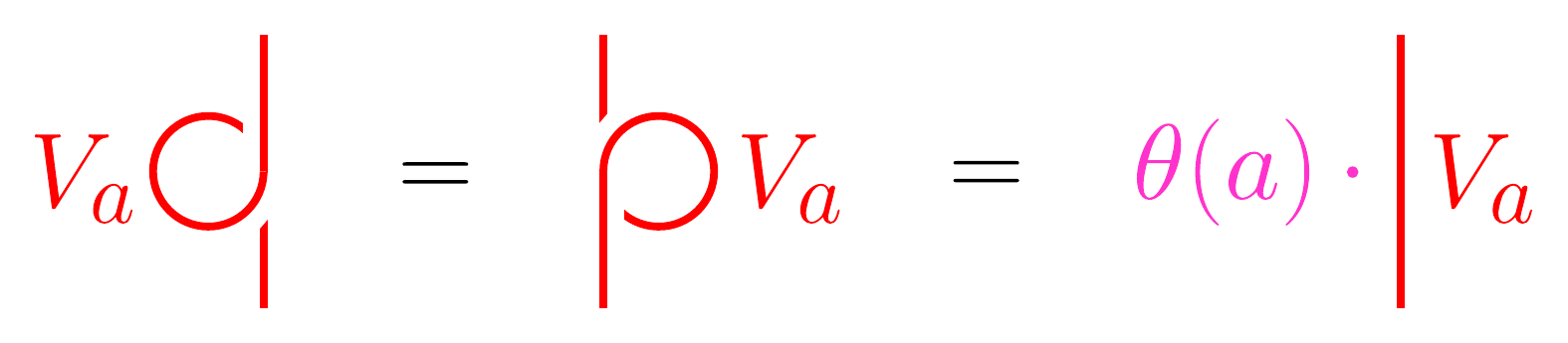}
	\vspace{-5pt}
	\caption{}
	\label{fig:1-form-anomaly-1}
\end{figure}

 Its 1-form symmetry $A$ may have a 't Hooft anomaly classified by the group\footnote{Here, we denote by $B^2\!A$ the classifying space of the 1-form symmetry $A$ (which is an Eilenberg-Maclane space of type $K(A,2)$) and by $\Gamma(A)$ the universal quadratic group of $A$.} 
\begin{equation}
    H^4(B^2\!A,U(1)) \; \cong \; \text{Hom}(\Gamma(A),U(1)) \, ,
\end{equation}
whose elements we can view as quadratic functions $\theta: A \to U(1)$. Physically, these capture the topological spin of the 1-form symmetry in the sense that rotating a line $a \in A$ by $360^{\circ}$ produces a phase $\theta(a) \in U(1)$ as illustrated in figure \ref{fig:1-form-anomaly-1}. 
Since all of the depicted configurations can be moved across any surface $g \in G$ without producing extra phases, the quadratic function $\theta$ must be $G$-invariant in the sense that
\begin{equation}
    \theta(\varphi_g(a)) \; \stackrel{!}{=} \; \theta(a)
\end{equation}
for all $g \in G$ and $a \in A$, where $\varphi: G \to \text{Aut}(A)$ denotes the wrapping action of the 0-form symmetry $G$ on the 1-form symmetry $A$.

\begin{figure}[h]
	\centering
	\includegraphics[height=1.7cm]{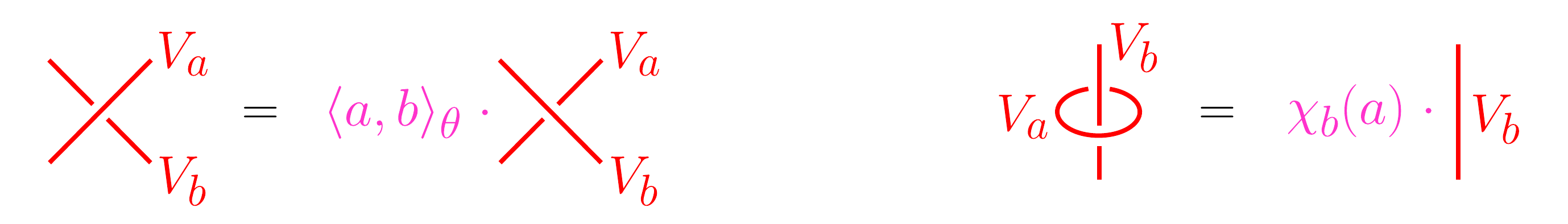}
	\vspace{-5pt}
	\caption{}
	\label{fig:1-form-anomaly-2}
\end{figure}

As a consequence of the topological spin $\theta$, crossing two 1-form symmetry defects $a,b \in A$ produces a phase 
\begin{equation}
    \braket{a,b}_{\theta} \; = \; \frac{\theta(a \! \cdot \! b)}{\theta(a) \! \cdot \! \theta(b)} \; \in \; U(1)
\end{equation}
as illustrated on the left-hand side of figure \ref{fig:1-form-anomaly-2}.
This induces a symmetric bilinear form $\braket{.,.}_{\theta}: A \times A \to U(1)$, which due to the $G$-invariance of $\theta$ is also $G$-invariant. Equivalently, we may regard $\braket{.,.}_{\theta}$ as a collection of characters $\chi_b := \braket{.,b}_{\theta} \in A^{\vee}$ indexed by group elements $b \in A$ that capture the linking of $b$ by lines $a \in A$ as illustrated on the right-hand side of figure \ref{fig:1-form-anomaly-2}. The $G$-invariance of $\braket{.,.}_{\theta}$ then translates into the condition
\begin{equation}\label{eq:anomalous-2-group-compatibility-1}
    \chi_{\varphi^{-1}_g(b)}(a) \; = \; \chi_b(\varphi_g(a)) \, .
\end{equation}

\begin{figure}[h]
	\centering
	\includegraphics[height=3.7cm]{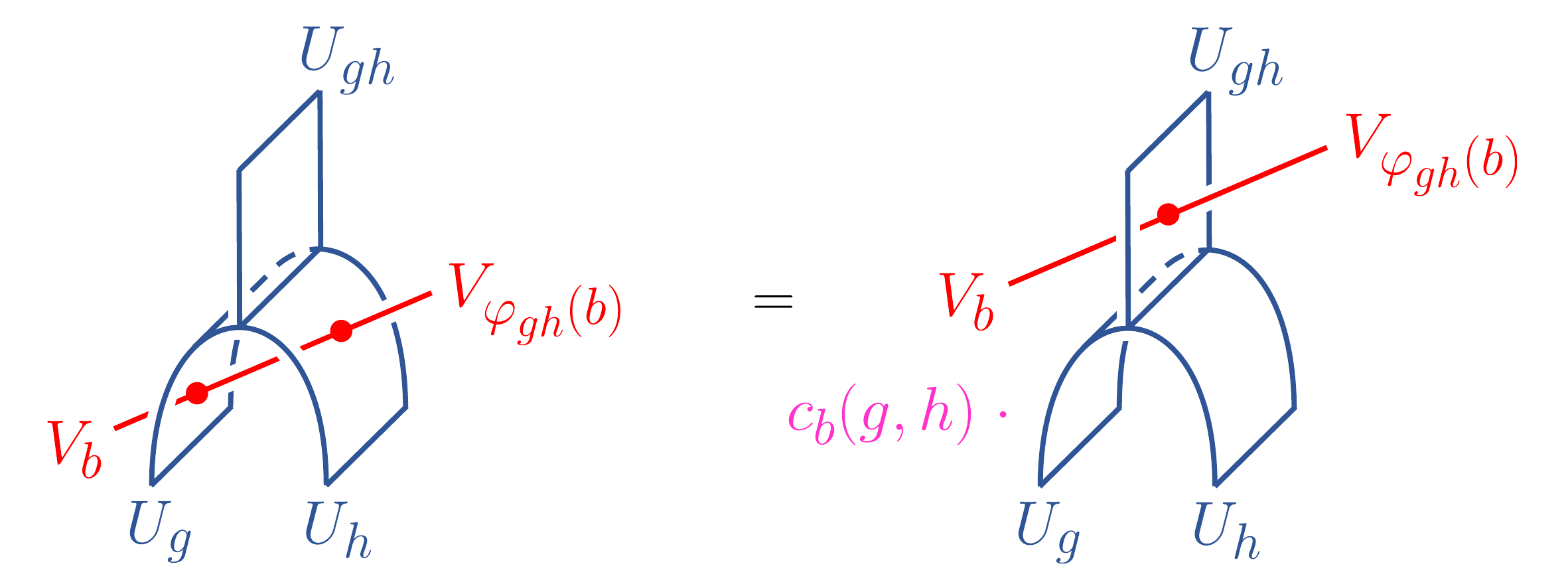}
	\vspace{-5pt}
	\caption{}
	\label{fig:2-group-mixed-anomaly}
\end{figure}

In addition, there may be a mixed anomaly between $G$ and $A$ capturing the fact that moving a 1-form defect $b \in A$ across the junction of two 0-form defects $g,h \in G$ may produce a non-trivial phase $c_b(g,h) \in U(1)$ as illustrated in figure \ref{fig:2-group-mixed-anomaly}.
The collection of these phases may be regarded as a map
\begin{equation}
    c : \; G \times G \, \to \, A^{\vee} \, ,
\end{equation}
which in order to be compatible with the fusion of three 0-form defects $g,h,k \in G$ needs to satisfy the condition
\begin{equation}
(d_{\varphi^{\vee}}c)(g,h,k) \; \stackrel{!}{=} \; \braket{\alpha(g,h,k), \, . \,}_{\theta} \, .
\end{equation}
Here, $\varphi^{\vee}$ denotes the Pontryagin dual action of $\varphi$ and $\alpha \in Z^3_{\varphi}(G,A)$ is the representative of the Postnikov class of the 2-group $\mathcal{G}$. Equivalently, we may regard $c$ as a map
\begin{equation}
    c : \; G \times G \, \to \, U(1)^A
\end{equation}
satisfying the compatibility condition
\begin{equation}\label{eq:anomalous-2-group-compatibility-2}
    (d_{\varphi}c)_b(g,h,k) \; \stackrel{!}{=} \; \chi_b(\alpha(g,h,k)) \, .
\end{equation}

In summary, the triple $(\varphi,c,\chi)$ together with the compatibility conditions (\ref{eq:anomalous-2-group-compatibility-1}) and (\ref{eq:anomalous-2-group-compatibility-2}) defines a a 2-representation of $\mathcal{G}$ that describes how $\mathcal{G}$ acts on the set $A$ of its own 1-form symmetry lines. The associated 2-representation data is induced by the 't Hooft anomalies for $\mathcal{G}$. Conversely, if the 1-form symmetry $A$ transforms in a non-trivial 2-representation of the 2-group $\mathcal{G}$, this indicates the presence of a 't Hooft anomaly.

\section{Surfaces and 3-representations}
\label{sec:surfaces}

In this section, we investigate how surface defects transform in 3-representations of a finite symmetry group $G$. Since the realm of 3d TQFTs is much richer than in one and two dimensions, we will uncover some novel features not present in the case of 1- and 2-representations. Since the mathematical literature on 3-representations is less developed, our exposition will be less systematic. However, we will make contact with the seminal work~\cite{etingof2010fusion}, which will appear as one-dimensional 3-representations of $G$.

We again begin with an elementary approach using properties of topological defects to construct the data of a 3-representation, before reformulating it in a way that uncovers the mathematical notion of induction of 3-representations. We then rephrase the problem in a more categorical context that manifests the abstract definition of 3-representations. 

\subsection{Preliminaries}

\begin{figure}[h]
	\centering
	\includegraphics[height=3.35cm]{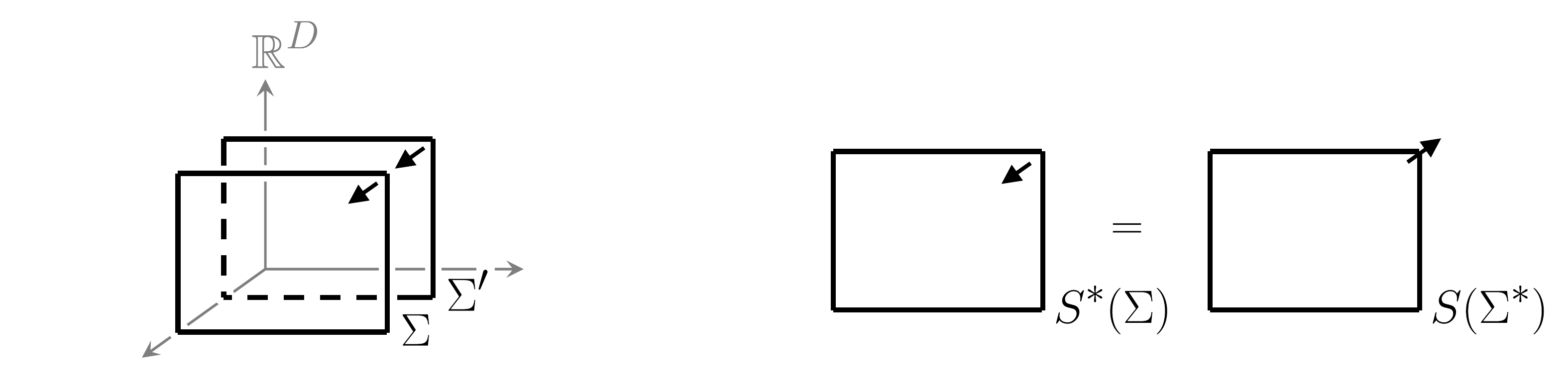}
	\vspace{-5pt}
	\caption{}
	\label{fig:surface-preliminaries}
\end{figure}

We will consider surface defects supported on oriented surfaces $\Sigma,\Sigma',...$ aligned along a common pair of axes in $D$-dimensional euclidean space-time $\mathbb{R}^D$. We draw these axes vertically and horizontally as illustrated on the left-hand side of figure \ref{fig:surface-preliminaries}.

Given a surface defect $S$, we may define a conjugate surface $S^*$ by
\be
S^*(\Sigma) \; := \; S(\Sigma^{\ast}) \, ,
\ee
where $\Sigma^{\ast}$ is the orientation reversal of $\Sigma$ as illustrated in the right-hand side of figure \ref{fig:line-preliminaries}. We will typically omit the orientation from figures, unless stated otherwise.

\begin{figure}[h]
	\centering
	\includegraphics[height=3.4cm]{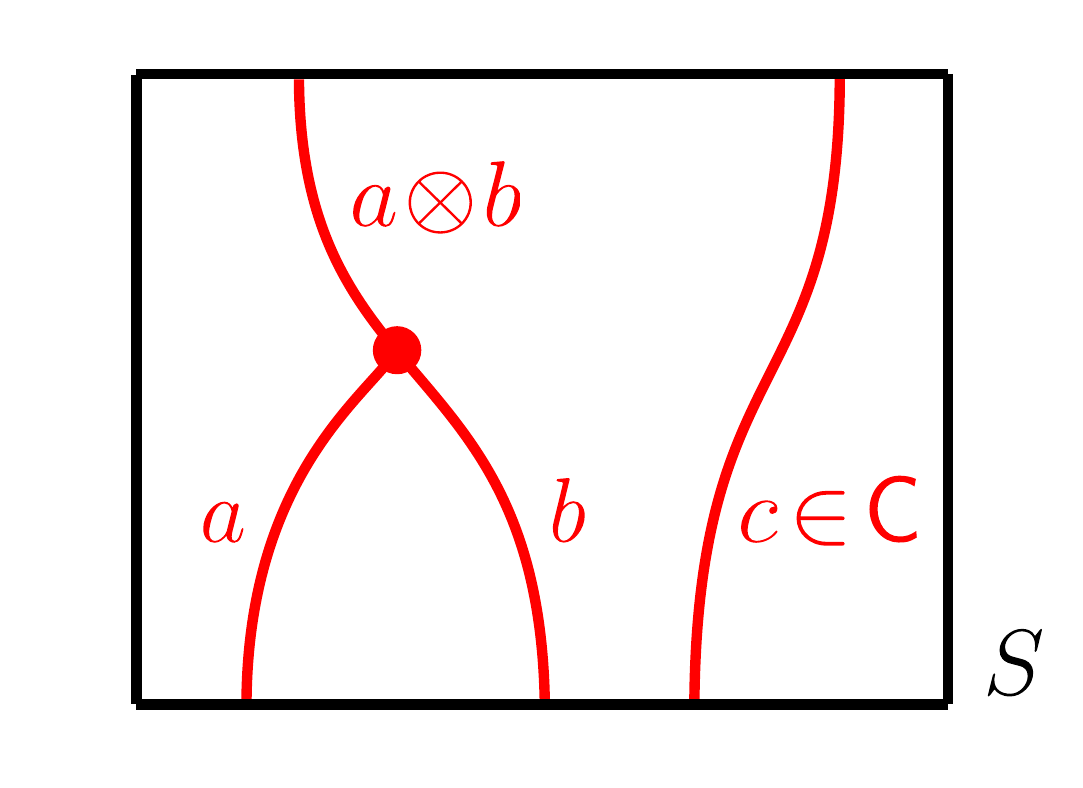}
	\vspace{-5pt}
	\caption{}
	\label{fig:surface-fusion-category}
\end{figure}

A surface defect $S(\Sigma)$ may support topological line defects and junctions, as illustrated schematically in figure~\ref{fig:surface-fusion-category}, which we will assume are captured by a multi-fusion category $\mathsf{C}_S$. As a consequence, it may be decomposed as a direct sum 
\be
 \mathsf{C}_S \; = \; \mathsf{C}_1 \oplus \cdots \oplus \mathsf{C}_n
\ee
for some collection of indecomposable multi-fusion categories $\mathsf{C}_1,\ldots, \mathsf{C}_n$. This motivates the introduction of simple and reduced line defects:

\begin{itemize}
\item A surface defect $S$ is \textit{simple} if $\mathsf{C}_S$ is an indecomposable multi-fusion category. If a surface defect $S$ is not simple, it admits a decomposition
\be
S \; = \; S_1 \, \oplus \, \cdots \, \oplus \, S_s
\ee
in terms of simple surface defects with $\mathsf{C}_{S_j} = \mathsf{C}_j$ indecomposable. We therefore restrict attention to simple surface defects in what follows.

\item A simple surface defect $S$ is \textit{reduced} if $\mathsf{C}_S$ is fusion. A simple surface defect $S'$ always admits a topological interface with a reduced simple surface $S$, which reflects the fact that any indecomposable multi-fusion category is Morita equivalent to a fusion category~\cite{davydov2011witt}.
\end{itemize}
We assume in what follows that surface defects are simple and reduced. 

\subsection{Elementary perspective}

\begin{figure}[h]
	\centering
	\includegraphics[height=4.1cm]{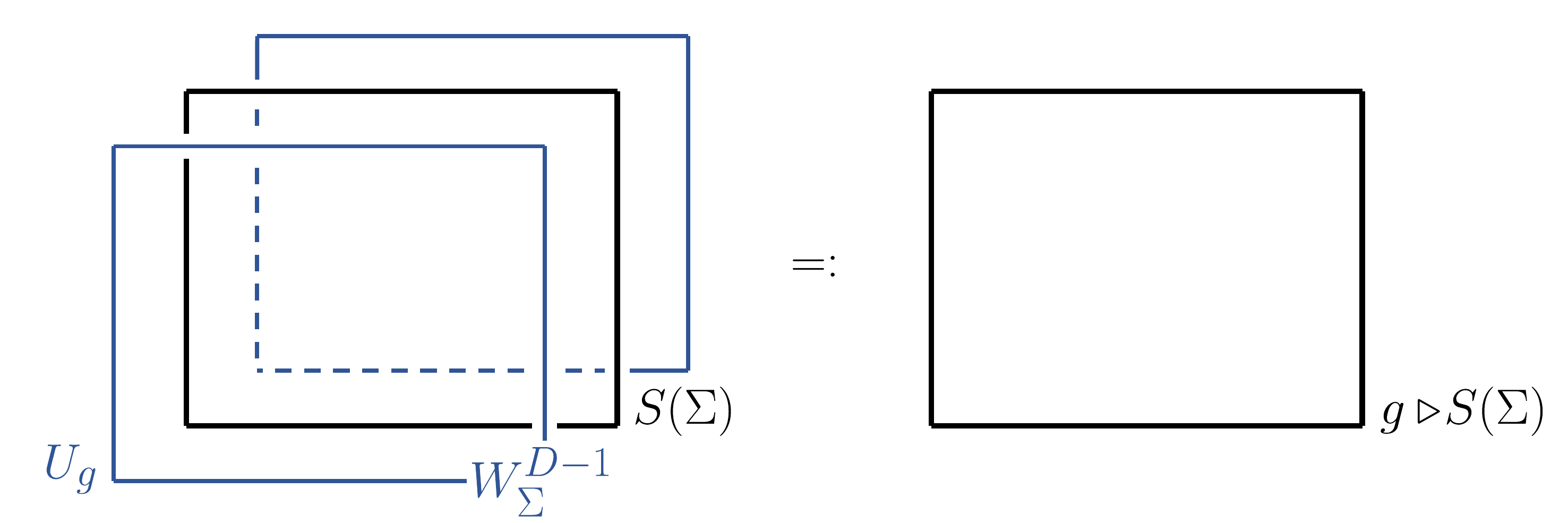}
	\vspace{-5pt}
	\caption{}
	\label{fig:surface-action}
\end{figure}

Let us begin by considering again a $D$-dimensional quantum field theory with a finite group symmetry $G$ implemented by codimension-one topological defects $U_g(\Sigma_{D-1})$ with $g \in G$. We define an action of elements $g \in G$ on surface operators $S$ supported on a surface $\Sigma$ by
\begin{equation}\label{eq:surface-action}
    g \,\triangleright\, S(\Sigma) \; := \; U_g(W_{\Sigma}^{D-1}) \, S(\Sigma) \, ,
\end{equation}
where $W_{\Sigma}^{D-1} = \mathbb{R}^2 \times S_{\Sigma}^{D-3}$ denotes a small sandwich wedging the surface $\Sigma$ as illustrated in figure \ref{fig:surface-action}.

Leveraging the results of previous sections, we expect this action to be described by 3-representations of the group $G$. However, we anticipate the latter to be much richer than 1- or 2-representations due to the possibly non-trivial fusion category $\mathsf{C}$ of topological lines supported on $S$ The latter may interact with the action of $G$ on $S$ in a non-trivial manner, leading to an increased richness of 3-representations compared to 1- and 2-representations of $G$.

To distinguish these novel features from the results of the previous sections, we will divide our analysis into two cases: In the first case, we will assume that the surface operators $S$ do not support any non-trivial topological lines, which will reproduce a classification of 3-representations that is analogous to the classification of 2-representations for line operators. In the second case, we will consider a surface operator $S$ that preserves $G$ but supports non-trivial topological lines described by a fusion category $\mathsf{C}$. This will lead to a classification of 3-representations by novel types of data.

\subsubsection{Case 1: trivial lines}
\label{sssec:trivial-lines}

First, let us consider surface operators $S$ that do not support any non-trivial topological line defects. The collection of such surface operators $S(\Sigma)$ supported on $\Sigma$ forms a discrete set which is typically of infinite cardinality. However, we may restrict ourselves to the study of finite subsets $\mathcal{S}$ by fixing $S(\Sigma) \neq 0$ to be non-zero and defining
\begin{equation}
    \mathcal{S} \; := \; \lbrace \, g \triangleright S(\Sigma) \; | \; g \in G \rbrace \, ,
\end{equation}
which due to the finiteness of $G$ is a finite set. As a consequence, we can label its elements $S_i(\Sigma)$ by a finite index $i = 1,...,n$, which sets up a bijection $\mathcal{S} \cong \lbrace 1,...,n \rbrace$. The action (\ref{eq:surface-action}) may then be seen as being implemented by permutations $\sigma_g \in S_n$ such that
\begin{equation}
    g \, \triangleright \, S_i(\Sigma) \; = \; S_{\sigma_g(i)}(\Sigma) \, .
\end{equation}
In particular, compatibility with the fusion of symmetry defects in the bulk requires these permutations to satisfy
\begin{equation}
    \sigma_g \, \circ \, \sigma_h \; = \; \sigma_{g \cdot h} \, ,
\end{equation}
so that $\sigma: G \to S_n$ defines a transitive permutation action of $G$ on the set of $n$ elements.

\begin{figure}[h]
	\centering
	\includegraphics[height=9cm]{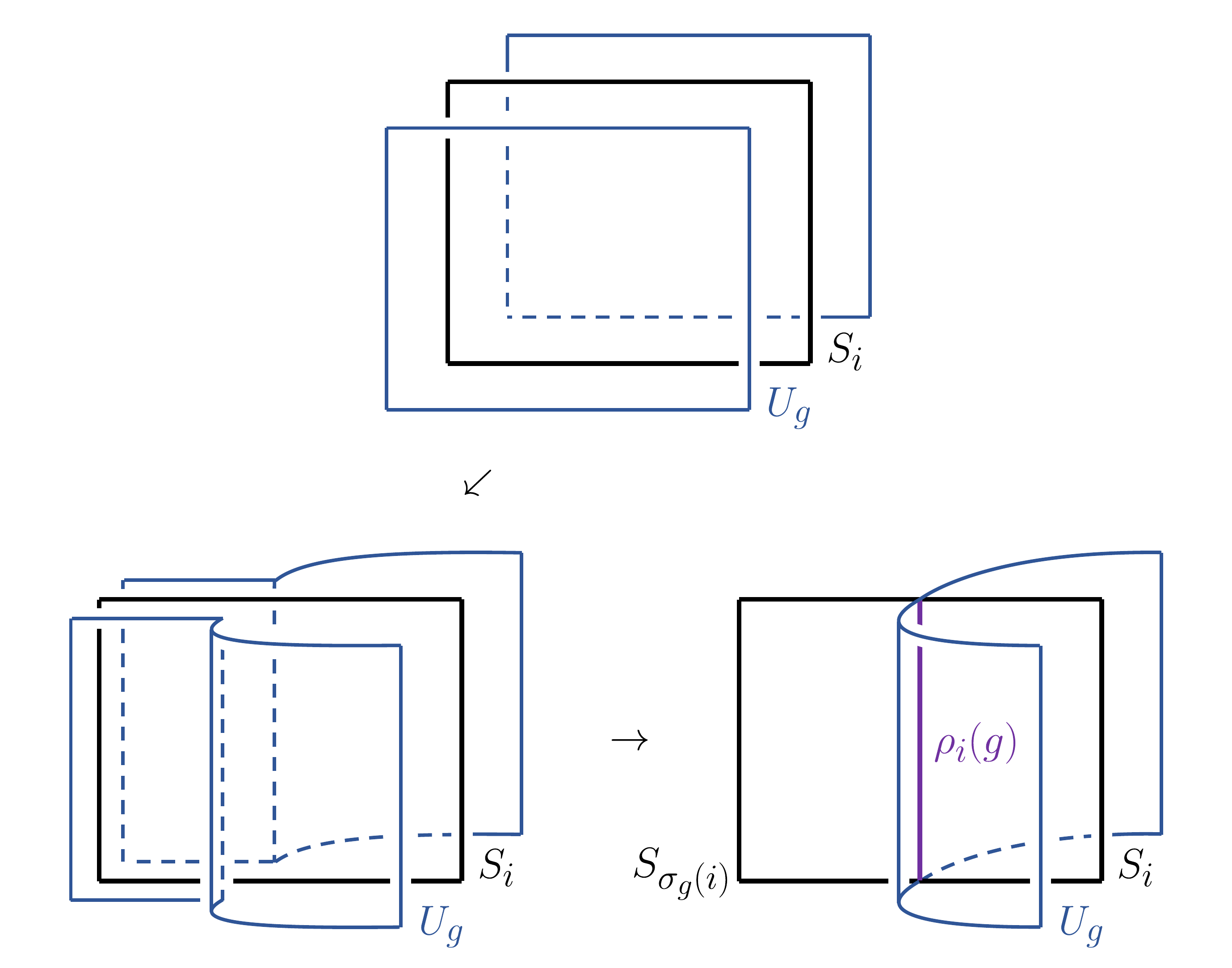}
	\vspace{-5pt}
	\caption{}
	\label{fig:intersection-lines}
\end{figure}

\begin{figure}[h]
	\centering
	\includegraphics[height=4.5cm]{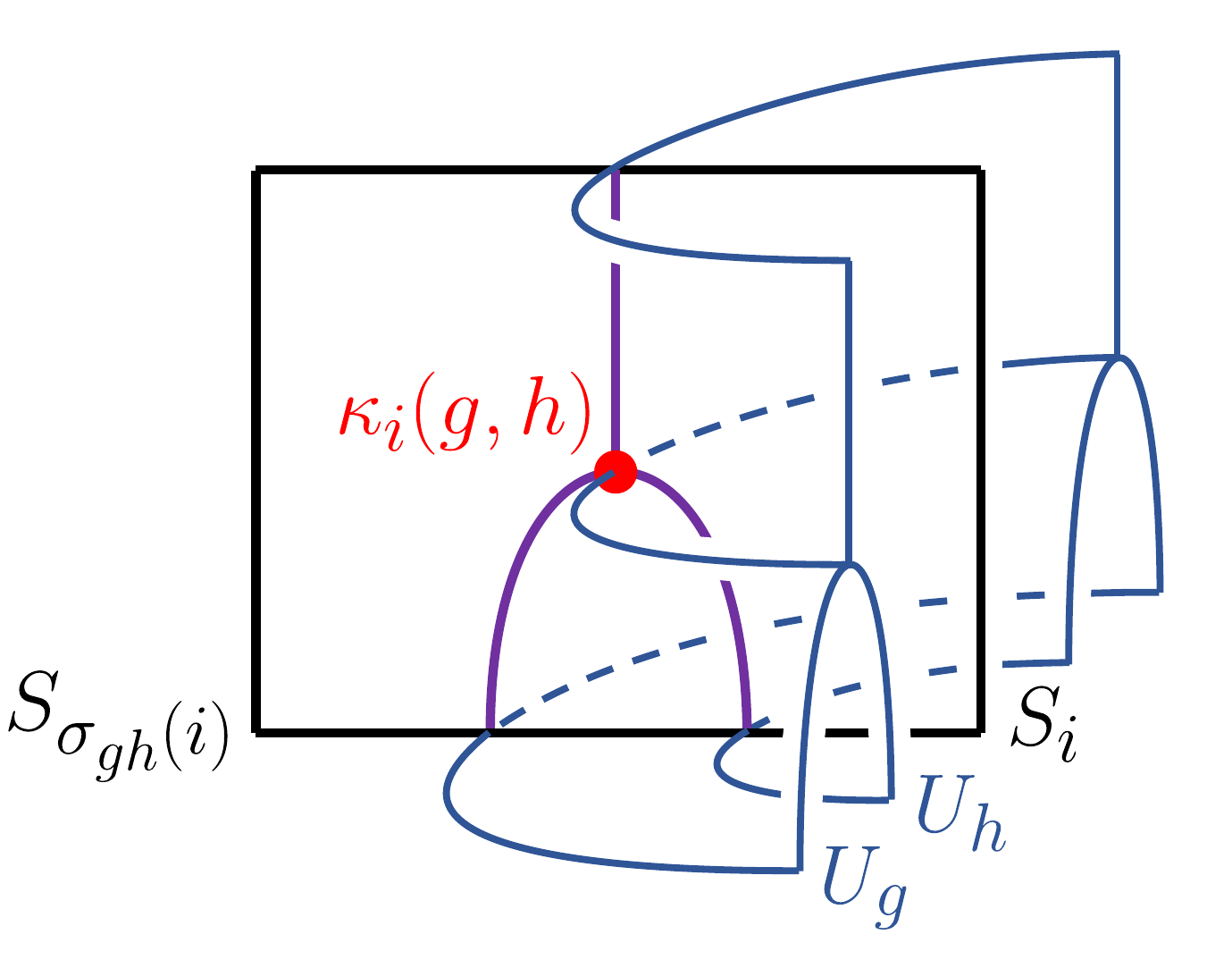}
	\vspace{-5pt}
	\caption{}
	\label{fig:intersection-junctions}
\end{figure}

\begin{figure}[h]
	\centering
	\includegraphics[height=4.6cm]{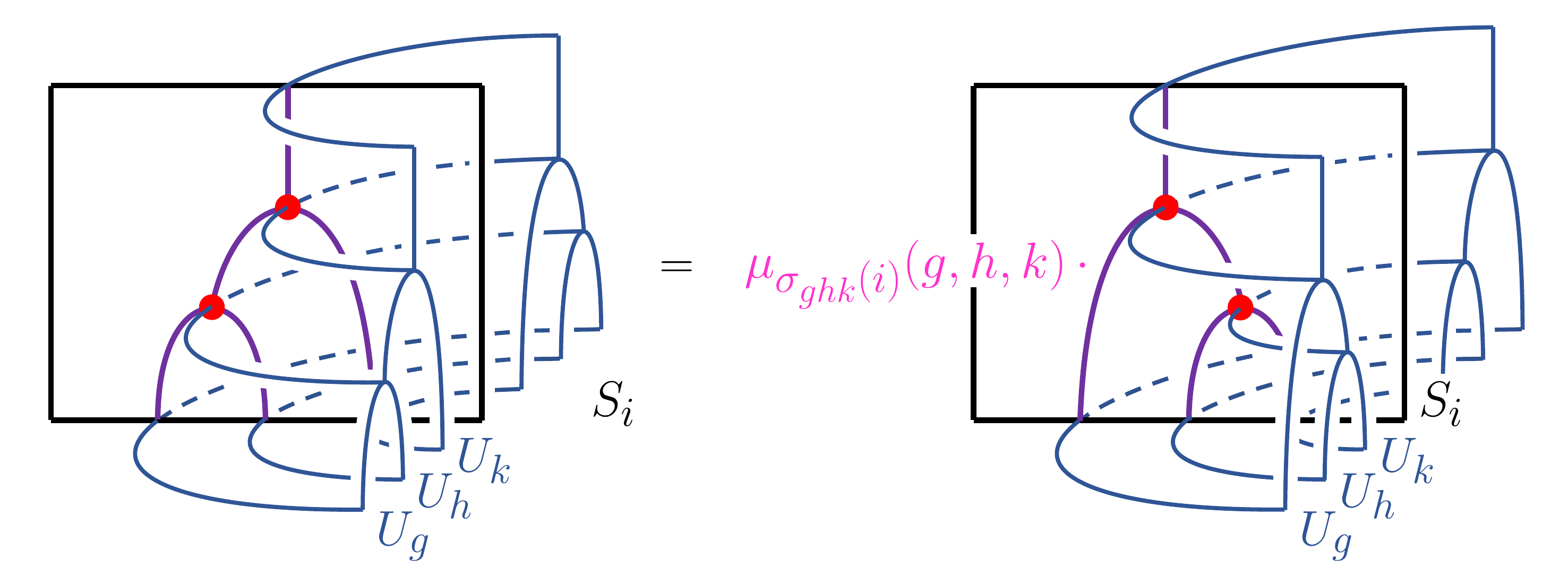}
	\vspace{-5pt}
	\caption{}
	\label{fig:intersection-anomaly}
\end{figure}

To proceed, we consider the sequence of topological operations shown in figure \ref{fig:intersection-lines}, which show that the codimension-one defects $g \in G$ intersect each surface $S_i$ at a unique topological line interface $\rho_i(g)$ connecting it to the transformed surface $S_{\sigma_g(i)}$. Furthermore, the fusion of two symmetry defects $g,h \in G$ in the bulk intersects each surface $S_i$ at a unique topological point-like junction $\kappa_i(g,h)$ as illustrated in figure \ref{fig:intersection-junctions}.
However, these junctions may only respect the group law of $G$ projectively in the sense that\footnote{Here, we denote by $\circ$ and $\star$ the vertical and horizontal composition of the junctions $\kappa_i(g,h)$, respectively.} 
\begin{equation}
\begin{aligned}
&\kappa_i(gh,k) \, \circ \, \big[ \kappa_{\sigma_k(i)}(g,h) \, \star \, \text{Id}_{\rho_i(k)} \big]  \\[2pt] 
= \;\; \mu_{\sigma_{ghk}(i)}(g,h,k) \;\; \cdot \;\; &\kappa_i(g,hk) \, \circ \, \big[ \text{Id}_{\rho_{\sigma_{hk}(i)}(g)} \, \star \, \kappa_i(h,k) \big]
\end{aligned}    
\end{equation}
for some multiplicative phase $\mu_i(g,h,k) \in U(1)$. Physically, this means that the two possible ways to intersect the surface $S_i$ with the fusion of three symmetry defects $g,h,k \in G$ are equivalent up to a multiplicative phase as illustrated in figure \ref{fig:intersection-anomaly}. 

The collection of phases $\mu_i(g,h,k) \in U(1)$ may be viewed as a map
\begin{equation}
    \mu: \; G \times G \times G \; \to \; U(1)^n \, ,
\end{equation}
which, in order to be compatible with the fusion of four symmetry defects $g,h,k,l \in G$, needs to form a twisted 3-cocycle $\mu \in Z^3_{\sigma}(G,U(1)^n)$, where $U(1)^n$ is regarded as a $G$-module via the permutation action $\sigma$.

In summary, we can label the action of the symmetry group $G$ on genuine surface operators $S_i \in \mathcal{S} \cong \lbrace 1, ..., n \rbrace$ by pairs $(\sigma,\mu)$ consisting of
\begin{enumerate}
    \item a permutation action $\sigma: G \to S_n$,
    \item a twisted 3-cocycle $\mu \in Z^3_{\sigma}(G,U(1)^n)$.
\end{enumerate}
Mathematically, this is a special case of a 3-representation of $G$ that is analogous to the classification of 2-representations of $G$ we encountered in section \ref{sec:lines}. Similarly to before, two such 3-representations $(\sigma,\mu)$ and $(\sigma',\mu')$ of $G$ are considered equivalent if there exists a permutation $\tau \in S_n$ such that
\begin{equation}
    \sigma' \; = \; \tau \circ \sigma \circ \tau^{-1} \qquad \text{and} \qquad [\mu'] \; = \; [\tau \triangleright \mu ] \, ,
\end{equation}
where $[.]: Z^3_{\sigma}(G,U(1)^n) \to H^3_{\sigma}(G,U(1)^n)$ denotes the projection into twisted group cohomology. A more mathematical treatment of 3-representations of groups can be found in appendix \ref{app:3-representations}.

\subsubsection{Case 2: non-trivial lines}
\label{sssec:non-trivial-lines}

\begin{figure}[h]
	\centering
	\includegraphics[height=3.4cm]{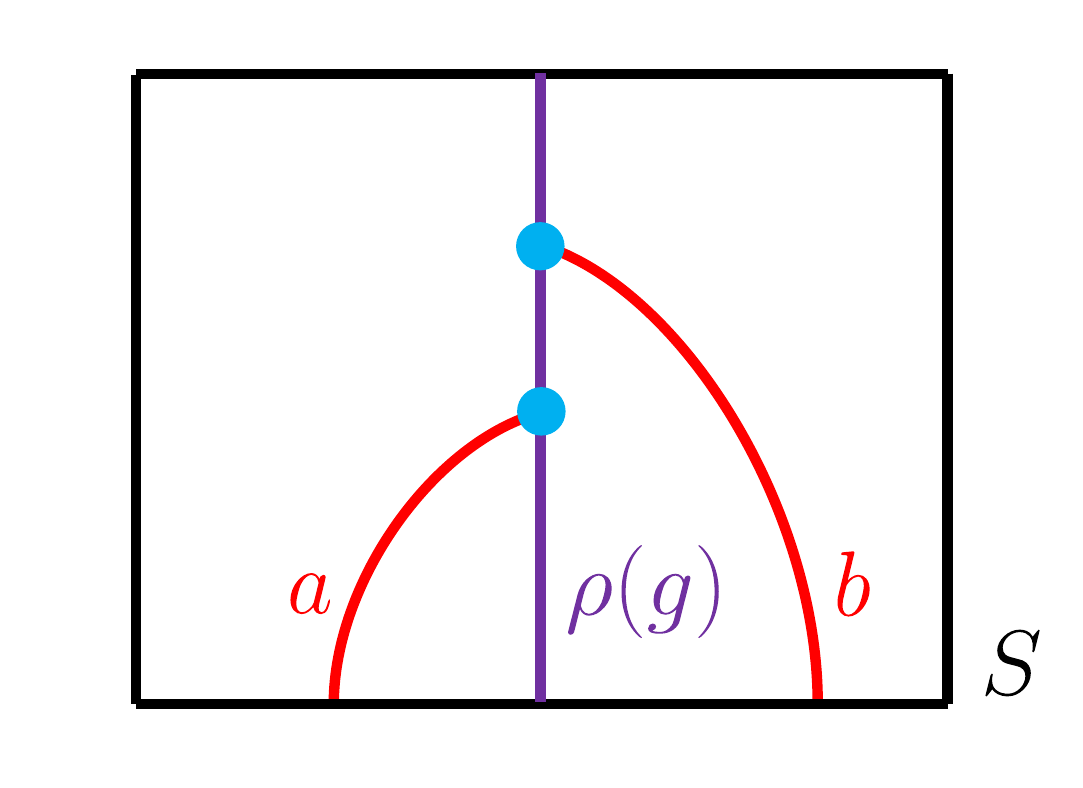}
	\vspace{-5pt}
	\caption{}
	\label{fig:left-right-module}
\end{figure}

Second, let us consider a surface operator $S$ that is left invariant by the sandwich action of $G$ (i.e. $g \triangleright S = S$ for all $g \in G$) but supports non-trivial topological line defects described by a fusion category $\mathsf{C}$. The topological interfaces $\rho(g)$ describing how symmetry defects $g \in G$ intersect the surface $L$ will then need to be equipped with instructions for how defect lines $a,b \in \mathsf{C}$ can end on it consistently from the left and from the right. This is illustrated in figure \ref{fig:left-right-module}.

Mathematically, this turns the interface $\rho(g)$ into a $\mathsf{C}$-bimodule. These $\mathsf{C}$-bimodules respect the group multiplication of $G$ up to specified bimodule-equivalences
\begin{equation}
    \kappa(g,h) \, : \;\; \rho(g) \otimes \rho(h) \;\; \to \;\; \rho(g h) \, ,
\end{equation}
which in particular implies that, up to equivalence, each $\rho(g)$ defines an invertible $\mathsf{C}$-bimodule. We can thus think of $\rho$ as inducing a group homomorphism
\begin{equation}
    \rho : \; G \; \to \; \text{BrPic}(\mathsf{C}) \, ,
\end{equation}
where $\text{BrPic}(\mathsf{C})$ denotes the \textit{Brauer-Picard group} of equivalence classes of invertible $\mathsf{C}$-bimodules.

\begin{figure}[h]
	\centering
	\includegraphics[height=3.4cm]{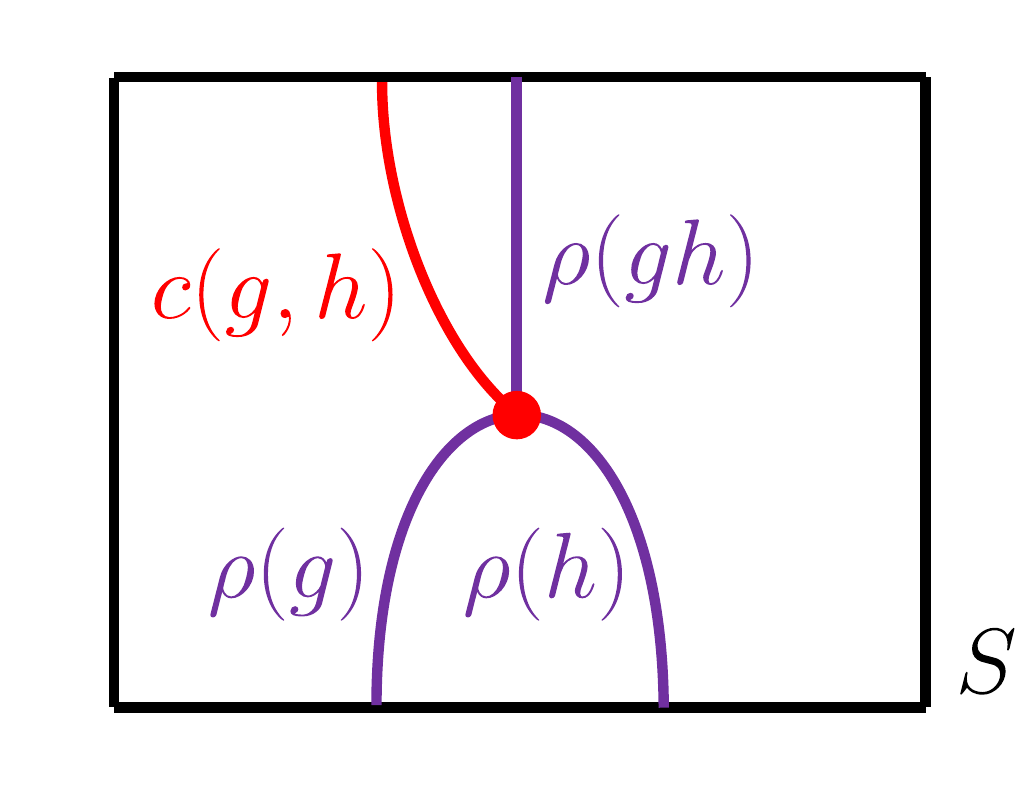}
	\vspace{-5pt}
	\caption{}
	\label{fig:intersection-extension}
\end{figure}

\begin{figure}[h]
	\centering
	\includegraphics[height=3.4cm]{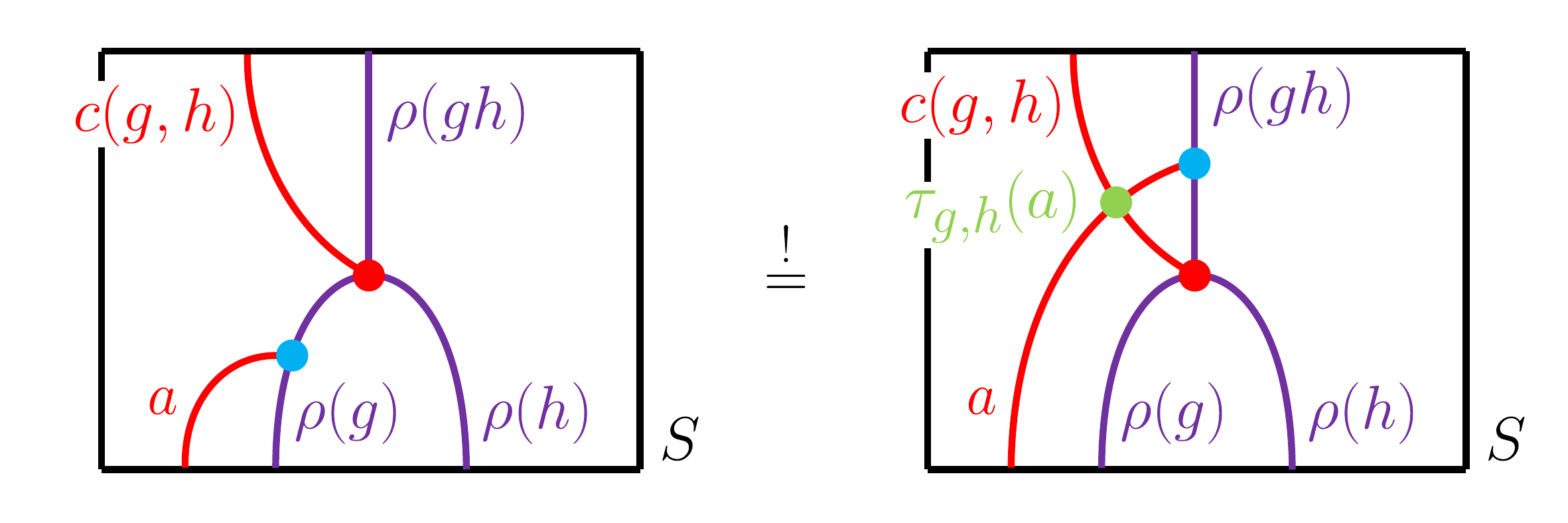}
	\vspace{-5pt}
	\caption{}
	\label{fig:intersection-extension-2}
\end{figure}

In order to classify the data associated to the bimodule equivalences $\kappa(g,h)$, we take the bimodules $\rho(g) \otimes \rho(h)$ and $\rho(gh)$ to be equal, so that $\kappa(g,h)$ can be seen as the left action of an invertible object $c(g,h) \in \mathsf{C}^{\times}$, i.e.
\begin{equation}
    \kappa(g,h) \; = \; c(g,h) \triangleright (.) \, .
\end{equation}
Physically, this means that when fusing the topological interfaces $\rho(g)$ and $\rho(h)$ together on the surface $S$, an invertible line $c(g,h) \in \mathsf{C}^{\times}$ emanates from their junction as illustrated in figure \ref{fig:intersection-extension}. 
Furthermore, these lines need to be compatible with the possibility to move a generic line $a \in \mathsf{C}$ ending on $\rho(g)$ or $\rho(h)$ across their junction as illustrated in figure \ref{fig:intersection-extension-2}. This requires an additional choice of crossing morphism
\begin{equation}
    \tau_{g,h}(a) \, : \;\; a \otimes c(g,h) \,\; \to \;\, c(g,h) \otimes a
\end{equation}
that describes how the line $a$ intersects the line $c(g,h)$ emanating from the junction of $\rho(g)$ and $\rho(h)$ as illustrated in figure \ref{fig:intersection-extension-2}. 

Mathematically, the line $c(g,h)$ together with the collection $\lbrace \tau_{g,h}(a) \rbrace_{a \in \mathsf{C}}$ of crossing morphisms defines an invertible object of the Drinfeld center $\mathsf{Z(C)}$ of $\mathsf{C}$, so that the collection of bimodule-equivalences $\kappa(g,h)$ can be seen as inducing a map
\begin{equation}
    \kappa : \; G \times G \; \to \; \mathsf{Z(C)^{\times}} \, ,
\end{equation}
where $\mathsf{Z(C)}^{\times}$ denotes the abelian group of isomorphism classes of invertible objects of $\mathsf{Z(C)}$. Compatibility with the fusion of three symmetry defects then requires this map to define a 2-cocycle $\kappa \in Z^2(G,\mathsf{Z(C)^{\times}})$.

More concretely, the compatibility with the fusion of three symmetry defects $g,h,k \in G$ in the bulk is implemented by natural isomorphisms
\begin{equation}
\mu(g,h,k) \, : \;\;\; \kappa(gh,k) \, \circ \, \big[ \kappa(g,h) \otimes \text{Id}_{\rho(k)} \big]  \;\;\; 
\Rightarrow \;\;\; \kappa(g,hk) \, \circ \, \big[ \text{Id}_{\rho(g)} \otimes \kappa(h,k) \big]  \, , 
\end{equation}
which we take to be given by multiplication by a phase $\mu(g,h,k) \in U(1)$ as before. Compatibility with the fusion of four symmetry defects in the bulk then requires the collection of these phases to define a 3-cocycle $\mu \in Z^3(G,U(1))$.

In summary, the action of the symmetry group $G$ on the surface operator $S$ supporting a fusion category $\mathsf{C}$ can be described by triples $(\rho,\kappa,\mu)$ consisting of
\begin{enumerate}
    \item a group homomorphism $\rho: G \to \text{BrPic}(\mathsf{C})$,
    \item a 2-cocycle $\kappa \in Z^2(G,\mathsf{Z(C)}^{\times})$,
    \item a 3-cocycle $\mu \in Z^3(G,U(1))$.
\end{enumerate}
Mathematically, this is a special case of a one-dimensional 3-representation of $G$ on the fusion category $\mathsf{C}$. We will comment on more general 3-representations of $G$ in the following subsection.

\subsection{Induction perspective}

So far we considered two special cases of 3-representations of the group $G$: On the one hand, we considered 3-representations on a collection of surface operators that support only trivial topological line defects. On the other hand, we considered 3-representations on a single surface operator supporting non-trivial line defects described by a fusion category $\mathsf{C}$. A general 3-representation of $G$ will be a combination of the above two cases in the sense that it may contain a collection of surface operators $S_i$ each of which supports non-trivial topological lines captured by a fusion category $\mathsf{C}_i$. 

The collection of fusion categories $\mathsf{C}_i$ may be combined into a \textit{multifusion category}\footnote{Here, the prefix ``multi" describes the fact that the monoidal unit of $\mathsf{B}$ is given by the direct sum $\bigoplus_i \mathbf{1}_i$ of the monoidal units of the components $\mathsf{C}_i$ and is hence not a simple object in $\mathsf{B}$.}
\begin{equation}
    \mathsf{B} \; := \; \mathsf{C}_1 \, \oplus \, ... \, \oplus \, \mathsf{C}_n \, .
\end{equation}
A generic 3-representation of $G$ on $\mathsf{B}$ may then be constructed by studying the topological interfaces that arise from intersecting the surfaces $S_i$ with symmetry defects $g \in G$ in the bulk. However, analogously to the case of irreducible 2-representations, the associated data may be reduced to the data of a one-dimensional 3-representation of a subgroup $H$ of $G$. Concretely, let $H$ be the subgroup of $G$ that preserves the surface operator $S := S_1$ under the sandwich action, i.e.
\begin{equation}
    H \; := \; \lbrace \, g \in G \; | \; g \triangleright S \, = \, S \, \rbrace \; \subset \; G \, .
\end{equation}
By repeating the analysis performed in subsection \ref{sssec:non-trivial-lines}, the action of $H$ on the surface $S$ can then be described by a one-dimensional 3-representation $(\rho,\kappa,\mu)$ of $H$ on the fusion category $\mathsf{C} := \mathsf{C}_1$. The action of the whole group $G$ on the collection of surface operators $S_i$ can be reconstructed from this data as the induction $\text{Ind}_H^G(\rho,\kappa,\mu)$ of the 3-representation $(\rho,\kappa,\mu)$ from $H$ to $G$.

In summary, the irreducible 3-representations of $G$ are labelled by quadruples $(H,\rho,\kappa,\mu)$ consisting of
\begin{enumerate}
    \item a subgroup $H \subset G$,
    \item a group homomorphism $\rho: H \to \text{BrPic}(\mathsf{C})$,
    \item a 2-cocylce $\kappa \in Z^2(H,\mathsf{Z(C)}^{\times})$,
    \item a 3-cocycle $\mu \in Z^3(H,U(1))$.
\end{enumerate}

This includes the types of 3-representations of $G$ discussed in subsection \ref{sssec:trivial-lines} by choosing $\mathsf{C} = \mathsf{Vec}$ as well as the ones discussed in subsection \ref{sssec:non-trivial-lines} by choosing $H=G$. Physically, the subgroup $H$ corresponds to the unbroken symmetry on the surface operator $S$. This may form a non-trivial group extension and admit a non-trivial mixed 't Hooft anomaly with the group of invertible lines on $S$, both of which are captured by the  class $[\kappa]$. Furthermore, $H$ may have an 't Hooft anomaly on $S$ determined by $[\mu]$.

\subsection{Categorical perspective}

\begin{figure}[h]
	\centering
	\includegraphics[height=4.9cm]{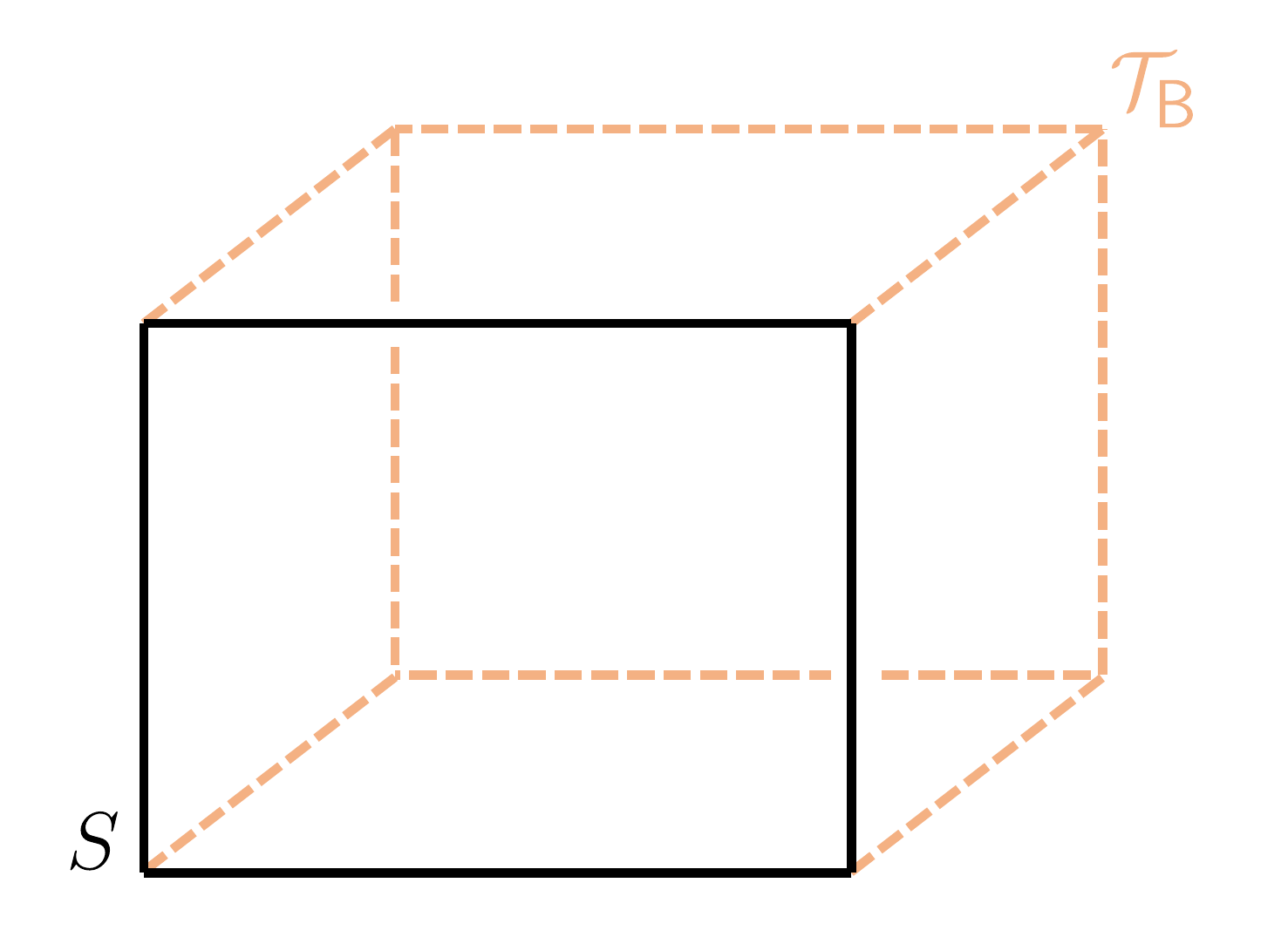}
	\vspace{-5pt}
	\caption{}
	\label{fig:3d-tqft-1}
\end{figure}

Similar to local and line operators, it is convenient to reformulate the considerations of the previous subsection in a more categorical manner. To do this, we again identify surface operators $S_i$ supporting fusion categories $\mathsf{C}_i$ as gapped boundary conditions for an attached auxiliary three-dimensional TQFT $\mathcal{T}_{\mathsf{B}}$, where $\mathsf{B}$ denotes the multifusion category $\mathsf{B} = \mathsf{C}_1 \oplus ... \oplus \mathsf{C}_n$. This is illustrated in figure \ref{fig:3d-tqft-1}.

The collection of framed fully-extended 3d TQFTs are captured by a fusion 3-category $3\mathsf{Vec}$ of multi-fusion categories and bi-module categories.
In this description, objects are 3d TQFTs with a choice of gapped boundary condition that supports the multi-fusion category of topological lines. This is natural for a correspondence between gapped boundary conditions and surface defects: the latter may support a multi-fusion category of topological lines.

Intrinsically, 3d TQFTs are in 1-1 correspondence with equivalence classes of objects in $3\mathsf{Vec}$, which forgets the choice of gapped boundary condition. Equivalence in $3\mathsf{Vec}$ is Morita equivalence of multi-fusion categories and any multi-fusion category is Morita equivalent to a sum
\be
\mathsf{B} = \mathsf{C}_1 \oplus \cdots \oplus \mathsf{C}_n
\ee
where $\mathsf{C}_j$ are fusion. 

We then set up a correspondence between a collection of simple surface defects $S_i(\Sigma)$ indexed by $i = 1,\ldots,n$ supporting fusion categories of topological lines $\mathsf{C}_{S_i} = \mathsf{C}_j$ and gapped boundary conditions for an auxiliary 3d TQFT $\cT_{\mathsf{B}}$ (by an abuse of notation, this depends only on the Morita equivalence class of $\mathsf{B}$). 

In this spirit, it is again convenient to restrict attention to simple line defects and work with a formulation of $3\mathsf{Vec}$ where equivalent objects are identified:
\begin{itemize}
    \item Objects are 3d TQFTs, determined by their 2-categories of boundary conditions $\mathsf{Mod}(\mathsf{B})$ for some multi-fusion category $\mathsf{B} = \mathsf{C}_1 \oplus ... \oplus \mathsf{C}_n$.
    \item 1-morphisms are are topological surface interfaces between 3d TQFTs and correspond to (pseudo-)2-functors $A: \mathsf{Mod}(\mathsf{B}) \to \mathsf{Mod}(\mathsf{B}')$ between the corresponding 2-categories of boundary conditions.

    \item 2-morphisms are topological line-like interfaces between topological surface interfaces and correspond to (pseudo-)natural transformation $\Phi: A \Rightarrow B$ between the corresponding 2-functors of 2-categories of boundary conditions.

    \item Its 3-morphisms are topological pointlike junctions between line-like topological interfaces. As such, they correspond to modifications $\varphi: \Phi \Rrightarrow \Psi$ between the corresponding (pseudo-)natural transformations.
\end{itemize}
The fusion of objects corresponds to stacking the associated 3d TQFTs.

\begin{figure}[h]
	\centering
	\includegraphics[height=5.25cm]{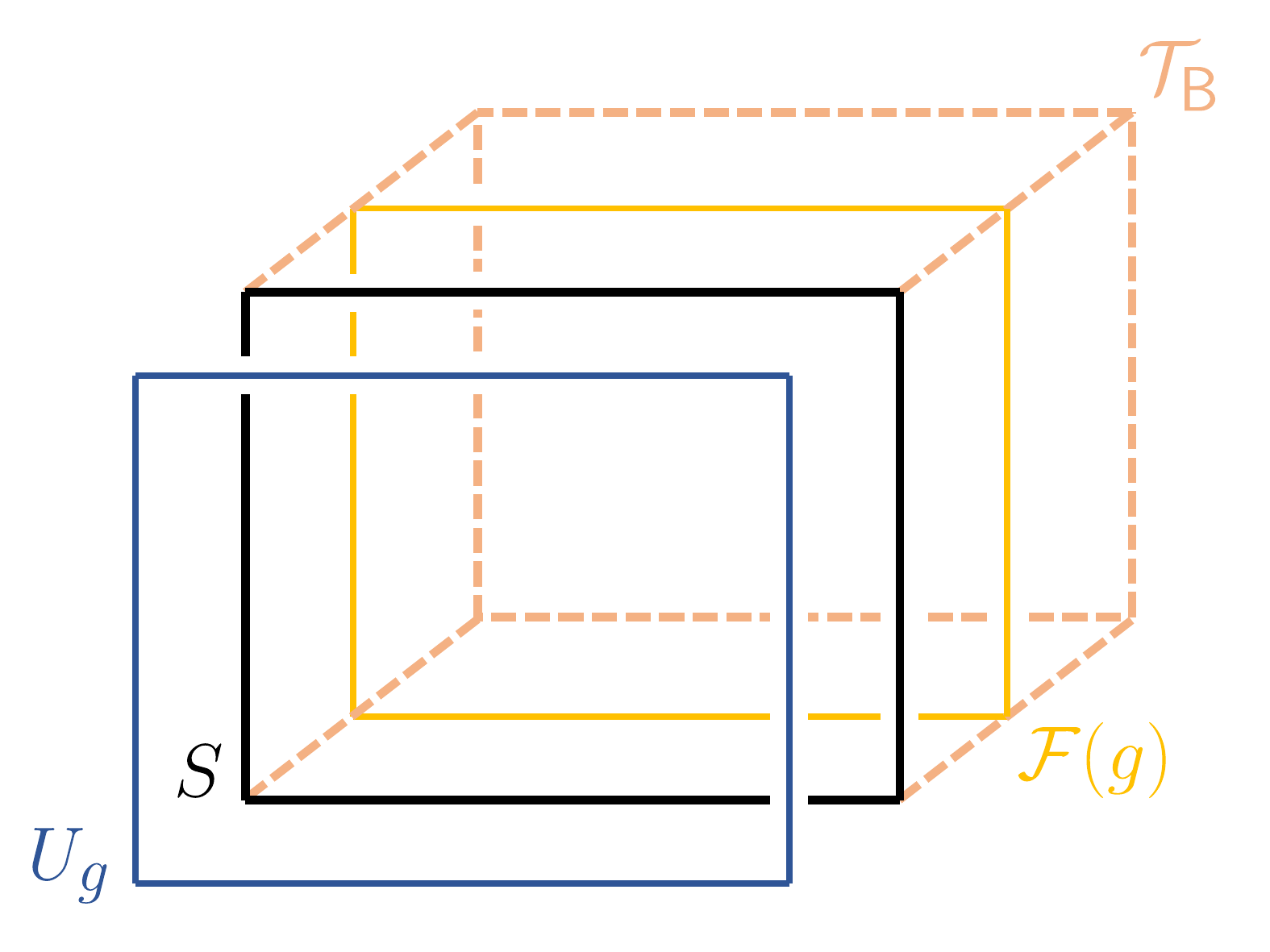}
	\vspace{-5pt}
	\caption{}
	\label{fig:3d-tqft-2}
\end{figure}

Using this picture, the action of the symmetry group $G$ on surface defects $S$ can now be translated into the $G$-equivariance of the attached TQFT $\mathcal{T}_{\mathsf{B}}$. Concretely, consider wedging $S$ between a small $(D-1)$-sandwich labelled by $g \in G$ as before. Due to the attached 3d TQFT, this now requires a choice of surface-like intersection $\mathcal{F}(g)$ between the defect $U_g$ and $\mathcal{T}_{\mathsf{B}}$, which we regard as a topological interface between $\mathcal{T}_{\mathsf{B}}$ and itself as illustrated in figure \ref{fig:3d-tqft-2}. 
As such, they can naturally be identified with 2-functors 
\begin{equation}
    \mathcal{F}(g)\, : \; \mathsf{Mod}(\mathsf{B}) \; \to \; \mathsf{Mod}(\mathsf{B})
\end{equation}
from the 2-category of boundary conditions of $\mathcal{T}_{\mathsf{B}}$ to itself, so that symmetry defects $g \in G$ can act on the simple boundary conditions $S \in \mathsf{Mod}(\mathsf{B})$ via
\begin{equation}
    S \; \mapsto \; \mathcal{F}(g)(S) \, .
\end{equation}
Physically, this corresponds to shrinking down the $(D-1)$-sandwich formed by $g$ towards $S$ and thus implements the action of $g$ on the corresponding surface operator.

\begin{figure}[h]
	\centering
	\includegraphics[height=5.3cm]{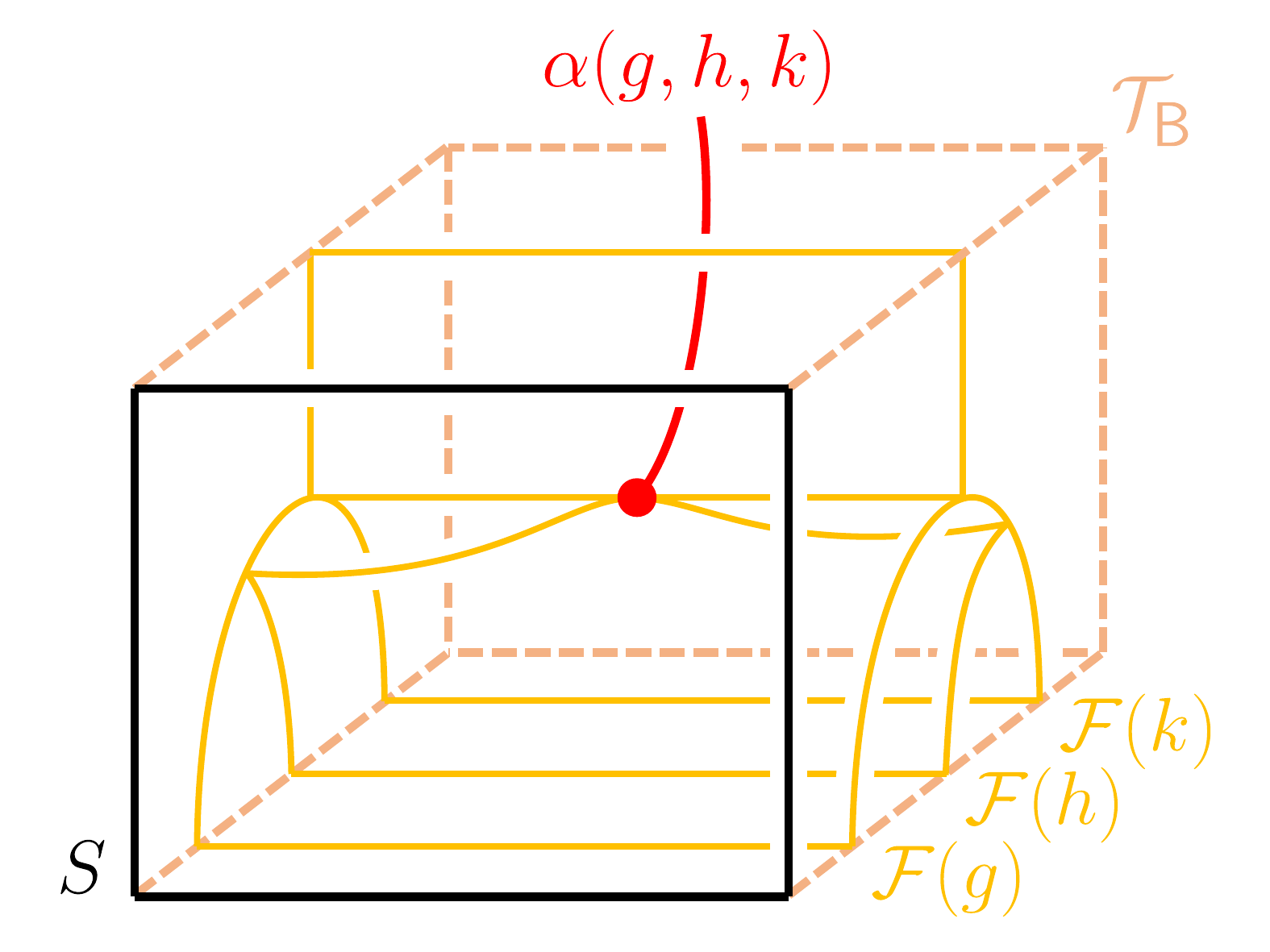}
	\vspace{-5pt}
	\caption{}
	\label{fig:3d-tqft-4}
\end{figure}

In general, the composition of these 2-functors may be controlled by a non-trivial associator
\begin{equation}
    \alpha(g,h,k) \, : \;\; \big[ \mathcal{F}(g) \circ \mathcal{F}(h) \big] \circ \mathcal{F}(k) \;\; \Rightarrow \;\; \mathcal{F}(g) \circ \big[ \mathcal{F}(h) \circ \mathcal{F}(k) \big] \, .
\end{equation}
Physically, this means that when crossing the fusion of three surface interfaces $\mathcal{F}(g)$, $\mathcal{F}(h)$ and $\mathcal{F}(k)$, an invertible topological line defect $\alpha(g,h,k) \in \mathsf{Z(B)}^{\times}$ in $\mathcal{T}_{\mathsf{B}}$ emanates from their junction as illustrated in figure \ref{fig:3d-tqft-4}. 
Compatibility with the fusion of four surface interfaces then restrains the collection of these lines to form a 3-cocycle $\alpha \in Z^3(G,\mathsf{Z(B)}^{\times})$. The associated class $[\alpha] \in H^3(G,\mathsf{Z(B)}^{\times})$ represents the obstruction for the functors $\mathcal{F}(g)$ to define consistent intersections that are compatible with the fusion of symmetry defects in the bulk. It can be interpreted as the Postnikov class of a non-trivial 2-group formed by the intersections surfaces $\mathcal{F}(g)$ and the invertible lines of $\mathcal{T}_{\mathsf{B}}$.

\begin{figure}[h]
	\centering
	\includegraphics[height=4.9cm]{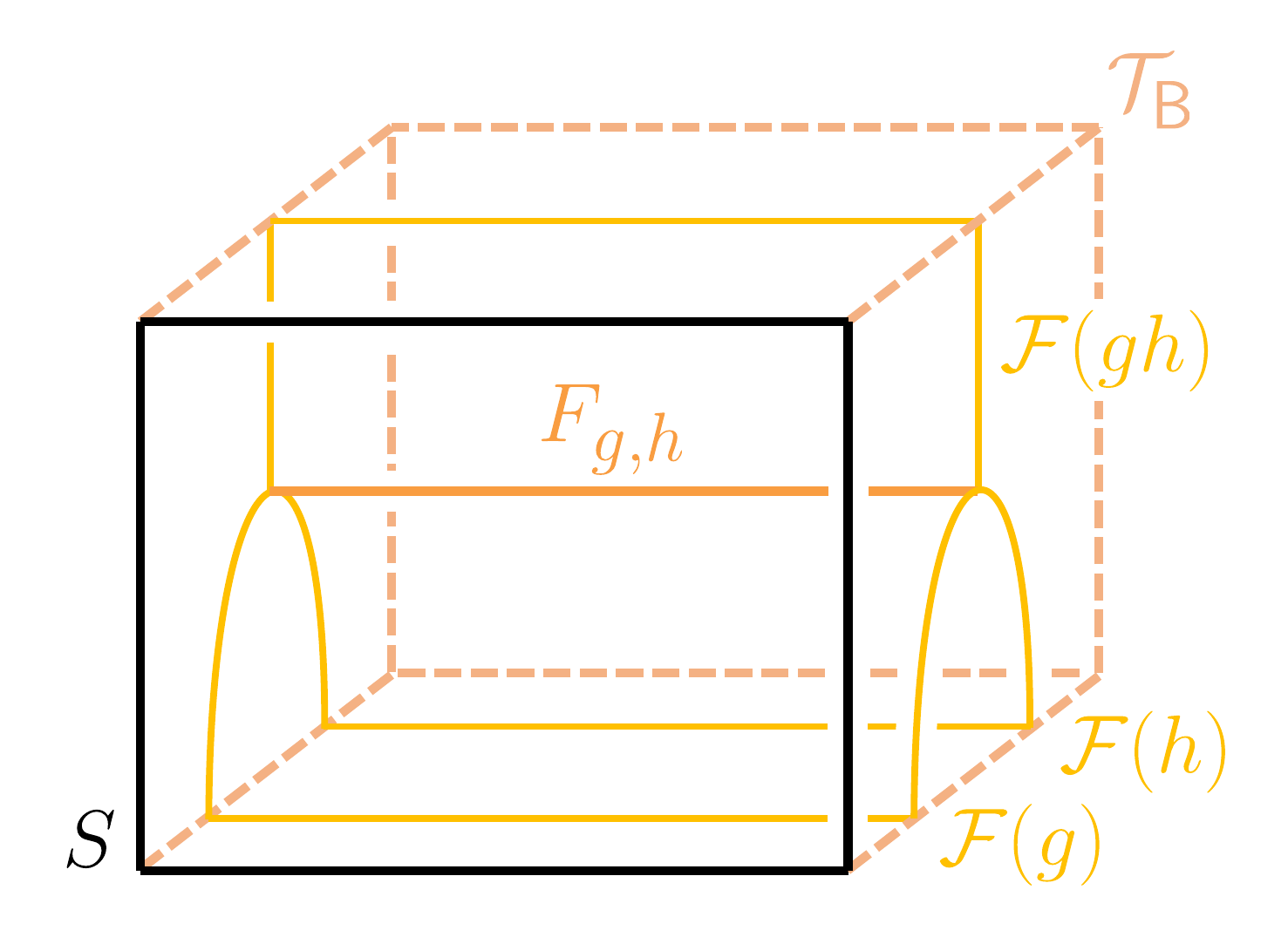}
	\vspace{-5pt}
	\caption{}
	\label{fig:3d-tqft-3}
\end{figure}

\begin{figure}[h]
	\centering
	\includegraphics[height=5.3cm]{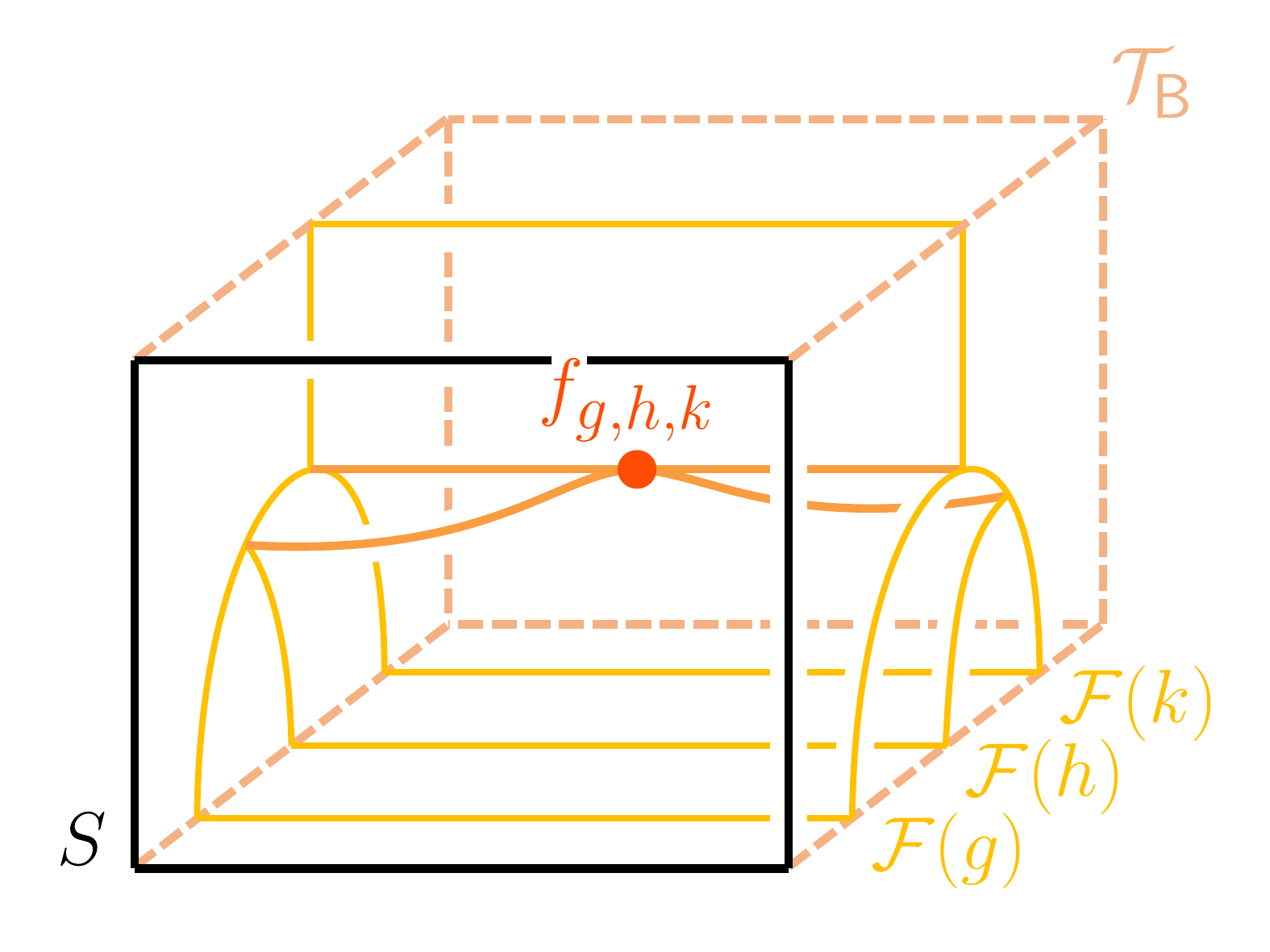}
	\vspace{-5pt}
	\caption{}
	\label{fig:3d-tqft-5}
\end{figure}

\begin{figure}[h]
	\centering
	\includegraphics[height=9cm]{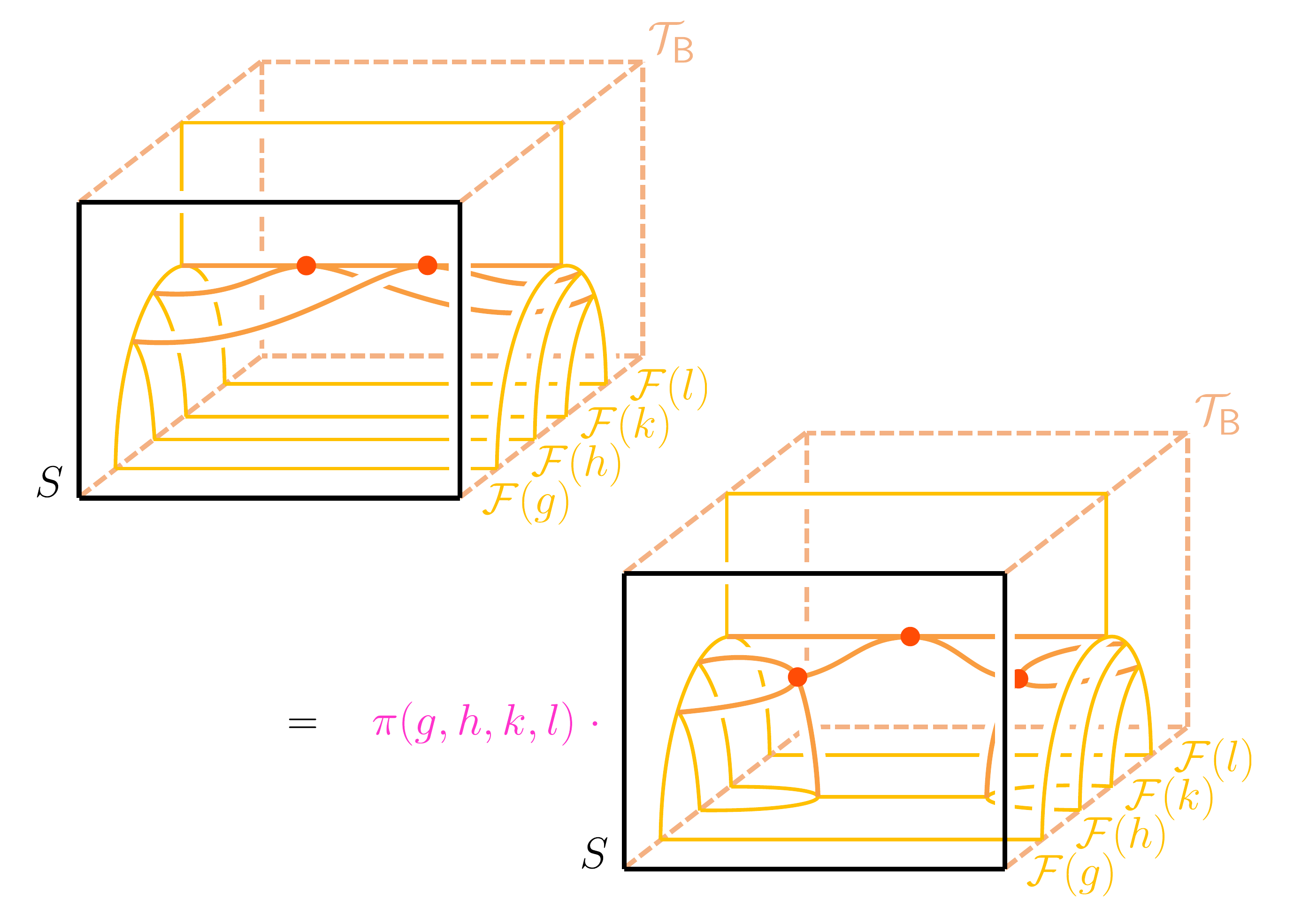}
	\vspace{-5pt}
	\caption{}
	\label{fig:3d-tqft-6}
\end{figure}

If $[\alpha] = 1$, we can introduce natural equivalences
\begin{equation}\label{eq:3d-tqft-line-junctions}
    F_{g,h} \, : \;\; \mathcal{F}(g) \, \circ \, \mathcal{F}(h) \;\, \Rightarrow \;\, \mathcal{F}(g \cdot h)
\end{equation}
that correspond to line-like junctions arising from the intersection of $\mathcal{T}_{\mathsf{B}}$ with the fusion of symmetry defects $g,h \in G$ in the bulk as illustrated in figure \ref{fig:3d-tqft-3}. 
Their compatibility with the fusion of three symmetry defects in the bulk is implemented by modifications
\begin{equation}\label{eq:3d-tqft-point-junctions}
f_{g,h,k} \, : \;\;\; F_{gh,k} \, \circ \, \big[ F_{g,h} \, \star \, \text{Id}_{\mathcal{F}(k)} \big]  \;\;\; 
\Rrightarrow \;\;\; F_{g,hk} \, \circ \, \big[ \text{Id}_{\mathcal{F}(g)} \, \star \, F_{h,k} \big]  \, , 
\end{equation}
which correspond to point-like junctions arising from the intersection of $\mathcal{T}_{\mathsf{B}}$ with the fusion of $g,h,k \in G$ as illustrated in figure \ref{fig:3d-tqft-5}.
In general, these junctions may respect the group law of $G$ only projectively\footnote{In this case, the 3d TQFT should itself be regarded as the boundary of a 4-dimensional theory.} in the sense that intersecting $\mathcal{T}_{\mathsf{B}}$ with the fusion of four symmetry defects $g,h,k,l \in G$ in two possible ways yields equivalent results up to a multiplicative phase $\pi(g,h,k,l) \in U(1)$ as illustrated in figure \ref{fig:3d-tqft-6}. 
The collection of these phases may be regarded as a map
\begin{equation}
    \pi : \; G \times G \times G \times G \; \to \ U(1) \, ,
\end{equation}
which, in order to be compatible with the intersection of five symmetry defects, needs to form a 4-cocycle $\pi \in Z^4(G,U(1))$. The associated class $[\pi] \in H^4(G,U(1))$ represents the obstruction for the modifications $f_{g,h,k}$ to define consistent intersections that are compatible with the fusion of symmetry defects in the bulk. If $[\pi] = 1$, they can be renormalised to satisfy the appropriate compatibility conditions.

In summary, the action of the symmetry group $G$ on genuine surface operators in this framework can be described by the following data:
\begin{enumerate}
    \item A 3d TQFT $\mathcal{T}_{\mathsf{B}}$ (an object in $\mathsf{3Vec}$),
    \item a topological surface interface $\mathcal{F}(g)$ between $\mathcal{T}_{\mathsf{B}}$ and itself (a 1-endomorphism of $\mathcal{T}_{\mathsf{B}}$ in $\mathsf{3Vec}$) for ech $g \in G$,
    \item a topological line-like junction $F_{g,h}$ (a 2-morphism in $\mathsf{3Vec}$) as in (\ref{eq:3d-tqft-line-junctions}) for each pair $g,h \in G$,
    \item a topological point-like junction $f_{g,h,k}$ (a 3-morphism in $\mathsf{3Vec}$) as in (\ref{eq:3d-tqft-point-junctions}) for each triple $g,h,k \in G$ .
\end{enumerate}
The collection of this data together with the compatibility conditions they satisfy can be recognised as the data of a (pseudo-)3-functor
\begin{equation}
    \mathcal{F} : \; \widehat{G} \; \to \; \mathsf{3Vec} \, ,
\end{equation}
where the group $G$ is regarded as a 3-category $\widehat{G}$ with a single object $\ast$, whose 1-endomorphisms are given by $\text{1-End}_{\widehat{G}}(\ast) = G$ and whose 2- and 3-morphisms are trivial. The collection of such 3-functors itself forms a 3-category whose
\begin{itemize}
    \item objects are 3-functors $\mathcal{F}: \widehat{G} \to \mathsf{3Vec}$,
    \item 1-morphisms are natural transformations $\eta: \mathcal{F} \Rightarrow \mathcal{F}'$,
    \item 2-morphisms are modifications $\Xi: \eta \Rrightarrow \eta'$,
    \item 3-morphisms are perturbations $\mathfrak{X}: \Xi \,\mathrel{\substack{\textstyle\Rightarrow\\[-0.5ex] \textstyle\Rightarrow}} \,\Xi'$.
\end{itemize}
We denote this category by $[\widehat{G},\mathsf{3Vec}]$ and recognise it as the fusion 3-category of finite 3-representations of $G$, 
\begin{equation}
    [\widehat{G},\mathsf{3Vec}] \; = \; \mathsf{3Rep}(G) \, .
\end{equation}
This reproduces the result from the previous subsection that surface operators transform in 3-representations of the symmetry group $G$. Equivalently, this may be regarded as the 3-category of $G$-equivariant 3d TQFTs.

\acknowledgments

The work of MB is supported by the EPSRC Early Career Fellowship EP/T004746/1 and the STFC Research Grant ST/T000708/1. The work of MB and AG is supported by the Simons Collaboration on Global Categorical Symmetry.

\appendix

\section{Higher representation theory}
\label{app:higher-representations}

In this appendix, we will review the higher representation theory of higher groups from an abstract categorical point of view. This should serve as a companion to the main body of the paper, where higher representations are constructed in a concrete physical context.

Generically, we will think of a linear $n$-representation of an $n$-group $\mathcal{G}$ as a functor
\begin{equation}
F : \; \widehat{\mathcal{G}} \, \to \, \mathsf{nVec}
\end{equation}
from the $n$-group thought of as an appropriate type of $n$-category $\widehat{\mathcal{G}}$ into the fusion $n$-category $\mathsf{nVec}$ of $n$-vector spaces. For $n>1$, there may be a multitude of models for $\mathsf{nVec}$ and different models may be natural for different applications\footnote{For a summary of models and relationships between them in the case $n=2$ see~\cite{bartlett2015modular}.}. For our purposes, the most natural choice is the recursive definition
\begin{equation}
\mathsf{nVec} \, := \, \Sigma^n\mathbb{C}
\end{equation}
introduced in~\cite{Gaiotto:2019xmp}, where $\Sigma\mathsf{C} := \text{Kar}(B\mathsf{C})$ denotes the Karoubi completion of the delooping of $\mathsf{C}$. Linear $n$-representations of $\mathcal{G}$ then form a fusion $n$-category which we denote by 
\begin{equation}
\mathsf{nRep}(\mathcal{G}) \, := \, [\widehat{\mathcal{G}},\mathsf{nVec}] \, .
\end{equation}
We will uncover this definition concretely in the cases $n=1,2,3$ below.

\subsection{1-representations}
\label{app:1-representations}

Let us begin by reformulating the representation theory of ordinary groups in a categorical framework. To do this, we fix a finite group $G$ and note that we can associate to it a category $\widehat{G}$ defined as follows: 
\begin{itemize}
\item Its set of objects contains a single element denoted by $\ast$.
\item The set of endomorphisms of $\ast$ is given by 
\begin{equation}
\text{End}_{\widehat{G}}(\ast) \, = \, G
\end{equation}
with composition given by group multiplication in $G$.
\end{itemize}
This construction makes it clear that we can think of groups as special types of categories with a single object, all of whose morphisms are invertible.

We would like to linearly represent the group $G$ on finite dimensional complex vector spaces. The latter form a fusion category $\mathsf{Vec}$ defined as follows:

\begin{itemize}
\item Its objects are finite-dimensional complex vector spaces $V \cong \mathbb{C}^n$. Up to equivalence, there is a single simple object corresponding to the one-dimensional vector space $\mathbb{C}$.

\item Its morphisms between objects $V$ and $W$ are given by linear maps
\vspace{-1pt}
\begin{equation}
\begin{gathered}
\includegraphics[height=0.63cm]{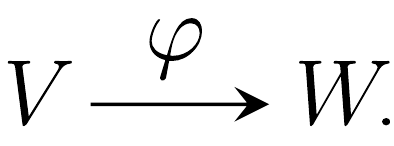}
\end{gathered}
\vspace{-4pt}
\end{equation} 
The composition of morphisms
\vspace{-1pt}
\begin{equation}
\begin{gathered}
\includegraphics[height=0.63cm]{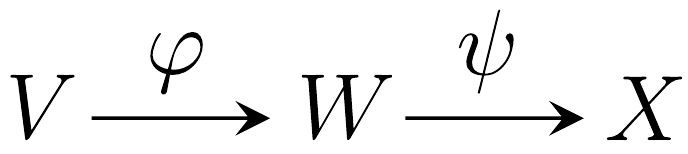}
\end{gathered}
\vspace{-4pt}
\end{equation}
is given by the composition of the corresponding linear maps.

\item The fusion of objects and morphisms
\vspace{-1pt}
\begin{equation}
\begin{gathered}
\includegraphics[height=1.28cm]{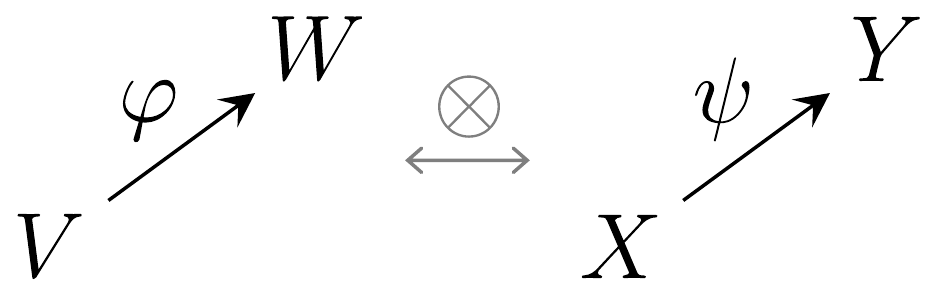}
\end{gathered}
\vspace{-4pt}
\end{equation}
is given by the tensor product of the corresponding vector spaces and linear maps.
\end{itemize}

Given the ingredients $\widehat{G}$ and $\mathsf{Vec}$, it is now straightforward to define the category $\mathsf{Rep}(G)$ of finite-dimensional representations of $G$:
\begin{itemize}
\item Its objects are functors $\mathcal{F}: \widehat{G} \to \mathsf{Vec}$. Concretely, this means that $\mathcal{F}$ assigns a finite dimensional complex vector space $V := \mathcal{F}(\ast) \in \mathsf{Vec}$ to the single object $\ast$ of $\widehat{G}$ and linear maps $\mathcal{F}(g) \in \text{End}(V)$ to the endomorphisms $g \in G$ of $\ast$ in $\widehat{G}$. Functoriality of $\mathcal{F}$ then ensures that the diagrams
\vspace{-1pt}
\begin{equation}\label{eq-rep-morphism}
\begin{gathered}
\includegraphics[height=2.3cm]{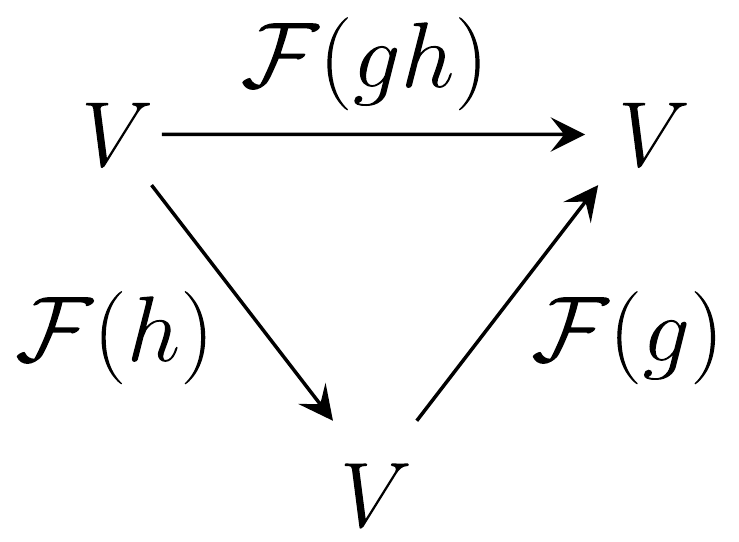}
\end{gathered}
\vspace{-4pt}
\end{equation}
commute for all $g,h \in G$, so that objects of $\mathsf{Rep}(G)$ correspond to ordinary representations of $G$ in the usual sense.

\item Its morphisms between objects $\mathcal{F}$ and $\mathcal{F}'$ are natural transformations $\eta: \mathcal{F} \Rightarrow \mathcal{F}'$. Concretely, this means that $\eta$ assigns a linear map $\varphi := \eta(\ast) \in \text{Hom}(V,V')$ to the single object $\ast$ of $\widehat{G}$ such that the diagrams
\vspace{-2pt}
\begin{equation}
\begin{gathered}
\includegraphics[height=2.1cm]{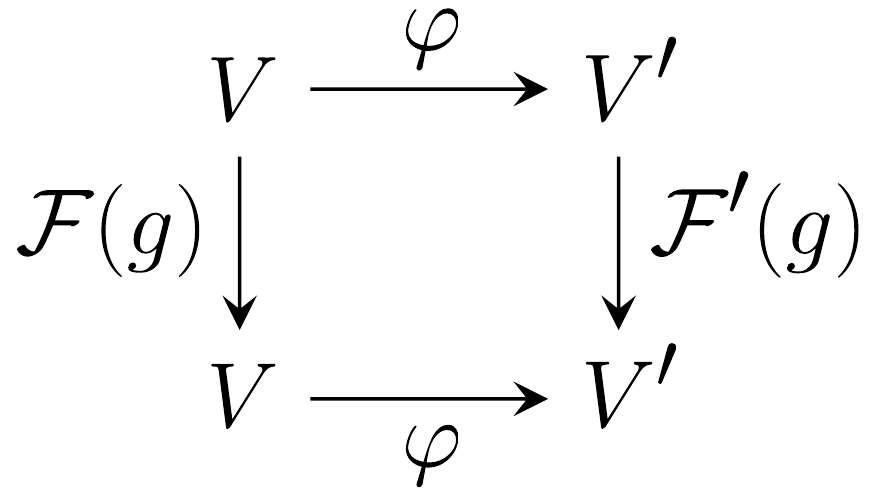}
\end{gathered}
\vspace{-4pt}
\end{equation}
commute for all $g \in G$. Morphisms in $\mathsf{Rep}(G)$ thus correspond to intertwiners between representations in the usual sense.
\end{itemize}

The fusion structure on $\mathsf{Vec}$ induces a fusion structure on $\mathsf{Rep}(G)$. In particular, this allows us to take direct sums and tensor products of representations in the usual sense.

Closely related is the category of projective representations of $G$, in which the diagram (\ref{eq-rep-morphism}) only commutes up to multiplicative phases $c(g,h) \in U(1)$. Pictorially, we denote this by adding a double arrow
\vspace{-1pt}
\begin{equation}
\begin{gathered}
\includegraphics[height=2.3cm]{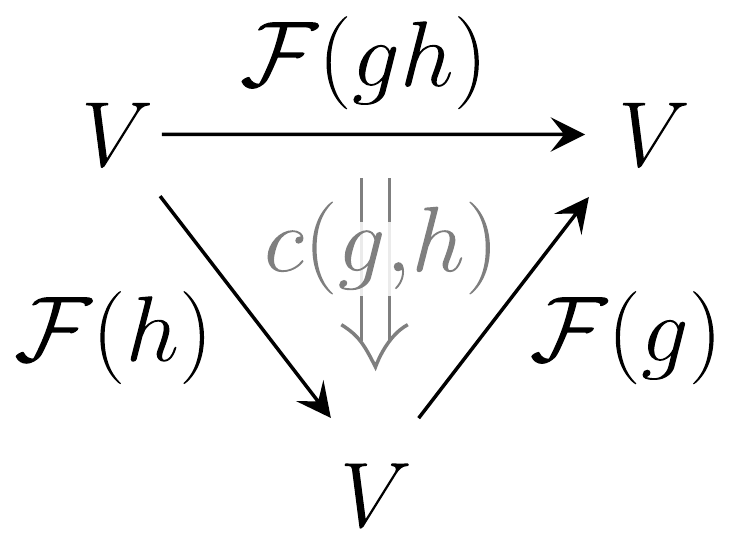}
\end{gathered}
\vspace{-4pt}
\end{equation}
to the diagram in (\ref{eq-rep-morphism}). These phases need to be compatible with the associativity of morphism composition in $\mathsf{Vec}$ in the sense that the diagram
\vspace{-1pt}
\begin{equation}
\begin{gathered}
\includegraphics[height=5.4cm]{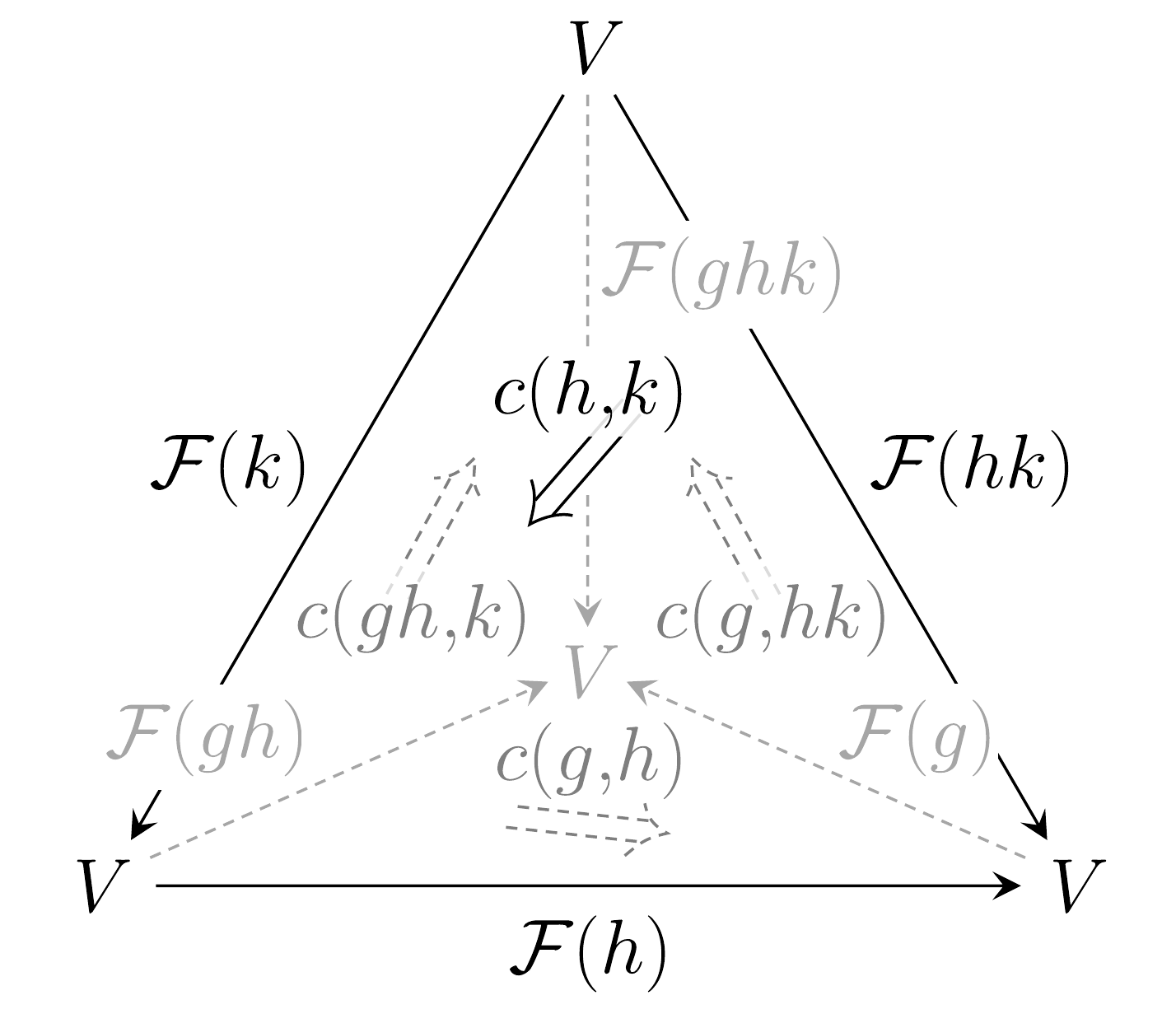}
\end{gathered}
\vspace{-4pt}
\end{equation}
2-commutes, which we can rewrite as the 2-cocycle condition
\begin{equation}
(dc)(g,h,k) \; := \; \frac{c(h,k) \cdot c(g,hk)}{c(gh,k) \cdot c(g,h)} \; \stackrel{!}{=} \; 1 \, .
\end{equation}
The collection of phases can thus be thought of as a 2-cocycle $c \in Z^2(G,U(1))$. We denote the corresponding category of projective representations by $\mathsf{Rep}^c(G)$. The category of ordinary representations of $G$ is recovered by choosing $c=1$.

\subsection{2-representations}
\label{app:2-representations}

In the following, we would like to emulate the discussion of the previous section to define the fusion 2-category of 2-representations of a finite 2-group. We will review the notions of 2-groups and 2-vector spaces and provide a classification of 2-representations in terms of elementary group-theoretical data. 

\subsubsection{2-groups}

In what follows, we will think of a 2-group $\mathcal{G}$ as a quadruple $\mathcal{G}=(G,A,\varphi,\alpha)$ consisting of
\begin{enumerate}
\item a finite group $G$,
\item a finite abelian group $A$,
\item a group action $\varphi: G \to \text{Aut}(A)$,
\item a Postnikov class representative $\alpha \in Z^3_{\varphi}(G,A)$.
\end{enumerate}

Similarly to the case of ordinary groups, we can associate to $\mathcal{G}$ a corresponding 2-category $\widehat{\mathcal{G}}$ that is defined as follows: 
\begin{itemize}
\item Its set of objects contains a single element denoted by $\ast$.

\item The set of 1-endomorphisms of $\ast$ is given by 
\begin{equation}
\text{1-End}_{\widehat{\mathcal{G}}}(\ast) \, = \, G
\end{equation}
with composition given by group multiplication in $G$.

\item The set of 2-morphisms between two 1-morphisms $g,h \in G$ is given by
\begin{equation}
\text{2-Hom}_{\widehat{\mathcal{G}}}(g,h) \; = \; \delta_{g,h} \cdot A
\end{equation}
with vertical composition given by group multiplication in $A$. The horizontal composition of two 2-morphisms $a \in \text{2-End}_{\widehat{\mathcal{G}}}(g)$ and $b \in \text{2-End}_{\widehat{\mathcal{G}}}(h)$ is given by
\begin{equation}
a \star b \; = \; a \cdot \varphi_g(b) \, \in \, \text{2-End}_{\widehat{\mathcal{G}}}(g \! \cdot \! h) \, .
\end{equation}

\item The 2-associator for the composition of 1-morphisms $g,h,k \in \text{1-End}_{\widehat{\mathcal{G}}}(\ast)$ is given by
\begin{equation}
\alpha(g,h,k) \, \in \, \text{2-End}_{\widehat{\mathcal{G}}}(g \! \cdot \! h \! \cdot \! k) \, .
\end{equation}
\end{itemize}
This construction makes it clear that we can alternatively think of 2-groups as special types of 2-categories with a single object, all of whose 1-morphisms and 2-morphisms are invertible.

\subsubsection{2-vector spaces}

Let us now consider the fusion 2-category $\mathsf{2Vec}$ of finite-dimensional 2-vector spaces, which can be described as follows:
\begin{itemize}
\item Its objects are finite-dimensional semi-simple associative algebras\footnote{Over a general field, this would be separable algebras. However, since we are working over $\mathbb{C}$, we can replace separable by semi-simple.}.
\item Its 1-morphisms are finite-dimensional bimodules.
\item Its 2-morphisms are bimodule maps.
\end{itemize}
As a consequence, equivalence between two objects $\mathcal{A}_1 \sim \mathcal{A}_2$ in $2\mathsf{Vec}$ corresponds to Morita equivalence of the corresponding finite-dimensional semi-simple algebras. Since any finite-dimensional semi-simple algebra $\mathcal{A}$ can be decomposed as a direct sum
\begin{equation}
\mathcal{A} \; \cong \; M_{r_1}(\mathbb{C}) \, \oplus \cdots \oplus \, M_{r_n}(\mathbb{C}) 
\end{equation}
of matrix algebras and any matrix algebra $M_r(\mathbb{C})$ is Morita equivalent to $\mathbb{C}$, any finite-dimensional semi-simple algebra is equivalent to a direct sum of $n$ copies of $\mathbb{C}$ in $\mathsf{2Vec}$. The equivalence classes of objects in $\mathsf{2Vec}$ are thus in 1-1 correspondence with natural numbers $n \in \mathbb{N}$, which reproduces the following description of the 2-category $\mathsf{2Vec}$ due to Kapranov and Voevodsky~\cite{bartlett2015modular}:
\begin{itemize}
\item Its objects are positive integers $n \in \mathbb{N}$. There is a single simple object corresponding to the positive integer $1$. 

\item Its 1-morphisms between objects $n$ and $m$ are given by $(m \times n)$-matrices
\vspace{-1pt}
\begin{equation}
\begin{gathered}
\includegraphics[height=0.63cm]{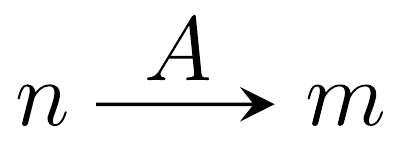}
\end{gathered}
\vspace{-4pt}
\end{equation} 
whose entries are finite-dimensional vector spaces. The composition of 1-morphisms
\vspace{-1pt}
\begin{equation}
\begin{gathered}
\includegraphics[height=0.63cm]{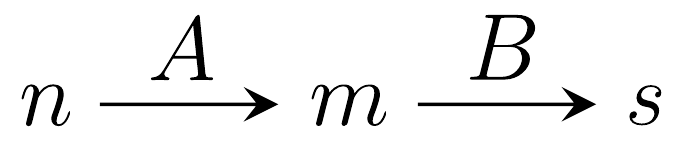}
\end{gathered}
\vspace{-4pt}
\end{equation}
is given by the multiplication of the corresponding matrices using tensor products and direct sums of vector spaces.

\item Its 2-morphisms between 1-morphisms $A$ and $B$ are given by $(m \times n)$-matrices
\vspace{-2pt}
\begin{equation}
\begin{gathered}
\includegraphics[height=2cm]{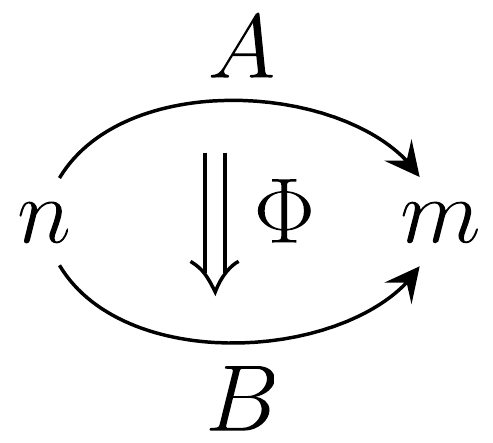}
\end{gathered}
\vspace{-4pt}
\end{equation}
whose entries are linear maps between the corresponding entries of $A$ and $B$. The vertical composition of 2-morphisms
\vspace{-4pt}
\begin{equation}
\begin{gathered}
\includegraphics[height=2.65cm]{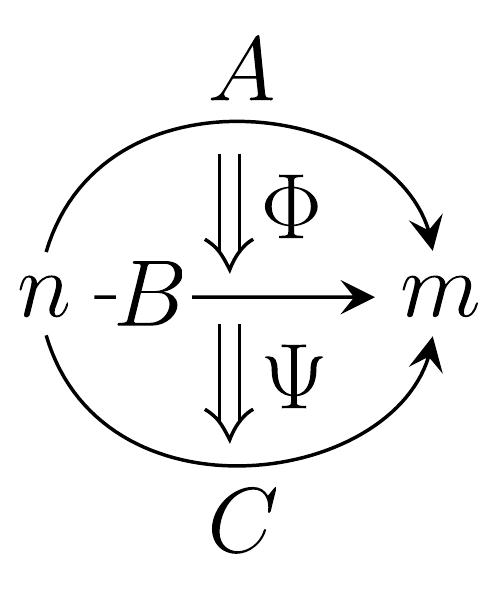}
\end{gathered}
\vspace{-6pt}
\end{equation}
is given by entry-wise composition of linear maps. The horizontal composition of 2-morphisms 
\vspace{-2pt}
\begin{equation}
\begin{gathered}
\includegraphics[height=2cm]{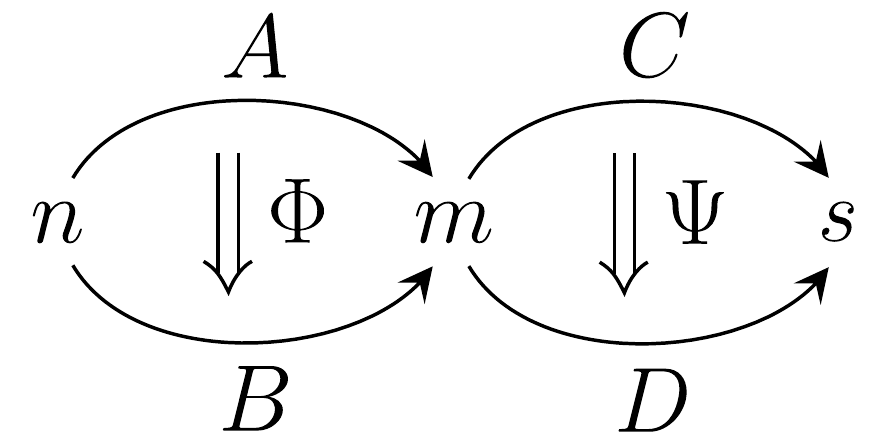}
\end{gathered}
\vspace{-4pt}
\end{equation}
is given by the multiplication of the corresponding matrices using tensor products and direct sums of linear maps.

\item The fusion of objects, 1-morphisms and 2-morphisms
\begin{equation}
\begin{gathered}
\includegraphics[height=2.25cm]{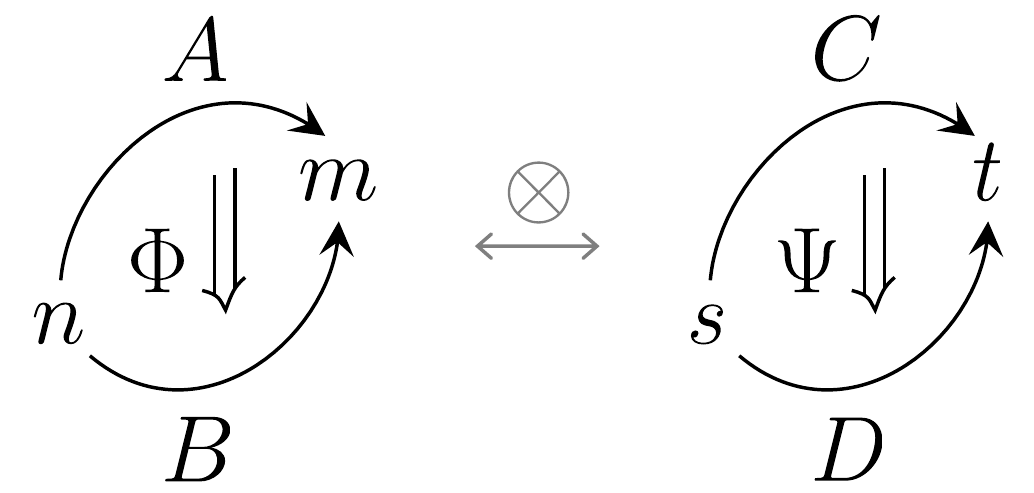}
\end{gathered}
\vspace{-2pt}
\end{equation}
is given by the tensor product of the corresponding matrices using tensor products and direct sums of vector spaces and linear maps, respectively.
\end{itemize}

\subsubsection{2-representations \& their classification}

In analogy to the case of ordinary groups, it is now straightforward to define the 2-category $\mathsf{2Rep}(\mathcal{G})$ of finite-dimensional 2-representations of the 2-group $\mathcal{G}$:
\begin{itemize}
\item Its objects are (pseudo-)-2-functors $\mathcal{F}: \widehat{\mathcal{G}} \to \mathsf{2Vec}$.
\item Its 1-morphisms are natural transformations $\eta: \mathcal{F} \Rightarrow \mathcal{F}'$.
\item Its 2-morphisms are modifications $\Xi: \eta \Rrightarrow \eta'$.
\end{itemize}

This rather abstract definition can be reformulated in terms of more elementary data by breaking down the definitions of (pseudo-)-2-functors, natural transformations and modifications. For instance, consider a fixed 2-representation $\mathcal{F}: \widehat{\mathcal{G}} \to \mathsf{2Vec}$. This then amounts to the following:
\begin{itemize}
\item To the single object $\ast$ of $\widehat{\mathcal{G}}$ the 2-functor $\mathcal{F}$ assigns a natural number $n := \mathcal{F}(\ast) \in \mathbb{N}$ which we call the dimension of the 2-representation.

\item To the 1-endomorphisms $g \in G$ of $\ast$ the 2-functor $\mathcal{F}$ assigns an invertible $(n \times n)$-matrix $\mathcal{F}(g)$ whose entries are finite dimensional complex vector spaces. Each row and column in $\mathcal{F}(g)$ can thus only contain one non-vanishing entry isomorphic to the one-dimensional vector space $\mathbb{C}$, so that up to isomorphism we can think of $\mathcal{F}(g)$ as a $(n \times n)$-permutation matrix. The 2-functor $\mathcal{F}$ thus induces a permutation action 
\begin{equation}
\sigma: \; G \, \to \, S_n
\end{equation}
of the group $G$ on the set of $n$ elements.

\item To the 2-endomorphisms $a \in A$ of a 1-morphism $g \in G$ the 2-functor $\mathcal{F}$ assigns a $(n \times n)$-matrix whose entries are linear maps between the corresponding entries of $\mathcal{F}(g)$. Since the latter only contains one non-vanishing entry isomorphic to $\mathbb{C}$ in each row and column, we can view $\mathcal{F}(a)$ as a collection of $n$ phases $\chi_i(a) \in U(1)$. The 2-functor $\mathcal{F}$ thus induces a homomorphism
\begin{equation}
\chi: \; A \, \to \, U(1)^n \, ,
\end{equation}
which we can think of as a collection of $n$ characters $\chi \in (A^{\vee})^n$. 2-functoriality of $\mathcal{F}$ then implies that this collection satisfies
\begin{equation}
g \triangleright_{\sigma} \chi(a) \; = \; \chi(g \triangleright_{\varphi} a)
\end{equation}
for all $g \in G$ and $a \in A$, where we denoted by $\triangleright_{\sigma}$ and $\triangleright_{\varphi}$ the action of $G$ on $U(1)^n$ and $A$ induced by $\sigma$ and $\varphi$, respectively.

\item For each pair $g,h \in G$ we have a 2-isomorphism
\vspace{-1pt}
\begin{equation}
\begin{gathered}
\includegraphics[height=2.25cm]{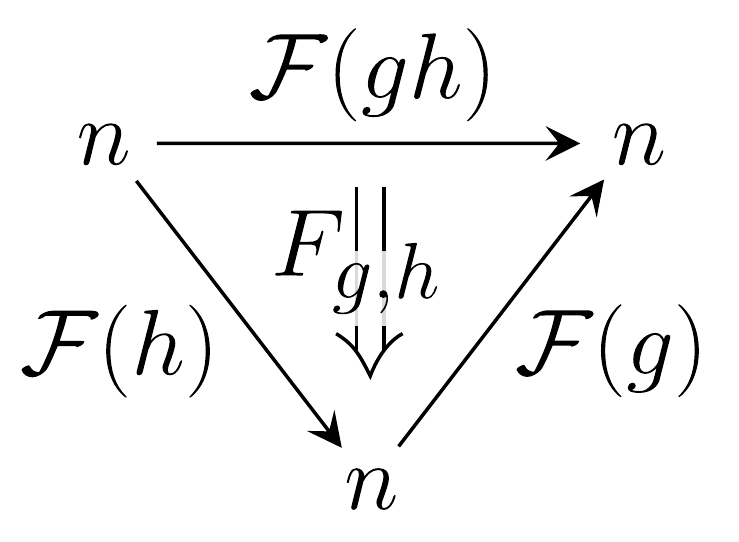}
\end{gathered}
\vspace{-4pt}
\end{equation}
controlling the composition of $\mathcal{F}(g)$ and $\mathcal{F}(h)$. As such, they comprise linear maps between the $n$ one-dimensional entries of $F(gh)$ and $F(g) \circ F(h)$, so that we can view $F_{g,h}$ as a collection of $n$ phases $c(g,h) \in U(1)^n$. These phases need to be compatible with the associativity constraints on 1-morphism composition in $\mathsf{2Vec}$ in the sense that the diagram
\vspace{-1pt}
\begin{equation}
\begin{gathered}
\includegraphics[height=5.4cm]{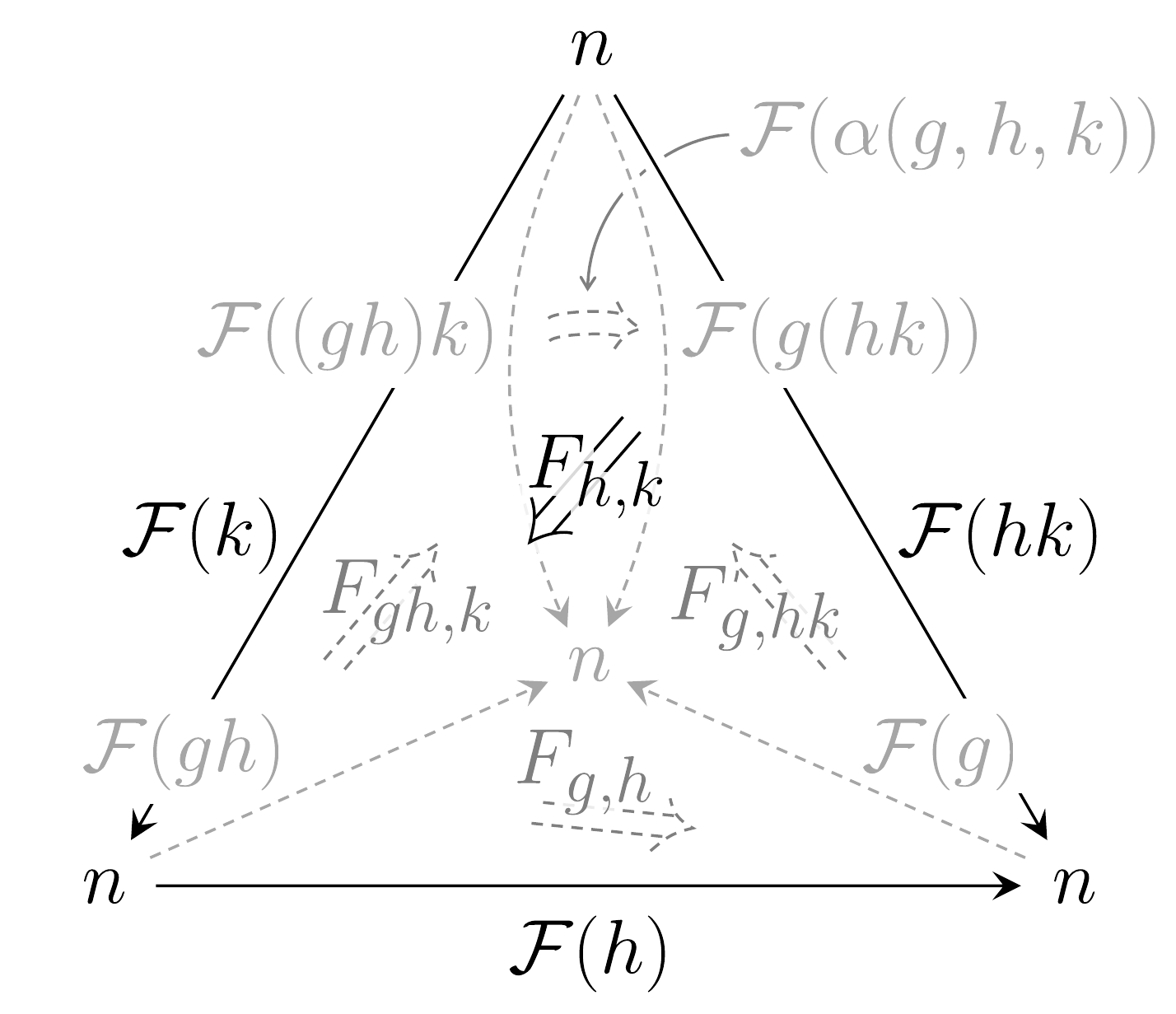}
\end{gathered}
\vspace{-4pt}
\end{equation}
2-commutes, which we can rewrite as the 2-cochain condition
\begin{equation}
(d_{\sigma}c)(g,h,k) \; = \; \frac{( g \triangleright_{\sigma} c(h,k) ) \cdot c(g,hk)}{c(gh,k) \cdot c(g,h)} \; \stackrel{!}{=} \; \braket{\chi, \alpha(g,h,k)} \, .
\end{equation}
The collection of phases can thus be thought of as a 2-cochain $c \in C^2(G,U(1)^n)$ that closes to the pairing $\braket{\chi,\alpha} \in Z^3(G,U(1)^n)$ of $\chi$ with the Postnikov representative $\alpha$.

\end{itemize}

In summary, we can think of 2-representations of the 2-group $\mathcal{G}$ as being labelled by quadruples $(n,\sigma,c,\chi)$ consisting of
\begin{enumerate}
\item a positive integer $n \in \mathbb{N}$,
\item a permutation action $\sigma: G \to S_n$,
\item a twisted 2-cochain $c \in C^2_{\sigma}(G,U(1)^n)$,
\item a collection of characters $\chi \in (A^{\vee})^n$,
\end{enumerate}
such that $g \triangleright_{\sigma}\chi(a) = \chi(g \triangleright_{\varphi} a)$ for all $g \in G$ and $a \in A$ and $d_{\sigma}c = \braket{\chi,\alpha}$. Two 2-representations labelled by quadruples $(n,\sigma,c,\chi)$ and $(n',\sigma',c',\chi')$ are considered equivalent if $n = n'$ and there exists a permutation $\tau \in S_n$ such that
\begin{equation}
\sigma' \, = \, \tau \circ \sigma \circ \tau^{-1} \, , \qquad [c'] \, = \, [\tau \triangleright c] \, , \qquad \chi' \, = \, \tau \triangleright \chi \, .
\end{equation}

\subsubsection{Direct sum and tensor product}

The fusion structure on $\mathsf{2Vec}$ induces a fusion structure on $\mathsf{2Rep}(G)$. In particular, this allows us to take direct sums and tensor products of 2-representations. In terms of the above labelling of 2-representations by elementary data, these can be described explicitly as follows: Consider two 2-representations $R=(n,\sigma,c,\chi)$ and $R'=(n',\sigma',c',\chi')$ of $\mathcal{G}$. Then, the following holds:
\begin{itemize}
\item Their direct sum is the 2-representation
\begin{equation}
R \oplus R' \; = \; \big(n + n', \, \sigma \oplus \sigma', \, c \oplus c', \, \chi \oplus \chi' \big) \, ,
\end{equation}
where the permutation action $\sigma \oplus \sigma': G \to S_{n + n'}$ is defined by
\begin{equation} 
(\sigma \oplus \sigma')_g(i) \;\; := \;\; 
\begin{cases}
\sigma_g(i) & \text{if} \;\; 1 \leq i \leq n \\
\sigma'_g(i-n) + n & \text{if} \;\; n+1 \leq i \leq n+n' 
\end{cases} \, ,
\end{equation}
the twisted 2-cochain $c\oplus c' \in C^2_{\sigma \oplus \sigma'}(G,U(1)^{n+n'})$ is given by
\begin{equation}
(c \oplus c')_i(g,h) \; := \; 
\begin{cases}
c_i(g,h) & \text{if} \;\; 1 \leq i \leq n \\
c'_{i-n}(g,h) & \text{if} \;\; n+1 \leq i \leq n+n' 
\end{cases} \, ,
\end{equation}
and the collection of characters $\chi \oplus \chi' \in (A^{\vee})^{n+n'}$ is taken to be
\begin{equation}
(\chi \oplus \chi')_i \; := \; 
\begin{cases}
\chi_i & \text{if} \;\; 1 \leq i \leq n \\
\chi'_{i-n} & \text{if} \;\; n+1 \leq i \leq n+n' 
\end{cases} \, .
\end{equation}

\item Their tensor product is the 2-representation
\begin{equation}
R \otimes R' \; = \; \big(n \cdot n', \, \sigma \otimes \sigma', \, c \otimes c', \, \chi \otimes \chi' \big) \, ,
\end{equation}
where the permutation action $\sigma \otimes \sigma': G \to S_{n\cdot n'}$ is defined by
\begin{equation}
(\sigma \otimes \sigma')_g(i,j) \; := \; \big(\sigma_g(i),\sigma'_g(j)\big) \, ,
\end{equation}
the twisted 2-cochain $c\otimes c' \in C^2_{\sigma \otimes \sigma'}(G,U(1)^{n \cdot n'})$ is given by 
\begin{equation}
(c \otimes c')_{(i,j)}(g,h) \; := \; c_i(g,h) \cdot c'_j(g,h) \, ,
\end{equation}
and the collection of characters $\chi \otimes \chi' \in (A^{\vee})^{n \cdot n'}$ is taken to be
\begin{equation}
(\chi \otimes \chi')_{(i,j)} \, := \, \chi_i \cdot \chi'_j \, .
\end{equation}
\end{itemize}

\subsubsection{Irreducibility}

The irreducible 2-representations are those which cannot be written as a direct sum of other 2-representations. In particular, this means that the permutation action $\sigma: G \to S_n$ of an irreducible 2-representation $S = (n,\sigma,c,\chi)$ is transitive, so that
\begin{equation}
\lbrace 1,...,n \rbrace \; \cong \; G / H
\end{equation}
as $G$-sets, where we denoted by $H := \text{Stab}_{\sigma}(1) \subset G$ the stabiliser of the permutation action. The twisted 2-cochain $c \in C^2_{\sigma}(G,U(1)^n)$ then induces an ordinary 2-cochain
\begin{equation}
u \; := \; c_1|_H \; \in \; C^2(H,U(1))
\end{equation}
on the subgroup $H$, which closes to $du = \braket{\lambda,\alpha|_H}$, where we denoted $\lambda := \chi_1 \in A^{\vee}$. In summary, we can label the irreducible 2-representations of $\mathcal{G}$ by triples $(H,u,\lambda)$ consisting of
\begin{enumerate}
\item a subgroup $H \subset G$,
\item a cochain $u \in C^2(H,U(1))$,
\item a $H$-invariant character $\lambda \in A^{\vee}$,
\end{enumerate}
such that $du = \braket{\lambda,\alpha|_H}$. Two irreducible 2-representations labelled by triples $(H,u,\lambda)$ and $(H',u',\lambda')$ are equivalent if there exists a group element $g \in G$ such that
\begin{equation}
H' \, = \, {}^{g\!}H \, , \qquad [u'] \, = \, [{}^gu] \, , \qquad \lambda' \, = \, {}^{g\hspace{-0.5pt}} \lambda \, ,
\end{equation}
where the conjugation action of $g$ on the entries of the triple $(H,u,\lambda)$ is defined by
\begin{equation}
{}^{g\!}H \; = \; g H g^{-1} \, , \quad\; ({}^gu)(h,h') \; = \; u(g^{-1}hg, g^{-1}h'g) \, , \quad\; ({}^g\lambda)(a) \, = \, \lambda(g^{-1} \triangleright_{\rho} a) \, .
\end{equation}

The irreducible 2-representations can be thought of as elementary building blocks in the sense that any generic 2-representation of $\mathcal{G}$ can be written as a direct sum of irreducible ones. In particular, the fusion of two irreducible 2-representations labelled by triples $(H,u,\lambda)$ and $(K,v,\eta)$ must again decompose as a direct sum of other irreducible 2-representations, which is reflected in Mackey's decomposition formula
\begin{equation}
(H,u,\lambda) \, \otimes \, (K,v,\eta) \;\; = \bigoplus_{[g] \, \in \, H \backslash G / K} \big( H \cap {}^{g\!}K, \; u \cdot {}^gv, \; \lambda \cdot {}^g\eta \big) \, .
\end{equation}
Similarly, the 1-morphism categories between irreducible 2-representations allow for a Mackey-type decomposition and are given by
\begin{equation}
\text{1-Hom}\big( (H,u,\lambda), \,(K,v,\eta) \big) \;\; = \bigoplus_{\substack{[g] \, \in \, H \backslash G / K \\[2pt] \overline{\lambda} \, \cdot \, {}^{g\hspace{-0.5pt}}\eta \; = \; 1}} \mathsf{Rep}^{\, \overline{u} \, \cdot \, {}^{g\hspace{-0.5pt}}v}(H \cap {}^{g\!}K) \, .
\end{equation}

\subsubsection{Example}

\begin{figure}[h]
	\centering
	\includegraphics[height=5.7cm]{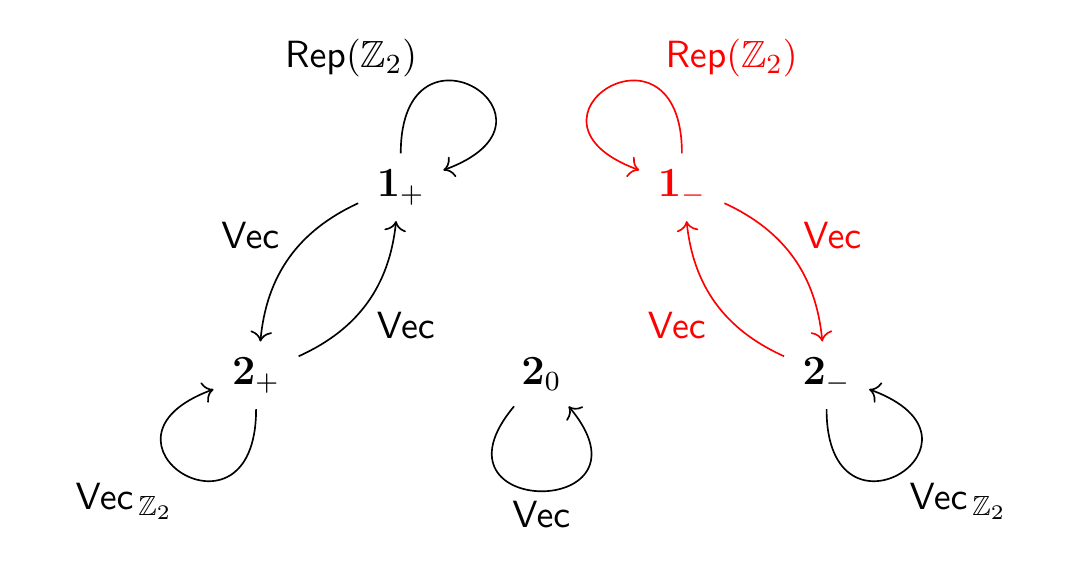}
	\caption{}
	\label{fig:example-2-rep}
\end{figure}

As an example, consider the 2-group $\mathcal{G} = (G,A,\varphi,\alpha)$ where
\begin{itemize}
\item $G= \mathbb{Z}_2$ is the cyclic group of order 2 with generator $x$,
\item $A= \mathbb{Z}_4$ is the cyclic group of order 4 with generator $y$,
\item $\varphi: G \to \text{Aut}(A)$ is defined on the generators by $\varphi_x(y) = y^3$,
\item $[\alpha] \in H^3_{\varphi}(G,A) = \mathbb{Z}_2$ is represented by a normalised 3-cocycle satisfying
\begin{equation}
\alpha(x,x,x) \, = \, 1 \qquad \text{or} \qquad \alpha(x,x,x) \, = \, y
\end{equation}
corresponding to a split or a non-split 2-group.
\end{itemize}
The irreducible 2-representations of $\mathcal{G}$ are then labelled by triples $(H,u,\lambda)$ as before, and can be summarised as follows:
\begin{itemize}
\item In the split case, there is no non-trivial choice of 2-cocycle $u$ since $H^2_{\varphi}(G,U(1)) = 1$. Up to equivalence, there are five irreducible 2-representations labelled by 
\begin{equation}
\renewcommand{\arraystretch}{1.8}
\setlength{\tabcolsep}{10pt}
\begin{tabular}{c | c c c}
& $H$ & $u$ & $\lambda$  \\
\hline
$\mathbf{1}_+$ & $\mathbb{Z}_2$ & $1$ & 1\\
$\mathbf{1}_-$ & $\mathbb{Z}_2$ & $1$ & $\widehat{y}^2$ \\
$\mathbf{2}_+$ & $\lbrace 1 \rbrace$ & $1$ & $1$ \\
$\mathbf{2}_0$ & $\lbrace 1 \rbrace$ & $1$ & $\widehat{y}$ \\
$\mathbf{2}_-$ & $\lbrace 1 \rbrace$ & $1$ & $\widehat{y}^2$
\end{tabular}_{\strut^{\scalebox{1}{,}}}
\end{equation}
where we denoted by $\widehat{y}$ the generator of the Pontryagin dual $A^{\vee} = \mathbb{Z}_4$ of $A$. Their fusion rules are summarised by the fusion table
\begin{equation}
\renewcommand{\arraystretch}{1.8}
\setlength{\tabcolsep}{10pt}
\begin{tabular}{c | c  c  c  c  c}
$\otimes$ & $\mathbf{1}_+$ & $\mathbf{1}_-$ & $\mathbf{2}_+$ & $\mathbf{2}_0$ & $\mathbf{2}_-$ \\
\hline
$\mathbf{1}_+$ & $\mathbf{1}_+$ & $\mathbf{1}_-$ & $\mathbf{2}_+$ & $\mathbf{2}_0$ & $\mathbf{2}_-$ \\
$\mathbf{1}_-$ & $\mathbf{1}_-$ & $\mathbf{1}_+$ & $\mathbf{2}_-$ & $\mathbf{2}_0$ & $\mathbf{2}_+$ \\
$\mathbf{2}_+$ & $\mathbf{2}_+$ & $\mathbf{2}_-$ & $\mathbf{2}_+ \oplus \mathbf{2}_+$ & $\mathbf{2}_0 \oplus \mathbf{2}_0$ & $\mathbf{2}_- \oplus \mathbf{2}_-$ \\
$\mathbf{2}_0$ & $\mathbf{2}_0$ & $\mathbf{2}_0$ & $\mathbf{2}_0 \oplus \mathbf{2}_0$ & $\mathbf{2}_+ \oplus \mathbf{2}_-$ & $\mathbf{2}_0 \oplus \mathbf{2}_0$ \\
$\mathbf{2}_-$ & $\mathbf{2}_-$ & $\mathbf{2}_+$ & $\mathbf{2}_- \oplus \mathbf{2}_-$ & $\mathbf{2}_0 \oplus \mathbf{2}_0$ & $\mathbf{2}_+ \oplus \mathbf{2}_+$
\end{tabular}_{\strut^{\scalebox{1}{.}}}
\end{equation}
The 1-morphism spaces between simple objects are illustrated in figure \ref{fig:example-2-rep}.
	
\item In the non-split case, the condition $du = \braket{\lambda, \alpha|_H}$ is non-trivial only when $H = \mathbb{Z}_2$ and $\lambda = \widehat{y}^2$. However, since no such 2-cochain $u$ exists, the corresponding 2-representation $\mathbf{1}_-$ no longer exists. This is indicated by the red colouring of the $\mathbf{1}_-$ and its attached morphism spaces in figure \ref{fig:example-2-rep}.
\end{itemize}

\subsection{3-representations}
\label{app:3-representations}

In the following, we would like to define a notion of 3-representations that is analogous to the construction of 1- and 2-representations discussed in the previous subsections. Since the associated realm of fusion 3-categories is much richer and more intricate than before, we will not try to be systematic but only briefly highlight some of the most salient features. In general, one may define the notion of a 3-representations for a generic 3-group, but for simplicity we will restrict our attention to the study of 3-representations of ordinary finite groups $G$ (regarded as a special type of 3-group).

\subsubsection{3-vector spaces}

Let us first consider the fusion 3-category $\mathsf{3Vec}$ of finite-dimensional 3-vector spaces, which can be described as follows: 
\begin{itemize}
    \item Its objects are multi-fusion categories.
    \item Its 1-morphisms are bimodule categories.
    \item Its 2-morphisms are bimodule functors.
    \item Its 3-morphisms are natural transformations.
\end{itemize}
As a consequence, equivalence of two objects $\mathsf{B}_1 \sim \mathsf{B}_2$ in $\mathsf{3Vec}$ corresponds to Morita equivalence of the corresponding multi-fusion categories. Since any multi-fusion category can be written as a direct sum of indecomposable ones and any indecomposable multi-fusion category is Morita equivalent to an ordinary fusion category, equivalence classes of objects in $\mathsf{3Vec}$ are in 1:1-correspondence with finite direct sums of fusion categories. In particular, there is certainly more than one simple object up to equivalence, which shows that objects in $\mathsf{3Vec}$ are now significantly richer than before.

\subsubsection{3-representations \& their classification}

Analogously to the cases of 1- and 2-representations, we can define the 3-category $\mathsf{3Rep}(G)$ of finite-dimensional 3-representations of an ordinary group $G$ as follows:
\begin{itemize}
    \item Its objects are (pseudo-)3-functors $\mathcal{F}: \widehat{G} \to \mathsf{3Vec}$.
    \item Its 1-morphisms are natural transformations $\eta: \mathcal{F} \Rightarrow \mathcal{F}'$.
    \item Its 2-morphisms are modifications $\Xi: \eta \Rrightarrow \eta'$.
    \item Its 3-morphisms are perturbations $\mathfrak{X}: \Xi \,\mathrel{\substack{\textstyle\Rightarrow\\[-0.5ex] \textstyle\Rightarrow}} \,\Xi'$.
\end{itemize}
Here, we regard the group $G$ as a special type of 3-category $\widehat{G}$ with a single object $\ast$, whose 1-endomorphisms are given by $\text{1-End}_{\widehat{G}}(\ast) = G$ and whose 2- and 3-morphisms are trivial.

This rather abstract construction of 3-representations can be reformulated in terms of more elementary data by breaking down the definitions of (pseudo-)3-functors, natural transformations, modification and perturbations. For instance, consider a fixed 2-representation $\mathcal{F}: \widehat{G} \to \mathsf{3Vec}$, which amounts to the following data:
\begin{itemize}
\item To the single object $\ast$ of $\widehat{G}$ the 3-functor $\mathcal{F}$ assigns a multifusion category $\mathsf{B}$. Up to equivalence, we can think of the latter as a direct sum $\mathsf{B} = \mathsf{C}_1 \oplus ... \oplus \mathsf{C}_n$ of ordinary fusion categories $\mathsf{C}_i$, whose number $n$ is called the dimension of the 3-representation. For simplicity, let us assume that the equivalence class of $\mathsf{B}$ is simply labelled by a single fusion category $\mathsf{C}$, corresponding to a one-dimensional 3-representation of $G$.

\item To the 1-endomorphisms $g \in G$ of $\ast$ the 3-functor assigns invertible $\mathsf{C}$-bimodule categories $\mathcal{F}(g)$, which induce a group homomorphism
\begin{equation}
    \rho: \; G \, \to \, \text{BrPic}(\mathsf{C})
\end{equation}
from $G$ into the Brauer-Picard group $\text{BrPic}(\mathsf{C})$ of isomorphism classes of invertible $\mathsf{C}$-bimodules.

\item For each pair $g,h \in G$ we have a 2-isomorphism
\begin{equation}
\begin{gathered}
\includegraphics[height=2.25cm]{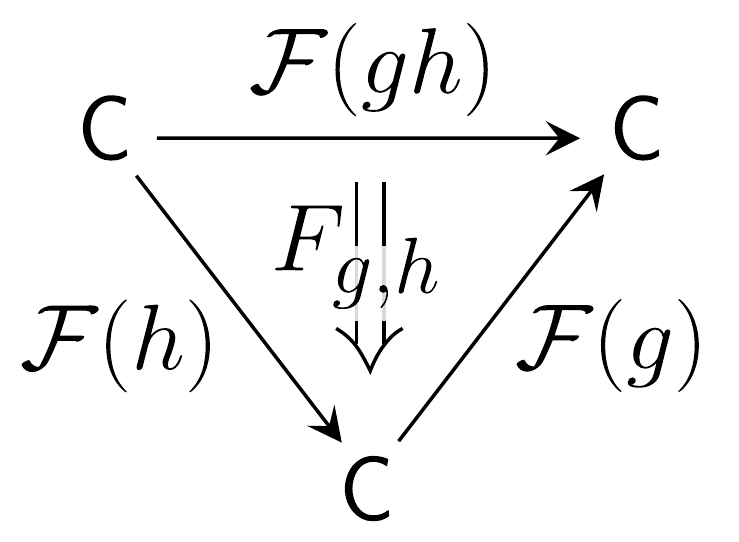}
\end{gathered}
\vspace{-4pt}
\end{equation}
controlling the composition of $\mathcal{F}(g)$ and $\mathcal{F}(h)$. As such, they correspond to invertible $\mathsf{C}$-bimodule functors which, up to equivalence, are given by the left action of a fixed invertible element $\kappa(g,h)$ of the Drinfeld centre $\mathsf{Z(C)}$ of $\mathsf{C}$. The collection of 2-isomorphisms $F_{g,h}$ then induces a 2-cocycle $\kappa \in Z^2(G,\mathsf{Z(C)}^{\times})$, where $\mathsf{Z(C)}^{\times}$ denotes the abelian group of isomorphism classes of invertible objects in $\mathsf{Z(C)}$.

\item For each triple $g,h,k \in G$ we have a 3-isomorphism
\vspace{-1pt}
\begin{equation}
\begin{gathered}
\includegraphics[height=5.4cm]{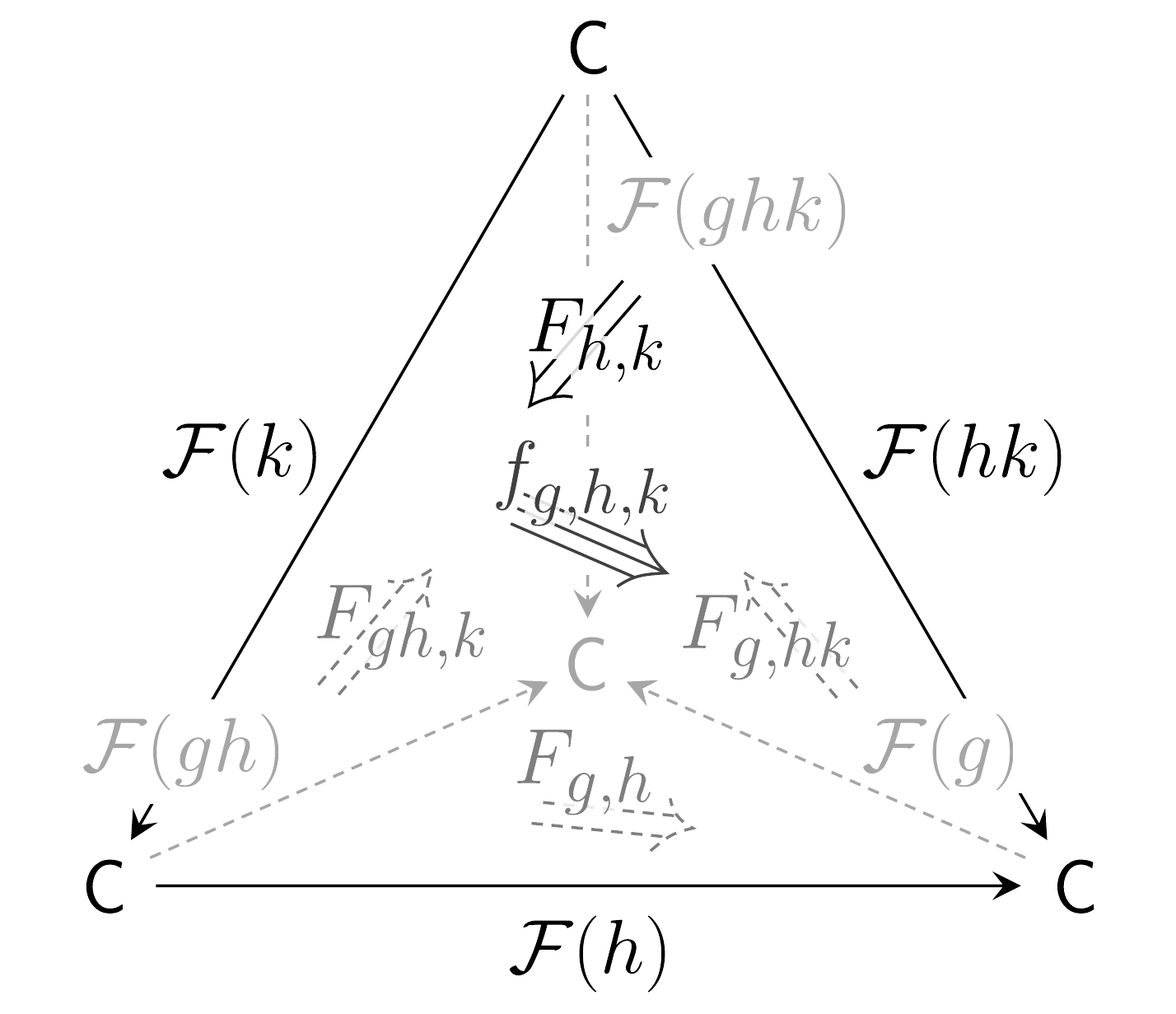}
\end{gathered}
\vspace{-4pt}
\end{equation}
controlling the composition of the 2-isomorphisms $F_{g,h}$. As such, they correspond to natural isomorphisms of $\mathsf{C}$-bimodule functors given by multiplication by a phase $\mu(g,h,k) \in U(1)$. The collection of 3-isomorphisms $f_{g,h,k}$ then induces a 3-cocycle $\mu \in Z^2(G,U(1))$.
\end{itemize}

In summary, we can think of the one-dimensional 3-representations of $G$ as being labelled by quadruples $(\mathsf{C},\rho,\kappa,\mu)$ consisting of
\begin{enumerate}
    \item a fusion category $\mathsf{C}$,
    \item a group homomorphism $\rho: G \to \text{BrPic}(\mathsf{C})$,
    \item a 2-cocycle $\kappa \in Z^2(G,\mathsf{Z(C)}^{\times})$,
    \item a 3-cocycle $\mu \in Z^3(G,U(1))$.
\end{enumerate}
Analogously to the case of 2-representations, a general irreducible $n$-dimensional 3-representation of $G$ on a multifusion category $\mathsf{B} = \mathsf{C}_1 \oplus ... \oplus \mathsf{C}_n$ may be obtained as the induction of a one-dimensional 3-representation of a subgroup $H \subset G$.

\bibliographystyle{JHEP}
\bibliography{gauging}

\end{document}

%% file: preamble.tex
\usepackage{amsfonts}
\usepackage{amscd}
\usepackage{amssymb}
\usepackage{amsmath,bbm}
\usepackage{graphicx}
\usepackage{epsfig}
\usepackage{latexsym}
\usepackage{mathtools}
\usepackage{hyperref}
\usepackage{tikz-cd}
\usepackage[vcentermath]{youngtab}
\def \be  {\begin{equation}}
\def \ee  {\end{equation}}
\def \bea {\begin{equation}\begin{aligned}}
\def \eea {\end{aligned}\end{equation}}
\def \ba  {\begin{eqnarray}}
\def \ea  {\end{eqnarray}}
\def \bb  {\begin{thebibliography}}
\def \eb  {\end{thebibliography}}
\def \lab #1 {\label{#1}}

\newcommand\cA{\mathcal{A}}
\newcommand\cB{\mathcal{B}}

\newcommand\cF{\mathcal{F}}
\newcommand\cG{\mathcal{G}}

\newcommand\cO{\mathcal{O}}

\newcommand\cS{\mathcal{S}}
\newcommand\cT{\mathcal{T}}

\newcommand\cW{\mathcal{W}}

\newcommand\C{\mathbb{C}}

\newcommand\bZ{\mathbb{Z}}
\newcommand\R{\mathbb{R}}

\definecolor{cardinal}{rgb}{0.6,0,0}
\definecolor{darkgreen}{rgb}{0,0.5,0}
\definecolor{golden}{rgb}{0.92, 0.7, 0}
\definecolor{midnight}{rgb}{0, 0, 0.5}
\definecolor{darkblue}{rgb}{0.2, 0, 0.8}